\documentclass[poms, nonblindrev]{informs3} % current default for manuscript submission

\OneAndAHalfSpacedXI
\usepackage{natbib}
 \bibpunct[, ]{(}{)}{,}{a}{}{,}%
 \def\bibfont{\small}%
 \def\bibsep{\smallskipamount}%
 \def\bibhang{24pt}%

\usepackage{mathtools}   % for the := sign, ceil
\DeclarePairedDelimiter{\ceil}{\lceil}{\rceil}

\usepackage{amsmath}
\usepackage{algorithm}
\usepackage[noend]{algpseudocode} % for algorithms 

\usepackage{multirow}
\usepackage{multicol}
\usepackage{dsfont} % indicatriz
\usepackage{bbm} % indicatriz 2
\usepackage{bbold}
\usepackage{verbatim} % bloques de comment
\usepackage{caption}
\usepackage{subcaption} %subtables 
\usepackage{booktabs}
\usepackage{url}
\usepackage{hyperref}

\newcolumntype{C}[1]{>{\centering\arraybackslash}p{#1}}
\newcolumntype{R}[1]{>{\raggedleft\arraybackslash}p{#1}}
\newcolumntype{L}[1]{>{\raggedright\let\newline\\\arraybackslash\hspace{0pt}}m{#1}}
\newcommand\norm[1]{\left\lVert#1\right\rVert}
\TheoremsNumberedThrough     % Preferred (Theorem 1, Lemma 1, Theorem 2)
\ECRepeatTheorems

\EquationsNumberedThrough    % Default: (1), (2), ...
\MANUSCRIPTNO{}

\begin{document}
%%%%%%%%%%%%%%%%

%\setcounter{page}{1}
%\DOI{10.1287/xxxx.0000.0000}%

\RUNAUTHOR{Morales, Sebasti\'an and Thraves, Charles}

\RUNTITLE{Resource Allocation for Political Campaigns}

% \TITLE{Bundling Information Goods of Decreasing Value}
\TITLE{On the Resource Allocation for Political Campaigns% under the Majority and the Electoral College Systems
}

\ARTICLEAUTHORS{%
\AUTHOR{Sebasti\'{a}n Morales}
\AFF{Department of Industrial Engineering, University of Chile, Santiago, Chile, \EMAIL{sebastian.morales.a@uchile.cl}}

\AUTHOR{Charles Thraves}
\AFF{Department of Industrial Engineering, University of Chile, Santiago, Chile, \EMAIL{cthraves@dii.uchile.cl}} %, \URL{}}
} % end of the block

\ABSTRACT{%
% Enter your abstract
In an election campaign, candidates must decide how to optimally allocate their efforts/resources optimally among the regions of a country. As a result, the outcome of the election will depend on the players' strategies and the voters' preferences. In this work, we present a zero-sum game where two candidates decide how to invest a fixed resource in a set of regions, while considering their sizes and biases. We explore the Majority System (MS) as well as the Electoral College (EC) voting systems. We prove equilibrium existence and uniqueness under MS in a deterministic model; in addition, their closed form expressions are provided when fixing the subset of regions and relaxing the non-negative investing constraint. For the stochastic case, we use Monte Carlo simulations to compute the players' payoffs, together with its gradient and hessian. For the EC, given the lack of Equilibrium in pure strategies, we propose an iterative algorithm to find Equilibrium in mixed strategies in a subset of the simplex lattice. We illustrate numerical instances under both election systems, and contrast players' equilibrium strategies. Finally, we show that polarization induces candidates to focus on larger regions with negative biases under MS, whereas candidates concentrate on swing states under EC.
}%

%\KEYWORDS{deterministic inventory theory; infinite linear programming duality;
%  existence of optimal policies; semi-Markov decision process; cyclic schedule}

 \HISTORY{}

\maketitle
%%%%%%%%%%%%%%%%%%%%%%%%%%%%%%%%%%%%%%%%%%%%%%%%%%%%%%%%%%%%%%%%%%%%%%

% Samples of sectioning (and labeling) in MNSC
% NOTE: (1) \section and \subsection do NOT end with a period
%       (2) \subsubsection and lower need end punctuation
%       (3) capitalization is as shown (title style).
%
%\section{Introduction.}\label{intro} %%1.
%\subsection{Duality and the Classical EOQ Problem.}\label{class-EOQ} %% 1.1.
%\subsection{Outline.}\label{outline1} %% 1.2.
%\subsubsection{Cyclic Schedules for the General Deterministic SMDP.}
%  \label{cyclic-schedules} %% 1.2.1
%\section{Problem Description.}\label{problemdescription} %% 2.

% Text of your paper here

\section{Introduction}

\subsection{Motivation}

Democratic election is the most prevalent mechanism for choosing a country's leader all over the world at this time. As a result of several events that have taken place during the 20\textsuperscript{th} and beginning of the 21\textsuperscript{th} century, there are now, as for 2020, more than $55\%$ of the world's countries that rely on a ballot system to choose their authorities (\cite{owid}). 

Before the election day, candidates who run for president (or similar positions, such as prime minister) hold an election campaign. During this, multiple events take place in different regions of the country where candidates promote their ideas and promises to the potential voters of the regions visited. Among the multiple challenges that arise in this context from a candidate's perspective, this paper focuses on the following research question: \textbf{Given a resource with a fixed budget, for example, time, what is the optimal allocation of this resource among the various regions of the country?} The motivation to focus on regions or parts of a country such as states, is: (i) in-person electoral campaign events can take place in, at most, one regions at a given time, (ii) regions are usually characterized by populations that share common traits, such as the their political preference. Indeed, inhabitants from a particular region might have a very different political position compared to voters from another region, and (iii) the number of potential voters (or electoral votes depending on the case) varies from region to region. 

Given the heterogeneity of the different regions, some of them are going to be more attractive to invest in than others. For example, although California is the state with the most electoral votes in the US electoral system, candidates usually prefer to focus their campaign efforts on other places (see~\cite{Eventos_US_2016}). The reason is probably that the existing political preference biases of CA residents leave little room for improving the chances of winning for either candidate compared to those in other states. On the contrary, the so-called \textit{swing states} are known to be the ones that are key for winning the election. \textit{Swing states} are characterized by having a significant population that is undecided, or at least, to some degree can be convinced to vote for either one candidate or the other.

As expected, the electoral outcome in each region or state will depend on the level of effort, (i.e. the resource investment) of the candidates. Therefore, it is reasonable to think that the turnout for a candidate in a region will increase as the more effort she invest, while it will decrease the more effort her contenders make. Thus, the candidates' resource allocation problem is modeled as a zero-sum game. For simplicity, we present a game theory formulation of the setting described above for two candidates, under the Majority System, and the Electoral College systems. The aim of this work is not only the modelling and resolution of each election system, but also analyzing the contrasts between the equilibrium strategies obtained in both. For example, how does polarization affect candidates' resource allocations? Also, what is the impact of voter uncertainty in the two systems? Under what circumstances do \textit{swing states} become attractive to candidates? We believe there are several such interesting questions that can be answered by using mathematical models able to capture the problem structure in order to analyze candidates' actions.

Although in reality an election outcome is a result of multiple factors, we provide a simplified model that is still capable of providing insightful results able to resemble candidates' decisions observed in reality. 
Some of the main challenges are: (i) create a modelling framework with the complexity that enables the representation of the agents' actions and payoffs of the setting, while also being tractable to solve, and (ii) the model resolution under the different cases which involves for instance the estimation of complex mathematical expressions, or the computation of mixed equilibria. We analyze the case of the two following election systems:
\begin{itemize}
    \item {\bf MS} ({\it Majority System}): The candidate with the most votes wins the election.
    \item {\bf EC} ({\it Electoral College}): Each region has a number of electoral votes. On each region, the candidate with the majority of votes wins all the electoral votes of the region. The candidate with more electoral votes wins the election. This is the electoral system in the US.
\end{itemize}
It is worth to clarify that in the actual EC system used in the US, in some states the Electors are free to vote their own choice (not necessarily matching the majority of the popular vote of the respective state). In 33 states the electors are obligated to vote for the popular vote winner candidate.

For ease of exposition, we present the strategy allocation problem from an election campaign setting. However, it is important to note that there are several other applications which share a similar game theoretic framework. For example, firms that compete for a market share within a set of localities, power control games in wireless network, and resource allocations in a battlefield.

\subsection{Contributions and Structure of the Paper}
The main contributions of this work can be summarized as the following three: 
\begin{enumerate}
    \item \textbf{Modelling:} The development of a game theory modelling framework able to capture candidates' resource allocation decisions, considering biases and abstention, under a Majority System and an Electoral College system, showing equilibrium existence and uniqueness for some particular cases under the Majority System, and developing closed form solutions for certain settings.
    \item \textbf{Algorithms:} The development of solution methods able to find the game equilibrium using Monte Carlo simulations to compute multidimensional integrals and its respective derivatives.
    \item \textbf{Numerics:} Solve numerical experiments providing insight into what equilibria arise under different settings. In addition, contrast the impact of voters' uncertainty as well as polarization in the candidates' strategy and the election outcome.%Bringing insights on candidates' equilibrium strategies in both election systems. 
    %In particular, we show how candidates' strategy profiles look for both electoral systems. In addition, we show that when polarization is: (i) low, candidates invest in most states in both MS and EC; (ii) medium, candidates focus mainly on states with larger populations in both MS and EC; (iii) high, candidates focus on large regions with a pre-existing negative bias under MS, whereas under EC they focus most of their efforts on \textit{swing states}.
    
    %1) Modeling: game theoretic models of how two candidates/parties/politicians compete, under two different electoral systems, and results characterizing the existence and uniqueness of equilibria.
%2) Algorithms: numerical algorithms for efficiently computing these equilibria.
%3) Numerics: numerical experiments providing insight into what equilibria arise under different settings.

\end{enumerate}
 %First, the development of a game theory modelling framework able to capture candidates' resource allocation decisions under a Majority System and an Electoral College system, showing equilibrium existence and uniqueness for some particular cases under the Majority System, and developing closed form solutions for certain settings. Second, the development of solution methods able to find the game equilibrium using Monte Carlo simulations to compute multidimensional integrals and its respective derivatives. Third, bringing insights on candidates' equilibrium strategies in both election systems. In particular, we show how candidates' strategy profiles look for both electoral systems. In addition, we show that when polarization is: (i) low, candidates invest in most states in both MS and EC; (ii) medium, candidates focus mainly on states with larger populations in both MS and EC; (iii) high, candidates focus on large regions with a pre-existing negative bias under MS, whereas under EC they focus most of their efforts on \textit{swing states}.

The paper structure is described as follows: The literature review is given in Section~\ref{sec:litreview}. In Section~\ref{sec:ms} the electoral model under MS is presented, followed by EC in Section~\ref{sec:ec}. The the numerical computations are shown in Section~\ref{sec:numresults}. Finally, conclusions and future work are given in Section~\ref{sec:conclusions}. All proofs are in the Appendix.

\section{Literature Review}\label{sec:litreview}

A basic model for the resource allocation problem was denoted as the ``Blotto game'' or ``Coronel Blotto''. In this game, two players decide how to allocate a finite resource among a finite set of objects (also denoted as \textit{battlefields}) where the player that allocates the most resources on an object wins it. The players' payoff results in the number of battles won (see \cite{borel1921theorie}). Since then, this problem has been studied under multiple variations; we refer the reader to \cite{kovenockconflicts}, \cite{duffy2017stochastic}, and \cite{thomas2018n} for more details on these variations.

An interesting setting of the Blotto game is the case with heterogeneous values (or weight) for each field. 
\cite{gross1950continuous} solved the game equilibrium where players maximize the total weighted battles won. Assuming symmetric budgets and heterogeneous weights, they solved the game for three fields. The case with more than three fields and homogeneous valuations was addressed in \cite{laslier2002distributive}. \cite{gross1950continuous} had pointed out the directions of the results for this case without providing technical details. Recently, the result has been generalized for heterogeneous valuations, allowing for more than three regions, by \cite{thomas2018n}. The solution of the problem with asymmetric budget and homogeneous battlefields has been characterized by \cite{roberson2006colonel} using \textit{n-copulas} on the marginal distribution of the players' strategies. This result has been extended by \cite{schwartz2014heterogeneous}, and \cite{kovenock2020generalizations}, for the heterogeneous valuations case. In all these settings, the outcome in each region is deterministic given the players' allocations. In our paper, we address a model where the result in each region is a probability that depends on the players' allocations, and pre-existing biases as well. %Other works have focused on experimental studies of the game, see for example \cite{avrahami2009weak}, and \cite{mago2017multi}. In another recent work, \cite{duffy2017stochastic}, performed experiments of the Blotto game for the cases where candidates maximize: the expected number of votes, and the probability of winning the majority of them. The authors conclude that individuals allocation decisions resemble the theoretical equilibria for each of the cases analyzed. 

An early work that introduces uncertainty in the outcome was developed by \cite{friedman1958game} who framed the problem as an advertising expenditure allocation. He was the first to find a closed form solution for the game equilibrium in which the players' chances of winning each region are proportional to the players' investments while they maximize the expected number of sales. \cite{brams19743} stated that under the Electoral College System, candidates invest in states in proportion to the power of $2/3$ of the state's weight. The author assumed that both candidates maximize the expected number of electoral votes, while assuming the resources allocated to each state are the same for both candidates. Although there is literature that supports the symmetry of candidates' allocation strategies to some degree (see \cite{shaw1999methods}), there is significant evidence for rejecting allocations to be proportional to the state's weight (see \cite{Eventos_US_2016}). In our work, we consider the probability of winning in the objective function, while allowing candidates' investments to differ within the same states. \cite{lake1979new} looked into the Electoral College where candidates maximize the chances of winning the majority of electoral votes. They found a procedure for computing the game equilibrium in closed form expressions using the Banzhaf Power Index (see \cite{banzhaf1964weighted}). More recently, \cite{duffy2015stochastic} extended the results of \cite{lake1979new} for the case of asymmetric budgets, and the results of \cite{friedman1958game} for more than two players. \cite{osorio2013lottery} generalized the closed form solution from \cite{friedman1958game} to the case of asymmetric players' valuations where candidates maximize the expected number of votes. Our work differs from these in the following aspects: (i) we incorporate states' biases, (ii) we allow for a more general representation of the stochastic voting outcome of each state (by using a Dirichlet distribution instead of a Bernoulli), and (iii) we explore equilibrium in mixed strategies. 

Another study closely related to ours is \cite{snyder1989election}. The author presents a model in which two parties compete in a legislative election, and analyzes the cases where parties maximize the expected number of elected seats, and the probability of winning a majority. They consider candidates' investments as a cost in their objective function. Similar to the work of \cite{snyder1989election}, \cite{klumpp2006primaries} studied a simultaneous and sequential equilibrium of the game, also considering the allocation cost in candidates' objective. The main differences of our work are that we study the majority and electoral college systems with a variable number of votes/electoral votes per region, and we consider the allocation cost as a budget constraint (also known as ``use it or lose it'').
%can be stated in the same points (i), (ii), (iii), and (iv) mentioned above.
%Our model differs in the following aspects: (i) we study the majority and electoral college systems with a variable number of votes/electoral votes per region, (ii) we provide algorithms to find the equilibria in each case, (iii) we explore equilibrium in mixed strategies for the Electoral College case, and (iv) we consider the allocation cost as a budget constraint.

Also under EC, \cite{stromberg2008electoral} studied a probabilistic model in which candidates allocate resources across states to maximize the probability of winning the election. Using a limiting approximation argument of the central limit theorem, the authors characterize conditions that must satisfy an interior equilibrium of the game. Our work differs since we do not use Gaussian approximations for the probability of winning; on the contrary, we use an exact method to compute this (see~\cite{kaplan2003new} where the authors reject the Gaussian distribution for the number of electoral votes).

%One of the closest works related to ours is \cite{snyder1989election}. The author presents a model where two parties compete in a legislative election, trying to maximize the expected number of elected seats, and the probability of winning a majority. They prove equilibrium existence and uniqueness under specific cases. Our model differs from theirs in the following aspects: (i) we study the majority and electoral college systems with a variable number of votes/electoral votes per region, (ii) we provide algorithms to find the equilibria in each case, and (iii) we explore equilibrium in mixed strategies for the Electoral College case. Other works have looked into the campaign allocation problem from different frameworks. For example \cite{kovenockconflicts}, and \cite{fu2015team} studied multi-battlefield games in which the success of the players depends on the result of multiple \textit{battles}, whereas \cite{rai2009generalized}, and \cite{shupp2013resource} have looked at the problem as a contest competition. The resource allocation has also been framed from the perspective of matching a political party's candidates to a set of districts to maximize the overall performance (see \cite{galasso2011competing}). 

Prediction of the election outcome has also captured the attention of researchers. Bayesian priors has been a technique widely used by researchers; applications of this in the EC can be found in \cite{kaplan2003new}, \cite{rigdon2009bayesian}, and \cite{rigdon2015forecasting}. Under the same electoral system, choice models have also been used for forecasting purposes (see \cite{wang2015forecasting}).

Finally, it is worth mentioning the connection of the presented problem with the market-share competition between two firms, see for example, \cite{bell1975market}, \cite{barnett1976more}, and \cite{monahan1987structure}. Similarly, framed as a multi-item contest problem, \cite{robson2005multi} found closed form expressions for the game equilibrium using a generalized version of the Tullock functional form (see \cite{buchanan1980efficient}). Our work differs on treatment of the bias parameters, while we also consider the possibility of abstention. In addition, we consider a stochastic version of the electoral game, analyzing the cases where candidates maximize the expected number of votes, as well as the probability of winning.

\section{Majority System}\label{sec:ms}

Two candidates, $A$ and $B$, compete on a political election campaign for president of a country. We assume throughout the paper that the election is for president, however this can be applied to any other election that shares the same settings of the model. The country is divided into a set of regions which will be denoted as $\mathcal{I}\coloneqq \{1,\dots,n\}$. Each region $i\in\mathcal{I}$ has $v_{i}$ voters. Both candidates are endorsed with a fixed campaign resource budget which they must allocate among the different regions. We will consider this resource to be the number of days of the campaign. Then, both candidates have a budget of $D$ days on which they are able to run their campaign events in the different regions. Consequently, candidates must decide how much effort ---how many days of campaigning--- they will put into each of the regions.
%As for this work, this resource will be refered to the days For this  have the same budget of days, $D$, in which they are allowed to run their campaign events within the different regions, thus candidates have to decide how much effort ---days of campaign--- they put in each of the regions.
Let $x_{i}\ge 0$ and $y_{i}\ge 0$ be the number of days inputted by candidates $A$ and $B$ respectively in region $i$. For simplicity, we normalize the budget to the unit value, i.e. $D=1$. It can be easily seen that we can use other limited \textit{resources}, besides days of campaigning, which candidates need to allocate strategically among the regions. For the sake of simplicity, the resource modeled in this work will be the days of campaigning. However, the model could easily be extended to incorporate additional resources, such as money or others, leading to a different polyhedral set as the strategy space. 

The strategy space for both candidates, denoted by $\Delta_n$, is the simplex in $\mathbb{R}^{n}$, namely $\Delta_n=\{{\bf x}\in\mathbb{R}^{n}|\sum_{i=1}^{n}x_{i}= 1,x_{i}\ge 0\}$. Intuitively, the more days that candidate $A$ invests in a region, the more votes she is likely to get from that particular region. Nonetheless, the more campaign her opponent ($B$) does in that region, the less the number of votes candidate $A$ will receive from that particular geographical area. Therefore, the outcome of votes from each particular region will depend on the political efforts of both contenders (see \cite{nagler1992presidential}). In addition, it is natural to think that some regions have an \textit{a-priori} bias towards one of the candidates. Put it differently, for the same level of efforts inputted by both candidates in a particular region, the outcome might favor one of the candidates over the other due to the already existing preferences of the population of the region. We assume both players play simultaneously. For each region $i$, let $s_{i}^{A}:\mathbb{R}\times \mathbb{R}\rightarrow [0,1]$ be the function that maps the efforts of candidates $A$ and $B$ ($x_i$ and $y_i$ respectively) into the fraction of the votes that candidate $A$ obtains in region $i$. Similarly define $s_{i}^{B}:\mathbb{R}\times \mathbb{R}\rightarrow [0,1]$ as the fraction of votes obtained by candidate $B$ in region $i$. We allow the possibility for abstention to happen, therefore, $0\leq s_{i}^{A}+s_{i}^{B}\leq1$. %We assume that the outcome of a region depends on the days of electoral campaign on that particular region, plus intrinsic parameters of the region that represent: voters \textit{a-priori} preferences of the population towards the candidates, and peoples' abstention. As the reader might note, the model can be easily extended in order to capture additional complexities.

We will first analyze a deterministic model of the problem, and then present a stochastic version. The voting system to be analyzed here is the Majority System. In MS, the candidate who obtains the most votes (nationwide) wins the election.

\subsection{Deterministic Game}\label{sec:PSdet}
In this case, the vote outcome of all regions is determined by the vectors of efforts, $\bf{x}$ and ${\bf y}$, of the candidates $A$ and $B$ respectively. For this setting, we will define the outcome function $s_{i}^{A}$ and $s_{i}^{B}$ for each region $i$ as $s_{i}^{A}(x_{i},y_{i}) = \frac{x_{i}+\alpha_{i}}{x_{i}+\alpha_{i}+y_{i}+\beta_{i}+\gamma_i}$ and $s_{i}^{B}(x_{i},y_{i}) = \frac{y_{i}+\beta_{i}}{x_{i}+\alpha_{i}+y_{i}+\beta_{i}+\gamma_i}$. 

\begin{comment}

\begin{eqnarray}\label{eq:si}
s_{i}^{A}(x_{i},y_{i}) &=& \frac{x_{i}+\alpha_{i}}{x_{i}+\alpha_{i}+y_{i}+\beta_{i}+\gamma_i}\label{eq:siA}\\
s_{i}^{B}(x_{i},y_{i}) &=& \frac{y_{i}+\beta_{i}}{x_{i}+\alpha_{i}+y_{i}+\beta_{i}+\gamma_i}\label{eq:siA}
\end{eqnarray}

\end{comment}

$\alpha_{i},\beta_{i}>0$ are the {\it bias parameters} towards candidates $A$ and $B$ respectively, and $\gamma_{i}\ge0$ is the \textit{abstention parameter} for region $i$. Bias parameters represent the intrinsic bias of the region towards a particular candidate. If $\alpha_{i}>\beta_{i}$, then people from region $i$ are leaning towards candidate $A$ since for the same levels of efforts, i.e. $x_{i}=y_{i}$, candidate $A$ gets more votes from the region than her contender. Vice-verse if $\alpha_{i}<\beta_{i}$. Also, note that high values of the bias parameters $\alpha_{i},\beta_{i}$ mean that the result of region $i$ is less sensitive with respect to the level of efforts $x_{i}$ and $y_{i}$, and therefore voters' preferences are highly polarized to change their votes given the candidates' campaigns. On the contrary, low levels of $\alpha_{i},\beta_{i}$ imply that the outcome of the people's votes is more sensitive to the candidates' efforts. The \textit{abstention parameter}, $\gamma_i$ for region $i$, is such that there is no abstention if $\gamma_i=0$. Otherwise abstention increases monotonically with the parameter. Note that the abstention can be seen as a ``third'' candidate option that does not campaign and has a \textit{bias} parameter equal to $\gamma_i$. Then, for a given region $i$, the total number of votes candidate $A$ ($B$) receives is $v_{i}s_{i}^{A}(x_{i},y_{i})$ ($v_{i}s_{i}^{B}(x_{i},y_{i})$); and the total number of votes candidate $A$ ($B$) receives is $\sum_{i\in \mathcal{I}}v_{i}s_{i}^{A}(x_{i},y_{i})$ $\left(\sum_{i\in \mathcal{I}}v_{i}s_{i}^{B}(x_{i},y_{i})\right)$. All parameters are public information. 

The objective of each player is to win the election. However, this can result in an infinite number of equilibria. Moreover, we can argue that some of these equilibria are more preferable than others. For instance, if candidate $A$ wins the election on one equilibrium with $51\%$ of the votes (between both candidates), whereas on another equilibrium wins with $68\%$ (between both candidates), there is no doubt that the second scenario is preferred by candidate $A$ (and especially the political parties behind the candidate). Then, as for the deterministic game, the objective of each candidate will be to maximize the number of votes obtained with respect to the total number of votes obtained between the two candidates. Note that an equilibrium---the formal definition of this will be given shortly---of this game is also an equilibrium of the game in which candidates aim to win the election regardless of the difference. 

The optimization problem candidates $A$ and $B$ solve are written as follows:
\vspace{-1.0cm}

\begin{multicols}{2}
    \begin{equation} \label{eq:det_max_A}
        \begin{aligned}
            & \underset{{\bf x} \in \Delta_n}{\text{\bf{max}}}
            & & Q^A({\bf x},{\bf y})\coloneqq \frac{\sum_{i\in\mathcal{I}} v_is_{i}^{A}}{\sum_{i\in\mathcal{I}}v_i(s_{i}^{A}+s_{i}^{B})}
        \end{aligned}
    \end{equation}
\break
    \begin{equation} \label{eq:det_max_B}
        \begin{aligned}
            & \underset{{\bf y} \in \Delta_n}{\text{\bf{max}}}
            & & Q^B({\bf x},{\bf y})\coloneqq \frac{\sum_{i\in\mathcal{I}} v_is_{i}^{B}}{\sum_{i\in\mathcal{I}}v_i(s_{i}^{A}+s_{i}^{B})}.
        \end{aligned}
    \end{equation}
\end{multicols}

\noindent The numerator of the candidates' objective function in~\eqref{eq:det_max_A} and \eqref{eq:det_max_B} has the total number of votes they get, while the denominator has the total number of votes obtained by both. This is clearly a zero-sum game since an increase in the percentage of votes of one candidate results in its loss from the opponent. 

\begin{definition}\label{def:1}
An equilibrium is a pair of effort vectors $({\bf x^*},{\bf y^*})\in\Delta_n\times\Delta_n$ such that each vector ${\bf x^{*}}$ and ${\bf y^*}$ is the optimal solution of the respective candidate's maximization problems given in Expressions~\eqref{eq:det_max_A} and \eqref{eq:det_max_B}.
\end{definition}

The following theorem states the existence of equilibrium of the game presented.
\begin{theorem}\label{the:det_ex}
There exists an equilibrium (in pure strategies) for the deterministic game. \end{theorem}
\proof{Proof.} See Appendix~\ref{proof:the:ex}.
\Halmos
\endproof 
Then next theorem states the uniqueness of the equilibrium.
\begin{theorem}\label{the:det_un}
The equilibrium of the deterministic game is unique \end{theorem}
\proof{Proof.} See Appendix~\ref{proof:the:un}.
\Halmos
\endproof 
Unfortunately, there is no closed form solution for the equilibrium of the game. Before showing a method to compute this, we will present a proposition that states a closed form solution for the equilibrium of an \textit{unbounded} version of the game. More precisely, consider the same election game as described above except that the candidates' efforts are allowed to take negative values (these efforts must still add up to one). Furthermore, consider that the efforts of both candidates are constrained to a particular subset of regions $\mathcal{I^*}\subseteq\mathcal{I}$. The latter subset represents the regions on which the candidates will focus their attention, whereas regions outside this set will have null investment. The resulting game defined with the given characteristics will be called an \textit{unbounded game constrained on the set of regions $\mathcal{I^*}$}. Existence and uniqueness of the equilibrium of this game can be shown using similar arguments to the ones used for the original game. The next proposition presents a closed form for its equilibrium.

\begin{proposition}\label{pro:CUB}
For any nonempty set $\mathcal{I^{*}}\subseteq\mathcal{I}$, the equilibrium of the unbounded game constrained in the set of regions $\mathcal{I^*}$ is given by
\begin{equation}\label{eq:x_CUB}
x_i^{UB(\mathcal{I^*})} = \frac{v_i}{v_{\mathcal{I^*}}} \left( (1 + \alpha_{\mathcal{I^*}}) + \frac{Q^A}{Q^A + Q^B} \gamma_{\mathcal{I^*}} \right) - \frac{Q^A}{Q^A + Q^B} \gamma_i - \alpha_i
\end{equation}
\begin{equation}\label{eq:y_CUB}
y_i^{UB(\mathcal{I^*})} = \frac{v_i}{v_{\mathcal{I^*}}} \left( (1 + \beta_{\mathcal{I^*}}) + \frac{Q^B}{Q^A + Q^B} \gamma_{\mathcal{I^*}} \right) - \frac{Q^B}{Q^A + Q^B} \gamma_i - \beta_i,
\end{equation}
for all $i\in \mathcal{I^*}$, where $\alpha_{\mathcal{I^*}} \coloneqq \sum_{j \in \mathcal{I^*}} \alpha_j$ and similarly with $\beta_{\mathcal{I^*}}$, $\gamma_{\mathcal{I^*}}$ and $v_{\mathcal{I^*}}$. %The amount of votes candidate $A$ gets
Candidates' votes can be computed as
 $Q^A = \frac{v_{\mathcal{I^*}}(1 + \alpha_{\mathcal{I^*}})}{2 + \alpha_{\mathcal{I^*}} + \beta_{\mathcal{I^*}} + \gamma_{\mathcal{I^*}}} + \sum_{j \notin {\mathcal{I^*}}} \frac{v_j \alpha_j}{\alpha_j + \beta_j + \gamma_j}$, and $Q^B = \frac{v_{\mathcal{I^*}}(1 + \beta_{\mathcal{I^*}})}{2 + \alpha_{\mathcal{I^*}} + \beta_{\mathcal{I^*}} + \gamma_{\mathcal{I^*}}} + \sum_{j \notin {\mathcal{I^*}}} \frac{v_j \beta_j}{\alpha_j + \beta_j + \gamma_j}$.
\end{proposition}
\proof{Proof.} See Appendix~\ref{proof:pro:CUB}.
\Halmos

From Proposition~\ref{pro:CUB}, we can obtain the equilibrium of the unbounded version of the game when fixing the set of regions where candidates can put their efforts. Note that the obtained equilibrium might have negative components, in which case it cannot be the equilibrium of the original game. Even if the unconstrained equilibrium quantities are all non-negative, this might not be the equilibrium of the original game. However, if $\mathcal{I^*}$ matches the set of regions with positive investment values in the equilibrium of the original game, then  $({\bf x
^*},{\bf y^*})=({\bf x^{UB(\mathcal{I^*})}},{\bf y^{UB(\mathcal{I^*})}})$. Equations~\eqref{eq:x_CUB} and~\eqref{eq:y_CUB} can be used to analyze the relation between the problem parameters and the game equilibrium (at least in a local neighborhood). A corollary that extends from Proposition
~\ref{pro:CUB} is for the particular case where $\mathcal{I^{*}}=\mathcal{I}$. It can be observed that in the latter case, $\frac{\partial x_{i}^{UB}}{\partial v_i},\frac{\partial y_{i}^{UB}}{\partial v_i}>0$, i.e., a region with a higher number of votes will induce more efforts by both candidates (see Appendix~\ref{proof:cor:UBA}). The interesting fact of the case in which $\mathcal{I^{*}}=\mathcal{I}$, is that if unbounded equilibrium quantities (from Proposition
~\ref{pro:CUB}) are non-negative, then this will also coincide with the equilibrium of the original game. %$(x^{*},y^{*})=(x^{UB(\mathcal{I})},y^{UB(\mathcal{I})})$. 
Also, in this particular case, it can be shown that, the fraction of votes obtained by candidate $A$ (with respect to both candidates) in each region is equal to $\frac{1+\sum_{j}\alpha_j}{2+\sum_{j}(\alpha_j+\beta_j)}$. Then, the total fraction of votes obtained by candidate $A$ equates the latter expression. Thus, the total fraction of votes obtained by candidate $A$ is independent of the abstention. Furthermore, the candidate who has the greater value of the sum of his bias parameters will win the election. Namely, if $\sum_{j}\alpha_j>\sum_{j}\beta_j$, then candidate $A$ wins the election. Another important observation that holds for the unbounded game constraint to $\mathcal{I^{*}}\subseteq\mathcal{I}$ in the case of no abstention is stated in the next corollary. Note that the unbounded version of the game can be interpreted as a hypothetical setting where candidates can lend-and-borrow efforts among the different regions, in which \textit{short positions} are possible.

\begin{corollary}\label{cor:UBG0}
If $\gamma_i=0$ for $i\in \mathcal{I^*}$, the equilibrium of the unbounded game constraint to $\mathcal{I^{*}}$ is such that the fraction of votes obtained by candidate $A$ in each region in $\mathcal{I^*}$ is the same. Specifically: 
\begin{equation}\label{eq:fracxG0}
\frac{x_i^{UB(\mathcal{I^*})}+\alpha_i}{x_i^{UB(\mathcal{I^*})}+\alpha_i+y_i^{UB(\mathcal{I^*})}+\beta_i} =\frac{1+\alpha_{\mathcal{I^*}}}{2+\alpha_{\mathcal{I^*}}+\beta_{\mathcal{I^*}}},\qquad\frac{y_i^{UB(\mathcal{I^*})}+\beta_i}{x_i^{UB(\mathcal{I^*})}+\alpha_i+y_i^{UB(\mathcal{I^*})}+\beta_i} =\frac{1+\beta_{\mathcal{I^*}}}{2+\alpha_{\mathcal{I^*}}+\beta_{\mathcal{I^*}}}.
\end{equation}
%for all $i\in\mathcal{I^*}$.
\end{corollary}
\proof{Proof.} See Appendix~\ref{proof:cor:UBG0}.
\Halmos

Under a compulsory voting system, if $\mathcal{I^*}$ matches the set of regions where the candidates' efforts are positive in the constraint game equilibrium (i.e. the original game), the result of Corollary~\ref{cor:UBG0} will hold. As a result, the fraction of votes each candidate obtains in each region of the set $\mathcal{I}^*$ %where candidate invest 
will be the same.

In order to compute the equilibrium of the original game, with the non-negativity constraints, we can iterate by solving a parametrized game with payoff function $Q_{t}^{A}({\bf x},{\bf y})\coloneqq t  Q^{A}({\bf x},{\bf y})+\sum_{j}\ln(x_{j})-\sum_{j}\ln(y_{j})$ for a fixed $t>0$, obtaining $({\bf x_t^{*}}, {\bf y_t^{*}})$ which denotes the equilibrium of this game. The proof of existence and uniqueness of equilibrium is analogous to that of the original game shown in Theorems~\ref{the:det_ex} and \ref{the:det_un}). Then, for a fixed value of $t$, we solve the game by using the infeasible start Newton method, and iterate until finding the equilibrium of the limit game.

Up to this point, we have assumed that for a given vector of (i) efforts, $\bf x$ and $\bf y$, (ii) bias parameters, {\boldmath $\alpha$} and {\boldmath $\beta$}, and (iii) abstention parameter, {\boldmath $\gamma$}; the outcome of the election is perfectly known and so can be computed exactly for every single region, and thus for the whole country. The latter is probably a strong assumption, since despite the amount of information we have on a particular region, we will probably not predict the result with $100\%$ accuracy. Therefore, in the next section we introduce a stochastic model that accounts for uncertainty in the vote outcomes.

\subsection{Stochastic Game}\label{sec:MS_sto}
For each region $i$, let $S_{i}^{A}$, $S_{i}^{B}$, and $S_{i}^{C}$ be the random variable of the fraction of votes received by candidate $A$, candidate $B$, and the abstained votes respectively, such that
\begin{eqnarray}\label{eq:dir_distribution}
(S^{A}_i,S^{B}_i,S^{C}_i) \sim \textbf{Dir}_3 \left( k (x_i + \alpha_i), k (y_i + \beta_i), k  \gamma_i \right),
\end{eqnarray}
where ${\bf Dir}_m$ is an m-dimensional Dirichlet distribution (in this case, three-dimensional). % and the parameter $k>0$ controls for the standard deviation. 

\begin{comment}

The PDF can then be written as
\begin{eqnarray}\label{eq:dir_pdf}
f_i(s_i^A, s_i^B, s_i^C) =   \frac{(s_i^A)^{k (x_i + \alpha_i) - 1}  (s_i^B)^{k (y_i + \beta) - 1}  (s_i^C)^{k \gamma_i - 1} }{\text{B}\left(k(x_i + \alpha_i),k(y_i + \beta_i), k\gamma_i\right)},
\end{eqnarray}
where $\text{B}(a,b,c) = \Gamma(a)\Gamma(b)\Gamma(c)/\Gamma(a+b+c)$.

\end{comment}
 Note that the expectation of $S_{i}^{A}$ and $S_{i}^{B}$ matches with the values of the analogous parameters ($s_{i}^{A}$ and $s_{i}^{B}$) in the deterministic model. Indeed, $\mathbb{E}(S^A_i) = \frac{x_i + \alpha_i}{x_i + \alpha_i + y_i + \beta_i + \gamma_i}$, $\mathbb{E}(S^B_i) = \frac{y_i + \beta_i}{x_i + \alpha_i + y_i + \beta_i + \gamma_i}$. The parameter $k>0$ regulates for the noise of the vote outcomes, so that higher values of $k$ represent settings with lower variability (and vice versa). In fact, the variance for the fraction of votes for candidate $A$ is $\text{Var}(S_i^A) = \frac{(x_i + \alpha_i)(y_i + \beta_i + \gamma_i)}{(x_i + \alpha_i + y_i + \beta_i + \gamma_i)^2 (1 + k(x_i + \alpha_i))}$, so $\lim_{k \rightarrow \infty}\text{Var}(S^A_i) =  0$ (analogous for $S_i^B$ and $S_i^C$).

In this case, where the vote outcomes are stochastic, candidates will exert their efforts in order to maximize the chances of getting elected, instead of maximizing the expected number of votes. Note that if candidates were to maximize the expected number of votes, it would result in a game that is equivalent to the one introduced in the deterministic section if there was no abstention. Let $R^A$ be the number of votes obtained by candidate $A$, and similarly for $R^B$. The probability that candidate $A$ wins the election can be computed as:
\begin{eqnarray}\label{eq:probRA}
\mathbb{P}\left(R^A>R^B\right)&=&\int_{\Delta_{3}}\dots \int_{\Delta_{3}}\mathbbm{1}_{\left\{\sum_{i\in\mathcal{I}}v_{i}s_{i}^{A}>\sum_{i\in\mathcal{I}}v_{i}s_{i}^{B}\right\}}\prod_{i\in\mathcal{I}}f_{i}\left(s_{i}^{A},s_{i}^{B},s_{i}^{C}\right)ds_i^A ds_i^B ds_i^C.
\end{eqnarray}

\begin{comment}

Then, the optimization problem of candidate $A$ can be written as 
\begin{equation} \label{eq:sto_max_A}
    \begin{aligned}
        & \underset{x}{\text{\bf{max}}}
        & & \mathbb{P}\left(R^A>R^B\right) \\
        & \text{\bf{s.t.}} & & \sum_{i\in\mathcal{I}} x_i = 1 \\
        & & & x_i \geq 0\qquad\forall i\in\mathcal{I},
    \end{aligned}
\end{equation}
while the optimization problem for candidate $B$ is
\begin{equation} \label{eq:sto_max_B}
    \begin{aligned}
        & \underset{x}{\text{\bf{max}}}
        & & \mathbb{P}\left(R^B>R^A\right) \\
        & \text{\bf{s.t.}} & & \sum_{i\in\mathcal{I}} y_i = 1 \\
        & & & y_i \geq 0\qquad\forall i\in\mathcal{I}.
    \end{aligned}
\end{equation}

\end{comment}

\noindent Then, the optimization problem of candidates $A$ and $B$ can be written as: 
\vspace{-1.0cm}

\begin{multicols}{2}
  \begin{equation} \label{eq:sto_max_A}
    \begin{aligned}
        & \underset{{\bf x} \in \Delta_n}{\text{\bf{max}}}
        & & \mathbb{P}\left(R^A>R^B\right)
    \end{aligned}
\end{equation}
\break
  \begin{equation} \label{eq:sto_max_B}
    \begin{aligned}
        & \underset{{\bf y} \in \Delta_n}{\text{\bf{max}}}
        & & \mathbb{P}\left(R^B>R^A\right).
    \end{aligned}
    \end{equation}
\end{multicols}

\begin{comment}
\begin{multicols}{2}
  \begin{equation} \label{eq:sto_max_A}
    \begin{aligned}
        & \underset{x}{\text{\bf{max}}}
        & & \mathbb{P}\left(R^A>R^B\right) \\
        & \text{\bf{s.t.}} & & \sum_{i\in\mathcal{I}} x_i = 1 \\
        & & & x_i \geq 0\qquad\forall i\in\mathcal{I},
    \end{aligned}
\end{equation}
\break
  \begin{equation} \label{eq:sto_max_B}
    \begin{aligned}
        & \underset{x}{\text{\bf{max}}}
        & & \mathbb{P}\left(R^B>R^A\right) \\
        & \text{\bf{s.t.}} & & \sum_{i\in\mathcal{I}} y_i = 1 \\
        & & & y_i \geq 0\qquad\forall i\in\mathcal{I}.
    \end{aligned}
    \end{equation}
\end{multicols}
\end{comment}

\begin{definition}\label{def:2}
An equilibrium for the stochastic game is a pair of effort vectors $({\bf x^*},{\bf y^*})\in\Delta_n\times\Delta_n$ such that each vector ${\bf x^{*}}$ and ${\bf y^*}$ is the optimal solution of the respective candidate maximization problems given in Expressions~\eqref{eq:sto_max_A} and \eqref{eq:sto_max_B}.
\end{definition}

Unlike the deterministic version of the game, in the stochastic version we have no guarantee of the existence of the equilibrium in pure strategies. However, we can state the existence of equilibrium in mixed strategies.
\begin{theorem}\label{the:sto_ex}
There exists an equilibrium in mixed strategies for the stochastic game. \end{theorem}
\proof{Proof.} See Appendix~\ref{proof:the:sto:ex}.
\Halmos
\endproof 
After several numerical computations, we observed that the objective function always happens to be quasi concave, which suggests that there might always exists an equilibrium in pure strategies. Therefore, we proceed to find an equilibrium in pure strategies as described in the following subsection.

\subsubsection{Computing Equilibrium}\label{sec:comp_eq_MS_sto}
To compute the equilibrium of the stochastic game, we use a gradient descent ascent method. Namely, we move the strategies of both players simultaneously according to their payoffs at the current solution and repeat until we reach a pair of strategies $({\bf x}, {\bf y})$ such that no player has an incentive to deviate. More specifically, we take a step $\rho>0$ in the chosen direction to update the new solution as $({\bf x},{\bf y}) \leftarrow ({\bf x}+\rho {\bf d^A}, {\bf y} + \rho  {\bf d^B})$, where ${\bf d^A}$ and ${\bf d^B}$ are the directions of maximum and minimum growth of $f({\bf x},{\bf y})\coloneqq \mathbb{P}(R^A>R^B)$ within $\Delta_n$. In order to compute these directions, the following proposition is introduced:

\begin{proposition}\label{pro:grad}
The direction of maximum growth of a function $f:\mathbb{R}^{n}\rightarrow \mathbb{R}$ within the simplex region, $\Delta_n$, is given by $d_{i}=x_{i}\left(\tau_{i}-\sum_{j}x_{j}\tau_{j} \right)$, where $\tau_i=x_{i}\left(\frac{\partial f}{\partial x_{i}}-\sum_{j}\frac{\partial f}{\partial x_{j}}x_{j} \right)$.
%The direction of maximum growth of the function $f$ on the variable $x$ in $\Delta_n$ on the $i^{th}$ component can be written as $d_{i}=x_{i}\left(\tau_{i}-\sum_{j}x_{j}\tau_{j} \right)$ where $\tau_i=x_{i}\left(\frac{\partial f}{\partial x_{i}}-\sum_{j}\frac{\partial f}{\partial x_{j}}x_{j} \right)$. 
\end{proposition}
\proof{Proof.} See Appendix~\ref{proof:pro:grad}.
\Halmos
\endproof 
It can be easily shown that the directions of Proposition~\ref{pro:grad} correspond to the complementary slackness of the optimization problem faced by both candidates. %Indeed, we can check the KKT conditions on the obtained strategy for each player. For the case of the first player, these can be written as:
%\begin{subequations} \label{KKT_SM_Est_X} \begin{align}
%    \frac{\partial}{\partial x_i} \mathbb{P}(R^A > R^B) + \mu_i = \lambda \quad \forall i \in \mathcal{I} \label{KKT_SM_Est_X_FOC} \\
%    x_i \mu_i = 0 \quad \forall i \in \mathcal{I} \label{KKT_SM_Est_X_CS} \\
%    x_i, \mu_i \geq 0 \quad \forall i \in \mathcal{I} \label{KKT_SM_Est_X_G0} \\
%    \sum_j x_j = 1. \qquad \qquad \label{KKT_SM_Est_X_S1}
%\end{align} \end{subequations}
%We can multiply each of the FOC in Equation~\eqref{KKT_SM_Est_X_FOC} by the respective $x_i$, add them up using complementary slackness, and get $\lambda=\sum_{j}x_{j}\frac{\partial}{\partial x_i} \mathbb{P}(R^A > R^B)$. Replacing this value of $\lambda$ on Equations~\eqref{KKT_SM_Est_X_FOC} in order to obtain $\mu_{i}$. Finally, we can multiply each $\mu_{i}$ by its respective $x_i$, and obtain the complementary slackness condition as $\xi_{i}^{A}\coloneqq x_i  \left( \sum_j x_j \frac{\partial}{\partial x_j} \mathbb{P}(R^A > R^B) - \frac{\partial}{\partial x_i} \mathbb{P}(R^A > R^B) \right) = 0$, $\forall i \in \mathcal{I}$. Analogous for candidate $B$, we define $\xi_{i}^{B}$. 
We use the complementary slackness as a stopping criteria. %Namely, we iterate until $\max\{\Vert\xi^{A}\Vert,\Vert\xi^{B}\Vert\} \leq \epsilon$. 

The algorithm we use is described as follows: 
\begin{algorithm}
\caption{Equilibrium for Stochastic Game}
\label{alg:1}
\begin{algorithmic}[1]
\State \bf{Input} $\alpha, \beta, \gamma, v \in \mathbb{R}_{+}^n$
 \State \bf{Set} $x,y = v/\left(\sum_{j}v_j\right)$, $\xi^A=\xi^B=\{\infty\}_{i=1}^{n}$
 \State \bf{While} $ \max\{\norm{\xi^A}\},\norm{\xi^B}\}\} > \epsilon $ {\bf{do}}
 \State \qquad  $x = x + \rho d^A$, $y = y - \rho d^B$
 % \State \qquad  $y = y - \rho d^B$
 \State \qquad {update} $\xi^A,\xi^B$
 \State \bf{End While}
 \State \bf{Return} $(x,y)$
\end{algorithmic}
\end{algorithm}
The starting point is set to be proportional to the regions' number of votes. Although the game is not convex-concave, the algorithm in practice always happens to converge at a stationary point.

The calculation of $f$ and its derivatives requires computing several composed integrals, see the expressions of these terms in Appendix~\ref{app:f_gradf}. We need to compute $3n$-dimension integrals for the terms mentioned, which is not possible in practice, not even for low values of $n$. Therefore, we use Monte Carlo simulation to approximate the value of these integrals. 

\subsubsection{Boosting}:
The need to compute several integrals by Monte Carlo simulation implies that at each pair of strategies $({\bf x},{\bf y})$ we evaluate when running Algorithm~\ref{alg:1}, we need to sample multiple Dirichlet random variables; one for every region and simulation. The next proposition states a result that helps us to re-use the simulations of the Dirichlet random variables for nearby points (i.e. candidate strategies that are close to the ones we are at). For ease of notation, let us denote $\textbf{S} \coloneqq\{\left(S_{i}^{A},S_{i}^{B},S_{i}^{C}\right)\}_{i=1}^{n}$.

\begin{proposition}\label{pro:close}
Let $g:\left(\Delta_{n}\right)^{n}\rightarrow \mathbb{R}$ be a scalar function. Then it holds that
\begin{eqnarray}\label{eq:dir_expectation}
\mathbb{E}\left(g(\textbf{S})|{\bf x}+{\bf \Delta x},{\bf y}+{\bf {\bf \Delta y}}\right) =  \mathbb{E}\left(K\times g(\textbf{S})\times\prod_{j}(S_{j}^{A})^{k\Delta x_{j}} (S_{j}^{B})^{k\Delta y_{j}}|{\bf x},{\bf y}\right),
\end{eqnarray}
where $K=\prod_{j}\frac{B(k(x_{i}+\alpha_i),k(y_{j}+\beta_{j}),k\gamma_{j})}{B(k(x_{i}+\Delta x_{j}+\alpha_i),k(y_{j}+\Delta y_{j}+\beta_{j}),k\gamma_{j})}$ and $B(\cdot,\cdot,\cdot)$ is the multivariate Beta function.
\end{proposition}
\proof{Proof.} See Appendix~\ref{proof:pro:close}.
\Halmos
\endproof 
Proposition~\ref{pro:close} allows us to reuse the simulations of the sampled Dirichlet distribution at a given point ${\bf x},{\bf y}$ in other points ${\bf x}+{\bf \Delta x},{\bf y}+{\bf \Delta y}$. 
In particular, we are interested in using Proposition~\ref{pro:close} with the function $g(\textbf{S})=\mathbbm{1}_{\left\{\sum_{i\in\mathcal{I}}v_{i}S_{i}^{A}>\sum_{i\in\mathcal{I}}v_{i}S_{i}^{B}\right\}}$, which takes value one in case the election is won by candidate $A$, and zero otherwise. As a result, we can sample the Dirichlet random variables once at a particular pair $({\bf x},{\bf y})$, and compute an (unbiased) estimate of the probability that candidate $A$ wins at any other point $({\bf x}+{\bf \Delta x},{\bf y}+{\bf \Delta y})$ by computing the RHS of Equation~\eqref{eq:dir_expectation} with $g(\textbf{S})=\mathbbm{1}_{\left\{\sum_{i\in\mathcal{I}}v_{i}S_{i}^{A}>\sum_{i\in\mathcal{I}}v_{i}S_{i}^{B}\right\}}$. However, as we move further from the point $({\bf x},{\bf y})$, it might turn out that the variance of the random variable in the RHS of Equation~\eqref{eq:dir_expectation} increases. (Note that the random variable of the RHS of Equation~\eqref{eq:dir_expectation}, unlike the rv in the LHS of Equation~\eqref{eq:dir_expectation}, does not follow a Bernoulli distribution.) The next proposition states a result with respect to the variances of the random variables of Equation~\eqref{eq:dir_expectation}.

\begin{proposition}\label{pro:close_var}
Let $g(\textbf{S})=\mathbbm{1}_{\left\{\sum_{i\in\mathcal{I}}v_{i}S_{i}^{A}>\sum_{i\in\mathcal{I}}v_{i}S_{i}^{B}\right\}}$. If the probability of winning for candidate $A$ is higher at point $({\bf x}+{\bf \Delta x},{\bf y}+{\bf \Delta y})$ than at point $({\bf x},{\bf y})$, i.e. $f({\bf x},{\bf y})>f({\bf x}+{\bf \Delta x},{\bf y}+{\bf \Delta y})$, then
\begin{eqnarray}\label{eq:dir_expectation2}
\text{Var}\left(g(\textbf{S})|{\bf x}+{\bf \Delta x},{\bf y}+{\bf \Delta y}\right) <  \text{Var}\left(K\times g(\textbf{S})\times\prod_{j}(S_{j}^{A})^{k\Delta x_{j}} (S_{j}^{B})^{k\Delta y_{j}}|{\bf x},{\bf y}\right),
\end{eqnarray}
where $K=\prod_{j}\frac{B(k(x_{j}+\alpha_j),k(y_{j}+\beta_{j}),k\gamma_{j})}{B(k(x_{j}+\Delta x_{j}+\alpha_j),k(y_{j}+\Delta y_{j}+\beta_{j}),k\gamma_{j})}$.
\end{proposition}
\proof{Proof.} See Appendix~\ref{proof:pro:close_var}.
\Halmos
\endproof 
Proposition~\ref{pro:close_var} provides a sufficient condition that implies that the variance of the rv $K\times g(\textbf{S})\times\prod_{j}(S_{j}^{A})^{k\Delta x_{j}} (S_{j}^{B})^{k\Delta y_{j}}|{\bf x}, {\bf y}$ is greater than the variance of $g(\textbf{S})|{\bf x},{\bf y}$. In other words, reusing the Dirichlet rv' simulations from a point $({\bf x},{\bf y})$ to estimate the probability that candidate $A$ wins at a point $({\bf x}+{\bf \Delta x},{\bf y}+{\bf \Delta y})$ will probably lead to more error than using the Dirichlet random variable at the point $({\bf x}+{\bf \Delta x},\bf{y}+{\bf \Delta y})$. It is interesting that there are cases in which this does not hold; namely, there are cases in which estimating the winning probability for candidate $A$ at a point $({\bf x}+{\bf \Delta x},{\bf y}+{\bf \Delta y})$ reusing the Dirichlet rv from point $({\bf x},{\bf y})$ has less variance than estimating this probability with the Dirichlet at the point $({\bf x}+{\bf \Delta x},{\bf y}+{\bf \Delta y})$. An example of an instance where the latter happens is ${\bf v^T}=(12, 3, 4)$, {\boldmath $\alpha^T$}$=(0.1, 0.5, 0.4)$, {\boldmath $\beta^T$}$=(0.2, 0.6, 0.3)$, {\boldmath $\gamma^T$}$=(0.1, 0.1, 0.1)$, ${\bf x}=(0.1, 0.35, 0.55)$, ${\bf y}=(0.4, 0.2, 0.4)$, ${\bf \Delta x}=(-0.07, -0.02, 0.09)$, and ${\bf \Delta y}=(0.15, -0.1, -0.05)$; the variance expressions of the LHS and RHS of Equation~\eqref{eq:dir_expectation2} result in $0.30$ and $0.23$ respectively. Finally, note that the variance of the rv $K\times g(\textbf{S})\times\prod_{j}(S_{j}^{A})^{k\Delta x_{j}} (S_{j}^{B})^{k\Delta y_{j}}|{\bf x},{\bf y}$ can be estimated simply by reusing the Dirichlet rvs from point $({\bf x},{\bf y})$, see Equation~\eqref{eq:var11} in Appendix~\ref{proof:pro:close_var}. All in all, we use the result of these propositions when applying Algorithm~\ref{alg:1} by re-using Dirichlet sampled rv at nearby points as long as the variance of the winning probability does not increases. Otherwise, we sample again.

\section{Electoral College}\label{sec:ec}

Under the Electoral College system, the candidates get all the electoral votes of the states where they have the majority of votes with respect to their contenders. Let $w_i\in\mathbb{Z}_{+}$ be the number of electoral votes of state $i\in\mathcal{I}$. Since the electoral college is used in the US, we prefer to denote \textit{regions} as \textit{states}. As in Section~\ref{sec:MS_sto}, we will assume that for each state $i\in\mathcal{I}$, the fraction of votes at each state $i$ received by candidates $A$ and $B$, and the fraction of abstention votes, $(S_{i}^A,S_{i}^B,S_{i}^{C})$, follows a Dirichlet distribution as in Equation~\eqref{eq:dir_distribution}. One of the consequences of using the Dirichlet distribution in the Electoral College system is the independence between the fraction of votes obtained by a candidate relative to the sum of the candidates' votes, and the abstention. This is formally stated in the following lemma:
\begin{lemma}\label{lem:Dirichlet_Abstencion}
If $(X, Y, Z) \sim \textbf{Dir}_3(a,b,c)$, then the relative value of $X$ with respect to $X+Y$ is independent of $Z$, that is, $\text{Cov}(\frac{X}{X+Y},Z)=0$. Moreover, it holds that $\frac{X}{X+Y}\sim \text{Beta}(a,b)$.
\end{lemma}
\proof{Proof.} See Appendix~\ref{proof:lem:DIR}.
\Halmos
\endproof 
Therefore, to determine the winner in each state $i$, we need to focus on the rv $S_{i}^{A}/(S_{i}^{A}+S_{i}^{B})$, which distributes as $\text{Beta}(k(\alpha_i+x_i),k(\beta_{i}+y_i))$. Let $G$ denote the event that candidate $A$ wins the election, namely $G\iff\sum_j w_j \mathbbm{1}_{\{S^A_j>S^B_j\}} > \sum_j w_j \mathbbm{1}_{\{S^B_j>S^A_j\}}$. Since we are using continuous distributions for the vote outcome in each state, the event of a draw has probability zero. However, there might be a non-zero probability of a draw between the candidates' electoral votes (when $\sum_{i} w_i$ is even and there is a subset of states whose electoral votes add up to half of the country's electoral votes). In this case the draw is broken by tossing a fair coin; we omit this in the given definition of $G$ in order to reduce notation, although we consider it in the computations performed. The optimization problems that candidates face are:
\vspace{-1.0cm}

\begin{multicols}{2}
    \begin{equation} \label{eq:mod_CE_X}
            \begin{aligned}
            & \underset{{\bf x} \in \Delta_n}{\text{\bf{max}}}
            & & \mathbb{P}(G)
        \end{aligned}
    \end{equation}
    \break
    \begin{equation} \label{eq:mod_CE_Y}
        \begin{aligned}
            & \underset{{\bf y} \in \Delta_n}{\text{\bf{max}}}
            & & 1-\mathbb{P}(G)
        \end{aligned}
    \end{equation}
\end{multicols}

\begin{comment}
\begin{multicols}{2}
    \begin{equation} \label{eq:mod_CE_X}
            \begin{aligned}
            & \underset{x}{\text{\bf{max}}}
            & & \mathbb{P}(G) \\
            & \text{\bf{s.t.}} & & \sum_j x_j = 1 \\
            & & & \boldsymbol{x} \geq 0,
        \end{aligned}
    \end{equation}
    \break
    \begin{equation} \label{eq:mod_CE_Y}
        \begin{aligned}
            & \underset{y}{\text{\bf{max}}}
            & & 1-\mathbb{P}(G) \\
            & \text{\bf{s.t.}} & & \sum_j y_j = 1 \\
            & & & \boldsymbol{y} \geq 0.
        \end{aligned}
    \end{equation}
\end{multicols}
\end{comment}

In order to compute $\mathbb{P}(G)$, we use a recursive procedure similar to that in~\cite{kaplan2003new} and \cite{rigdon2009bayesian}. Let $p_{i}$ be the probability that candidate $A$ wins state $i$, and $T_{k}$ be the rv of the number of electoral votes obtained by candidate $A$ from states $1$ to $k$. The recursion is given as
\begin{equation} \label{eq:recurrence} 
\begin{aligned}
   %\mathbb{P}(T_{k+1}=t)&=(1-p_{k+1})\mathbb{P}(T_{k}=t)+p_{k}\mathbb{P}(T_{k}=t-w_{k+1})\qquad\forall k\in\{1,\dots,N-1\}  \\
   \mathbb{P}(T_{k}=t)&=(1-p_{k})\mathbb{P}(T_{k-1}=t)+p_{k}\mathbb{P}(T_{k-1}=t-w_{k})\qquad\forall k\in\{2,\dots,N\}  \\
    \mathbb{P}(T_{1}=t)&=(1-p_{1})\mathbb{1}_{\{t=0\}}+p_{1}\mathbb{1}_{\{t=w_{1}\}}. 
\end{aligned} 
\end{equation}
Let $M\coloneqq \sum_{i=1}^{N}w_{i}$, i.e. the total number of electoral votes. Then, the probability of winning the election for candidate $A$ is given by $\mathbb{P}(G)=\sum_{t=\ceil{\frac{M}{2}}}^{M}\mathbb{P}(T_{N}=t)$ %-\mathbb{1}_{\{\bmod(M,2)=0\}}\frac{\mathbb{P}(T_{N}=t)}{2}$ 
(if $M$ is even $\mathbb{P}(T_{N}=\frac{M}{2}$) must be multiplied by half). As for the probability of winning for candidate $A$ in a state $i$ ($p_i$), this can be computed as the complement of the cdf of the distribution $\text{Beta}(k(\alpha_i+x_{i}),k(\beta_{i}+y_{i}))$ evaluated at $0.5$ (see Lemma~\ref{lem:Dirichlet_Abstencion}). Note that for the limit game where $k\rightarrow 0$, the Bernoulli parameter of each state can be computed in closed form as $p_{i}=(\alpha_i+x_i)/(\alpha_i+x_i+\beta_i+y_i)$. %Note that the higher the value of the parameter $k$, then the more concentrated the pdf of the Beta distribution becomes, which translates in higher sensitivity of the probability of winning $p_i$ with respect to the effort of the candidate for values where $p_{i}$ is close to $0.5$. 

We first explore the search for an equilibrium in pure strategies, and later on move to an equilibrium in mixed strategies.

\begin{definition}\label{def:EC_eq_sto}
An equilibrium in pure strategies under \textbf{EC} is a pair of effort vectors $({\bf x^*},{\bf y^*})\in\Delta_n^2$ such that each vector ${\bf x^{*}}$ and ${\bf y^*}$ is an optimal solution of the respective candidate maximization problems given in Expressions~\eqref{eq:mod_CE_X} and \eqref{eq:mod_CE_Y}.
\end{definition}

%\begin{definition}\label{def:EC_eq_sto_mixed}
%An equilibrium in mixed strategies under \textbf{EC} is a pair of effort vectors $({\bf x^*},{\bf y^*})\in\mathcal{X}\times\mathcal{Y}$ such that each vector ${\bf x^{*}}$ and ${\bf y^*}$ is an optimal solution of the respective candidate maximization problems given in Expressions~\eqref{eq:mod_CE_X} and \eqref{eq:mod_CE_Y}.
%\end{definition}
\subsection{Relationship between MS and EC}
It is interesting to note that there are some equivalences between the games under the MS and EC election systems for certain cases. The following theorems state two of these equivalences:
%The following two observations state some equivalences between the MS and EC under certain cases.
%As a side observation there is an equivalence between the EC and MS when maximizing the expected number of electoral and popular vote respectively.
\begin{theorem}\label{obs:exp_numb_votes}
If the number of electoral votes is proportional to the number of voters, then the two following games are the same:
\begin{itemize}
    \item MS where candidates maximize the expected number of votes with no abstention, i.e., ${\bf \gamma}=0$.
    \item EC where candidates maximize the expected number of electoral votes.
\end{itemize}

\proof{Proof.} See Appendix~\ref{proof:obs:exp_numb_votes}.
\Halmos
\endproof 
\end{theorem}
Although it is more natural that candidates maximize the probability of winning rather than the number of electoral votes obtained in the EC; in reality, a political party with almost no odds of winning might prefer the latter objective as a damage control strategy at the expense of the few chances of winning.

\begin{theorem}\label{obs:prob_winning}
If the number of electoral votes is proportional to the number of voters, then the two following games are equivalent in the limit where $k\rightarrow 0$, in the sense that players' utilities in both cases converge in probability:
\begin{itemize}
    \item MS where candidates maximize the probability of winning with no abstention, i.e., ${\bf \gamma}=0$.
    \item EC where candidates maximize the probability of winning.
\end{itemize}

\proof{Proof.} See Appendix~\ref{proof:obs:prob_winning}.
\Halmos
\endproof 
\end{theorem}
The limit case when $k$ approaches zero induces a ``U'' shaped density function on each state, leading to a highly correlated outcome among voters where either all of them support one candidate or the other. Such a setting is unlikely to be observed in reality.

\subsection{Computing Equilibrium in pure strategies}\label{sec:EC_pure}
\begin{figure}[H]
\begin{center}
\addtolength{\leftskip}{-0.5cm}
\includegraphics[scale = 0.38]{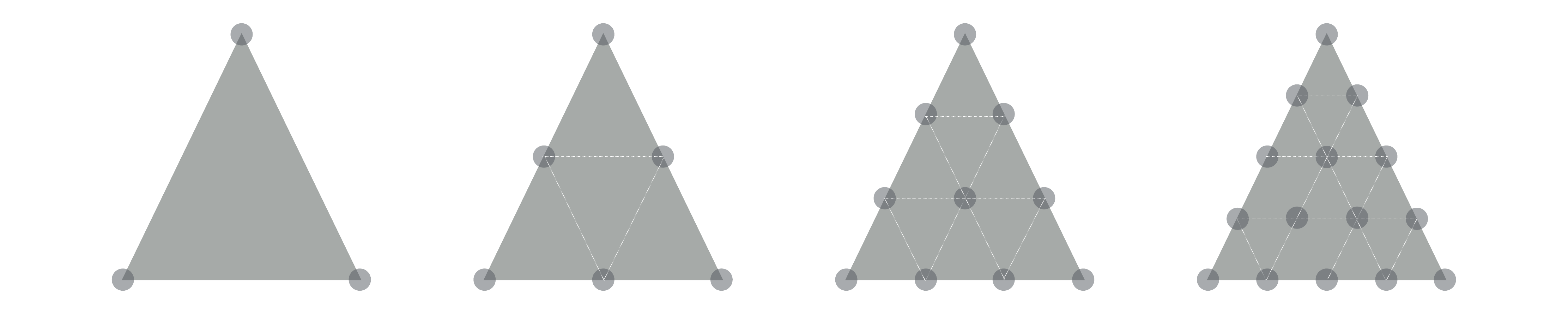}
\end{center}
\caption{Examples of $D^q(\Delta_3)$ for $q = 1, 2, 3, 4$.}
\label{fig:discrete_example}
\end{figure}

We apply a Gradient Descent Ascent method like the one used in Section~\ref{sec:comp_eq_MS_sto} but now with $\mathbb{P}(G)$ as the objective function of the zero-sum game. In this case, the derivative of the payoff function can be written as $\frac{\partial}{\partial x_{i}}\mathbb{P}(G) = \sum_{j}\frac{\partial }{\partial p_j}\mathbb{P}(G)\frac{\partial p_j}{\partial x_i} 
=\frac{\partial }{\partial p_i}\mathbb{P}(G)\frac{\partial p_i}{\partial x_i}$ since $\frac{\partial p_{j}}{\partial x_{i}}=0$ for $i\neq j$. $\frac{\partial p_i}{\partial x_i}$ is the derivative of the complementary cdf of the Beta distribution on $x_{i}$. Let $G_{i}$ be the event that candidate $A$ wins the electoral votes of state $i$, and let $G^{c}_i$ be the complement of this event. The Law of Total Probability implies that $\mathbb{P}(G)=\mathbb{P}(G|G_i)p_i+\mathbb{P}(G|G_i^c)(1-p_i)$, taking derivative with respect to $p_i$ results in $\frac{\partial }{\partial p_i}\mathbb{P}(G)=\mathbb{P}(G|G_i)-\mathbb{P}(G|G_i^c)$. $\mathbb{P}(G|G_i)$ and $\mathbb{P}(G|G_i^c)$ can be computed using the same recurrence as the one introduced in Equations~\eqref{eq:recurrence} but fixing the outcome of the $i^{th}$ state to winning ($p_i=1$) or losing ($p_i=0$) when the conditional event is $G_i$ or $G_i^c$ respectively. Then we get
\begin{equation}\label{eq:ec_grad_x}
\begin{aligned}
\frac{\partial}{\partial x_{i}}\mathbb{P}(G) =&\left(\mathbb{P}(G|G_i)-\mathbb{P}(G|G_i^c)\right)k\left(\mathbb{E}[\ln(\frac{S_i^A}{S_i^A+S_i^B})\mathbb{1}_{\{S_i^{A}\ge S_i^{B}\}}]+p_i\theta(k(\alpha_i+x_{i}),k(\beta_i+y_{i}))\right),
\end{aligned} 
\end{equation}
where $\theta(a,b)\coloneqq\psi(a+b)-\psi(a)$ and $\psi$ is the digamma function. Similar for candidate $B$.
%\begin{equation}\label{eq:ec_grad_y}
%\begin{aligned}
%\frac{\partial}{\partial y_{i}}\mathbb{P}(G) =&\left(\mathbb{P}(G|G_i)-\mathbb{P}(G|G_i^c)\right)k\left(\mathbb{E}[(1-\ln(\frac{S_i^A}{S_i^A+S_i^B}))\mathbb{1}_{\{(S_i^A\ge S_i^B\}}]+p_i\theta(k(\beta_{i}+y_i),k(\alpha_i+x_{i}))\right), 
%\end{aligned} 
%\end{equation}
With this, we can use the same procedure described in Section~\ref{sec:comp_eq_MS_sto}, in particular the use of Proposition~\ref{pro:grad}, and Algorithm~\ref{alg:1}. We find from numerical computations that the gradient descent ascent method converges to a point which, at least numerically, appears to be either an Equilibrium, or a Local Nash Equilibrium. In particular, the parameter $k$ (which controls for variability) seems to have a key role in this. Low values of $k$ lead to cases with existence of Equilibrium, whereas high values of $k$ tend to end up in a Local Nash Equilibrium. Intuitively, the latter case resembles a deterministic version of the game, where pure equilibrium does not seem plausible since in the extreme case (of $k\rightarrow\infty$), payoff functions are not even continuous. 
As a result of the lack of Equilibrium in pure strategies, we explore equilibrium in mixed strategies which indeed do exist.
%Back to the general game under \textbf{EC}, i.e. $k>0$, since the introduced algorithm does not always converge to an equilibrium in pure strategies, which is not guaranteed to exist, we explore equilibrium in mixed strategies which indeed exist.
\begin{theorem}\label{the:ec_ex_mix}
There exists an equilibrium in mixed strategies for the stochastic game under \textbf{EC}.
\end{theorem}
\proof{Proof.} The proof follows the same arguments given in Appendix~\ref{proof:the:sto:ex}.
\Halmos
\endproof 
Unfortunately, the search for a mixed equilibrium of the game is not a simple task. Furthermore, it might result in complicated strategies which might not be practical for the agents involved. As a result, we decide to look for mixed equilibria of the game in a finite subset of strategies. Note that since this is a zero-sum game, if the subset of strategies is finite, we can obtain a mixed equilibrium by simply solving an LP. More precisely, consider the \textit{$(n,q)$-simplex lattice} (as introduced in \cite{scheffe1958experiments}) as the set of points $D^q(\Delta_n)\coloneqq\{{\bf x}\in\mathbb{R}^{n}|x_i\ge0,\sum_{i=1}^nx_i=1,x_iq\in\mathbb{Z}_{+}\}$ where $q\in\mathbb{Z}_{+}$. The latter set represents a discretization of the simplex, (note also that we are considering just strategies where $\sum_{i}x_{i}=1$), in which the parameter $q$ controls for the refinement of the grid so that higher values of this parameter result in a more refined set, see Figure~\ref{fig:discrete_example}. Intuitively, the $(n,q)$-simplex lattice represents the players' strategies, so that each player is endorsed with $q$ indistinguishable \textit{balls} which they have to invest among the $n$ states. The number of elements in the $(n,q)$-simplex lattice is $\binom{n+q}{n}$, which unfortunately is exponential in the number of states and on the parameter $q$. However, we can think of a way to consider only a subset of strategies in $D^{q}(\Delta_n)$. Then, in order to find an equilibrium in mixed strategies, we consider the following an iteration procedure:
\begin{enumerate}
\item Start with a subset of strategies for both players of the set $D^q(\Delta_n)$
\item Find the equilibrium in mixed strategies
\item Explore the players' best responses in pure strategies in the action space $\Delta_n$.
\item For each player's best response at Step 3, obtain the vertices within the simplex lattice that generate the smallest convex hull that contains the best response. Add these vertices to the respective player strategy sets. If for both players there are no new strategies to be added, finish; otherwise, go to Step 2.
\end{enumerate}

In order to formalize the latter, consider the strategy sets $T^{A},T^{B}\subseteq D^{q}(\Delta_n)$ for each respective player. For Step 1, we start with a small set of strategies for both players. For example, these can be the $n$ canonical vectors ${\bf e_{i}}$ which have a 1 in the $i$\textsuperscript{th} component and 0 elsewhere, for $i=1,\dots,n$. For Step 2, define an equilibrium in mixed strategies where players have a finite set of strategies.
\begin{definition}\label{def:EC_eq_sto_finite}
An equilibrium in mixed strategies under \textbf{EC} for discrete sets of strategies $T^A$ and $T^{B}$ is a pair of vectors $({\boldsymbol \sigma_{A}^*},{\boldsymbol \sigma_B^*})\in \Delta_{|T^A|}\times\Delta_{|T^B|}$ such that ${\boldsymbol \sigma_{A}^{*}}\in\argmax_{{\boldsymbol \sigma}}\;\mathbb{E}_{{\boldsymbol \sigma},{\boldsymbol\sigma_B^{*}}}[\mathbb{P}(G)]$ and ${\boldsymbol \sigma_{B}^{*}}\in\argmax_{{\boldsymbol \sigma}}\;\mathbb{E}_{{\boldsymbol \sigma_{A}^{*}},{\boldsymbol\sigma}}[\mathbb{P}(G)]$.
\end{definition}
As mentioned, this equilibrium can be found by solving an LP due to the zero-sum game structure. Let $\boldsymbol{P}\in\mathbb{R}^{|T^{A}|\times|T^{B}|}$ be the payoff matrix of the first player, namely $P_{ij}=\mathbb{P}(G)$ where $\boldsymbol{x}$ is set to the $i$\textsuperscript{th} strategy of $T^{A}$, and similarly $\boldsymbol{y}$ with $T^{B}$. We can solve the following LPs
\vspace{-1.0cm}
\begin{multicols}{2}
    \begin{equation} \label{eq:gmaelp_a}
        \begin{aligned}
            \underset{v,\boldsymbol{\sigma_A}}{\max}
            & & v\qquad\qquad \\
            \text{s.t.} & & v\boldsymbol{e_B} -\boldsymbol{P^{T}} \boldsymbol{\sigma_B}\le 0 \\
            & & \boldsymbol{e^{T}_{A}}\boldsymbol{\sigma_{A}} = 1 \\
            & & \boldsymbol{\sigma_A} \geq 0,
        \end{aligned}
    \end{equation}
\break
    \begin{equation} \label{eq:gmaelp_b}
        \begin{aligned}
            \underset{u,\boldsymbol{\sigma_B}}{\min}
            & & u\qquad\quad \\
            \text{s.t.} & & u\boldsymbol{e_A} -\boldsymbol{P^{T}} \boldsymbol{\sigma_B}\ge 0 \\
            & & \boldsymbol{e^{T}_{B}}\boldsymbol{\sigma_{B}} = 1 \\
            & & \boldsymbol{\sigma_B} \geq 0,
        \end{aligned}
    \end{equation}
\end{multicols}
\noindent where $\boldsymbol{e_{A}}$ and $\boldsymbol{e_{B}}$ are vectors with $|T^{A}|$ and $|T^{B}|$ ones respectively.
\begin{comment}
\begin{equation} \label{eq:gmaelp_b}
    \begin{aligned}
        \underset{u,\boldsymbol{\sigma_B}}{\min}
        & & u\qquad\quad \\
        \text{s.t.} & & u\boldsymbol{e_A} -\boldsymbol{P^{T}} \boldsymbol{\sigma_B}\ge 0 \\
        & & \boldsymbol{e^{T}_{B}}\boldsymbol{\sigma_{B}} = 1 \\
        & & \boldsymbol{\sigma_B} \geq 0
    \end{aligned}
\end{equation}
\noindent where $\boldsymbol{e_{A}}$ and $\boldsymbol{e_{B}}$ are $|T^{A}|$ and $|T^{B}|$-dimensional vectors with ones respectively. From Optimization problem~\eqref{eq:gmaelp_b} we obtain the optimal probabilities for the second player $\boldsymbol{\sigma_{B}^{*}}$. The optimal probabilities for the first payer can be obtained solving the following analogous LP
\begin{equation} \label{eq:gmaelp_a}
    \begin{aligned}
        \underset{\boldsymbol{\sigma_A}}{\max}
        & & v \\
        \text{s.t.} & & v\boldsymbol{e_B} -\boldsymbol{P^{T}} \boldsymbol{\sigma_B}\le 0 \\
        & & \boldsymbol{e^{T}_{A}}\boldsymbol{\sigma_{A}} = 1 \\
        & & \boldsymbol{\sigma_A} \geq 0
    \end{aligned}
\end{equation}
\end{comment}
See Appendix~\ref{app:lpgame} for details.
%The details can be found in Appendix~\ref{app:lpgame}.

For Step 3, we need to find the player's best responses in pure strategies in $\Delta_n$ given a mixed strategy of their opponent. In order to do this, we use a gradient descent method for each player considering the expectation of the objective derivative according to the contender' mixture probabilities. Thus, 
%where the derivatives are the expectation of the expressions in Equations~\eqref{eq:ec_grad_x} and \eqref{eq:ec_grad_y} with respect to contender player' mixture probabilities (
candidate $A$ considers the expectation with respect to $\boldsymbol{\sigma_{B}^{*}}$, whereas candidate $B$ does it with respect to $\boldsymbol{\sigma_{A}^{*}}$. %Note that there is an optimal solution since the objective is a continuous function and the feasible space ($\Delta_n$) is compact. 
For Step 4, consider ${\bf x}\in\Delta_n$ as the best response of candidate $A$ (wlog). We would like to find the points in the simplex lattice that contain {\bf $x$} in its convex hull, while at the same time, being as small as possible (in the sense that there is no other subset of these points that also contains ${\bf x}$ in its convex hull). Recall that the grid refinement of the simplex lattice is given by the parameter $q$, which implies that all components of the points in $D^{q}(\Delta_n)$ are multiples of $1/q$. Then, we can remove from the analysis the fractional part of {\bf $x$} that is a multiple of $1/q$ and stay with the remainder; namely, we can define ${\bf x^r}\coloneqq {\bf x}-\frac{\lfloor q{\bf x} \rfloor}{q}\in[0,\frac{1}{q})^n$, where the floor function is applied for each component. Amplifying ${\bf x^r}$ by $q$ leads to $q{\bf x^r}\in[0,1)^n$. Let us define $m\coloneqq \sum_{i=1}^{n}qx^r_i$, then we can state the following claim:
\begin{claim}\label{cla:sumq}
It holds that $m\in\{0,1,\dots,n-1\}$.
\end{claim}
\proof{Proof.} See Appendix~\ref{proof:cla:sumq}.
\Halmos
\endproof 
Let $\mathcal{Y}\coloneqq\{{\bf y}\in\{0,1\}^n|\sum_{i=1}^{n}y_i=m\}$. Then, we aim to find a subset of vectors in $\mathcal{Y}$  
such that $q{\bf x^r}$ can be written as a convex combination of these. This results in $\binom{n}{m}$ vectors from which to choose. Thus, we use the following algorithm:
\begin{algorithm}[H] 
\caption{Finding binary vectors that contain fractional point}
\label{Alg_Discret}
\begin{algorithmic}[1]
\State \textbf{Input} $y \in [0,1)^n$, s.t. $m=\sum_{i=1}^n y_i\in\{0,1,\dots,n-1\}$
 \State \textbf{Set} $\mathcal{Z} = \varnothing$, $w = y$
 \State \textbf{While} $w \not\in \{ 0,1 \}^n$ \textbf{do}
 %\State \qquad \textbf{If } $w \in \{ 0,1 \}^n$\textbf{ do }
 %\State \qquad \qquad $\mathcal{Z} = \mathcal{Z} \cup \{w\}$
 %\State \qquad \qquad \textbf{Break}
 %\State \qquad \textbf{End If}
 \State \qquad $z = \arg\min_{v\in\mathcal{Y}}\;\|w-v\|_2^2\;$%  s.t. $\sum_{i=1}^{n}v_i=m$
 \State \qquad $\mathcal{Z} = \mathcal{Z} \cup \{z\}$
 \State \qquad $t = \min\left\{\min_{i: w_i < z_i} \left\{ \frac{w_i}{z_i-w_i} \right\},
             \min_{i: w_i > z_i}\left\{ \frac{1 - w_i}{w_i-z_i} 					 \right\}
             \right\}$
 \State \qquad $w = w + t \times (w-z)$
 \State \textbf{End While}
 \State $\mathcal{Z}=\mathcal{Z}\cup \{w\}$
 \State \textbf{Return} $\mathcal{Z}$
\end{algorithmic}
\end{algorithm}
Given a fractional point, Algorithm~\ref{Alg_Discret} returns a set with vectors in $\mathcal{Y}$ that contain the fractional point in its convex hull. In particular, at each iteration: In line 4, it finds the closest point in $\mathcal{Y}$ to the current point $w$. Note that the objective function in line 4 can be written as $\| {\bf w} - {\bf v} \|_2^2 = \sum_j w_j^2 + \sum_j v_j^2 - 2 \sum_j v_j w_j = \| {\bf w} \|^2_2 + m - 2 \sum_{j: v_j = 1} w_j$, which optimum is attained in the $n$-dimensional binary vector with ones in the $m$ largest components of ${\bf w}$, and $0$ elsewhere. Ties can be broken randomly. In line 5, we add ${\bf z}$ to the output set $\mathcal{Z}$. In lines 6 and 7, we move from the current point ${\bf w}$ along the direction ${\bf w-z}$ until the first component reaches $0$ or $1$ (from a different previous value), updating the current point ${\bf w}$. This cycle repeats until the current point is binary, in which case we stop iterating, and add this point to the set $\mathcal{Z}$ (line 9). The next Lemma states some properties of Algorithm~\ref{Alg_Discret}.
\begin{lemma}\label{lem:algp}
Denote ${\bf z^{(k)}}$ as the $k$\textsuperscript{th} point added in the set $\mathcal{Z}$ in Algorithm~\ref{Alg_Discret}, and $t^{(k)}$ as the value of the scalar $t$ in the $k$\textsuperscript{th} iteration of Algorithm~\ref{Alg_Discret}.
\begin{enumerate}
\item [(i)] The algorithm finishes in at most $n$ iterations, i.e., $|\mathcal{Z}|\le n$.
\item [(ii)] The weights of the convex combination of the algorithm input ${\bf y}$ can be computed as $\lambda_{k}=\frac{t^{(k)}}{1+t^{(k)}}\prod_{j=1}^{k-1}\frac{1}{1+t^{(j)}}\;\forall k<|\mathcal{Z}|$, $\lambda_{|\mathcal{Z}|}= \prod_{j=1}^{|\mathcal{Z}|}\frac{1}{1+t^{(j)}}$ where $\lambda_{k}>0$ for all $k$.
\end{enumerate}
\end{lemma}
\proof{Proof.} See Appendix~\ref{proof:lem:algp}.
\Halmos
\endproof 

We can see from Lemma~\ref{lem:algp} that the output of points to be returned is linear (at most $n$). Indeed, at each iteration, the algorithm fixes one component of ${\bf w}$ to $0$ or $1$, keeping its value fixed for the rest of the iterations. The latter is consequent with the second claim of Lemma~\ref{lem:algp}. It is interesting to observe that Algorithm~\ref{Alg_Discret} not only returns the vertices from which the input point ${\bf y}$ is a convex combination, but also, the weights of the convex combination by using the intermediate computations of the algorithm. Algorithm~\ref{Alg_Discret}] runs in $\mathcal{O}(n^2)$.

\begin{figure}[H]
\begin{center}
\addtolength{\leftskip}{-0.5cm}
\includegraphics[scale = 0.38]{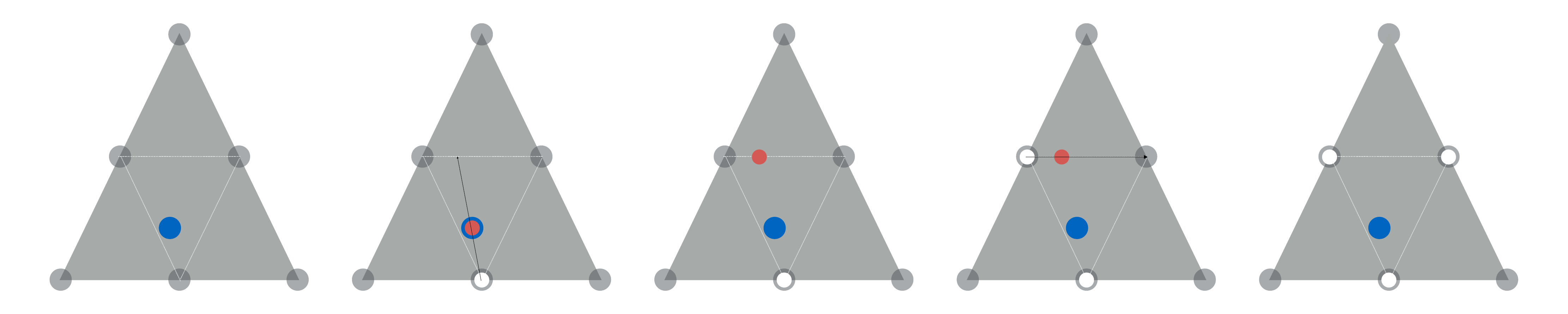}
\end{center}
\caption{Algorithm~\ref{Alg_Discret}. The blue point is the best response in $\Delta_n$ to discretize. The white dot of the second panel represents $z$ of the first iteration, while the black black segment the ray $w-z$. In the third panel, the red dot represents the new point $w$. The new point $z$ is depicted in red in the fourth panel, and finally the last point $w$ is obtained in the fifth panel in white.}\label{fig:discrete}
%Algorithm~\ref{Alg_Discret} from top left panel to top right, and then from bottom left to bottom right. The blue point is the best response on $\Delta_n$ to discretize. The white dot of the top right panel represents $z$ of the first iteration, while the black arrow denotes the ray $w-z$. The red dot represents in the bottom left panel is new point $w$. The bottom middle panel shows the the new point $z$ as the upper white point, and finally we get in the bottom right panel the last point (the white bottom right point) that represents the last visited point $w$.
%\floatfoot{Secuencia del algoritmo para $n=3$, $q = 5$. Punto azul es el vector input $x$. Punto rojo simboliza el vector $w$ dentro del algoritmo. Puntos en blanco indican conjunto de vectores $\mathcal{Z}$, que aparecen según función $h(w,m)$. Flechas negras indican el movimiento según lo estipulado por el escalar $t$, en dirección .  Fuente: Elaboración Propia. } 
\end{figure}

Then, we can use Algorithm~\ref{Alg_Discret} with ${\bf y}=q{\bf x^r}$ in order to obtain at most $n$ points in $\mathcal{Y}$, and transform these points to the original scale (recall that Algorithm~\ref{Alg_Discret} works in the $[0,1]$ hypercube, independent of $q$). More precisely, if ${\bf x}\in\Delta_n$ is the best response of one of the players, then the simplex lattice points to be added to the player strategy set are $\overline{\mathcal{Z}}({\bf x})\coloneqq\{\frac{\lfloor q{\bf x} \rfloor}{q}+\frac{{\bf z^{k}}}{q}|{\bf z^{k}}\in\mathcal{Z}(q{\bf x^r})\}$ (note the cardinality is at most $n$), where we use the notation $\mathcal{Z}(q{\bf x^r})$ to denote the output from Algorithm~\ref{Alg_Discret} with input ${\bf y}=q{\bf x^r}$. Figure~\ref{fig:discrete} shows how Algorithm~\ref{Alg_Discret} iterates in the original space $\Delta_n$ with $n=3$ and the points in the grid that conform the simplex lattice.
Finally, it can be stated that added points in set $\overline{\mathcal{Z}}$ are \textit{minimal} in the sense that there is no other set in $D^{q}(\Delta_n)$, for example $\mathcal{W}$, that generates {\bf $x$} as a convex combination such that $\text{Conv}(\mathcal{W})\subsetneq\text{Conv}(\mathcal{\overline{Z}})$. The formal proof is given in Appendix~\ref{app:minimal}. In summary, the full algorithm for finding a mixed equilibrium on subsets of the simplex lattice as strategy sets is given below.
\begin{algorithm}[H] 
\caption{Algorithm for mixed equilibrium in subsets of the simplex lattices}
\label{Alg_CE}
\begin{algorithmic}[1]
\State \textbf{Input} $\alpha, \beta \in \mathbb{R}_{+}^n$, $k \in \mathbb{R}_{+}$, $w \in \mathbb{Z}_+^n$, $q \in \mathbb{Z}_+$
 \State \textbf{Set} $T^A, T^B = \{ e_i \in \mathbb{R}^n, i \in \mathcal{I} \}$, $P_{ij} = \mathbb{P}(G|x^{(i)},y^{(j)}) \, \forall (x^{(i)},y^{(j)}) \in T^A \times T^B$
 \State \textbf{While} \text{True do:}
 \State \qquad $(\sigma_A,\sigma_B) = \text{solve}(P)$
 \State \qquad $x^{BR} = \arg\max_{x\in\Delta_n} \mathbb{P}(G|x,\sigma_{B})$, $y^{BR} = \arg\min_{y\in\Delta_n} \mathbb{P}(G|\sigma_{A},y)$
 \State \qquad \textbf{If } $\mathcal{\overline{Z}}(x^{BR}) \subseteq T^A$ \textbf{ and } $\mathcal{\overline{Z}}(y^{BR}) \subseteq T^B$ \textbf{ :  Break}
 %\State \qquad \qquad  
 %\State \qquad  \textbf{End If}
 \State \qquad $T^A = T^A \cup \mathcal{\overline{Z}}(x^{BR})$, $T^B = T^A \cup \mathcal{\overline{Z}}(y^{BR})$
 \State \qquad \textbf{Update } $P$
 \State \textbf{End While}
 \State \textbf{Return} $(T^A,\sigma_A,T^B,\sigma_B)$
\end{algorithmic}
\end{algorithm}
In line 2 of Algorithm~\ref{Alg_CE} the strategy sets are initialized with the canonical vectors, and the payoff matrix, $P$, is computed under the pairs of these strategies. Line 4 computes the mixed equilibrium by solving Optimization problems~\eqref{eq:gmaelp_b} and \eqref{eq:gmaelp_a}. Lines 5 and 6 compute each candidate best response on the continuous space $\Delta_n$. In line 7, we compute the discretized vectors in the simplex lattice for the players best responses; if these sets, $\mathcal{\overline{Z}}({\bf x^{BR})}$ and $\mathcal{\overline{Z}}({\bf y^{BR}})$, are already contained in the respective players strategy sets ($T^A$ and $T^B$), the algorithm finishes. If this is not the case, the new discretized strategies are added to the player's strategy sets in lines 11 and 12. Finally, in line 12 the payoff matrix is updated to include the payoffs for the pairs of players' strategies that involve new strategies.

\section{Numerical Results}\label{sec:numresults}
In this section, we show numerical computations of the game equilibria under the different settings introduced under the Majority and Electoral College voting systems. Then we analyze the effect of polarization for both election systems.

\subsection{Majority System}

\subsubsection{Deterministic case:}

\begin{comment}

We consider the instance with parameters according to Table~\ref{tab:instance_PS}. The instance consist in ten regions, each with different votes, biases, and abstention parameters. 

\begin{table}[h]
\begin{center}
\begin{tabular}{|c|c|c|c|c|}
\hline
\textbf{Region} & \textbf{v {[}\%{]}} &  \textbf{$\boldsymbol{\alpha}$ {[}\%{]}} & \textbf{$\boldsymbol{\beta}$ {[}\%{]}} & \textbf{$\boldsymbol{\gamma}$ {[}\%{]}} \\ \hline
1 & 23.2 & 45 & 71 & 94 \\ 
2 & 18.5 & 68 & 37 & 67 \\ 
3 & 14.4 & 32 & 24 & 121 \\ 
4 & 8.9 & 43 & 39 & 89 \\ 
5 & 8.2 & 76 & 65 & 92 \\ 
6 & 8.2 & 36 & 61 & 143 \\ 
7 & 6.8 & 51 & 54 & 45 \\ 
8 & 6.2 & 42 & 41 & 79 \\ 
9 & 3.4 & 85 & 31 & 102 \\ 
10 & 2.1 & 37 & 69 & 68 \\ \hline
\end{tabular}
\end{center}
\caption{Instance with 10 regions. Votes are in percentage units so $\sum_{j}v_{j}=1$. Bias and abstention parameters are in percentage units (i.e. values are multiplied by $100$) in order to have them in the same units as the effort vectors $x$ and $y$ which will be expressed in percentage units.}
\label{tab:instance_PS}
\end{table}

\end{comment}

%The unbounded equilibrium has negative components, and therefore it can not be the equilibrium of the original game (of plurality system under the deterministic model). Then, we proceed to solve the equilibrium for the game with non-negative constraints, see details at the end of Section~\ref{sec:PSdet}.

We focus our analyses on cases where strategies are non-negative. Numerical results of the unbounded game (i.e., without non-negativity constraints) are shown in Appendix~\ref{app:num_res_ub}. Table~\ref{tab:eqPSdet} shows the equilibrium for the deterministic game under MS in an instance composed of ten regions. The region's vote-share, biases, and abstention parameters are given in columns 2, 3-4, and 5 respectively. We observe that only the first three regions, which are the ones with the largest vote-share, are chosen to invest in by both candidates. Adding up the results from all regions, we see that candidate $A$, candidate $B$, and the abstentions are $30.0\%$, $28.9\%$, and $41.1\%$ respectively. Thus, candidate $A$ wins by obtaining $50.9\%$ of the votes in the election between $A$ and $B$. Among the regions that candidates invest in, it can be seen from Table~\ref{tab:eqPSdet} that they allocate most of their resources into regions in which the bias is leaning towards their contender. For example, candidate $A$ has a less favorable bias in region $1$ compared to candidate $B$ (i.e., $\alpha_1<\beta_1$); consequently, in equilibrium candidate $A$ ends up investing 
%candidate $B$ has a more favorable bias in region $1$ than candidate $A$ ($\beta_{1}>\alpha_{1}$)%, thus at same levels of effort in this region, candidate $B$ will obtain the majority of its votes. 
%, while in the equilibrium, candidate $A$ ends up investing 
$68.3\%$ of its resources into this region (versus $36.4\%$ for candidate $B$). As a result, the number of votes from the turnover in region 1 that candidate $A$ ends up with is slightly more than half. (See last column of Table~\ref{tab:eqPSdet}). As for the second region, the biases and candidate investments are in the opposite direction compared to the first region. Also, note that the two first regions end up getting the highest turnout nationwide. This is explained by the high number of votes there, which induce candidates' efforts to be focused on them.

The equilibrium shown in Table~\ref{tab:eqPSdet} is computed using procedure described at the end of Section~\ref{sec:PSdet}. Nonetheless, since we know \textit{ex post} that candidates focus exclusively on the first three regions, the equilibrium could have been computed using the closed form expression from Proposition~\ref{pro:CUB} with $\mathcal{I}^*=\{1,2,3\}$.
%Note that, had we know that the first three regions of the country would receive all the efforts from both candidates, we could have use Proposition~\ref{pro:CUB} to calculate the outcome without needing any algorithm. 

\begin{table}[h]
\begin{center}
\begin{tabular}{|c|c|c|c|c|c|c|c|c|c|}
\hline
\textbf{Region} & \textbf{v {[}\%{]}} & \textbf{$\boldsymbol{\alpha}$ {[}\%{]}} & \textbf{$\boldsymbol{\beta}$ {[}\%{]}} & \textbf{$\boldsymbol{\gamma}$ {[}\%{]}} & \textbf{$\boldsymbol{x}$ {[}\%{]}} & \textbf{$\boldsymbol{y}$ {[}\%{]}} & \textbf{Turnout {[}\%{]}} & \textbf{VFT $A$ {[}\%{]}} \\ \hline
1 & $23.3$ & $45$ & $71$ & $94$ & $68.3$ & $36.4$ & $70.1$ & $51.3$ \\ 
2 & $18.5$ & $68$ & $37$ & $67$ & $25.8$ & $52.1$ & $73.2$ & $51.3$ \\ 
3 & $14.4$ & $32$ & $24$ & $121$ & $5.9$ & $11.5$ & $37.8$ & $51.7$ \\ 
4 & $8.9$ & $43$ & $39$ & $89$ & $0.0$ & $0.0$ & $48.0$ & $52.4$ \\ 
5 & $8.2$ & $76$ & $65$ & $92$ & $0.0$ & $0.0$ & $60.5$ & $53.9$ \\ 
6 & $8.2$ & $36$ & $61$ & $143$ & $0.0$ & $0.0$ & $40.4$ & $37.1$ \\
7 & $6.8$ & $51$ & $54$ & $45$ & $0.0$ & $0.0$ & $70.0$ & $48.6$ \\ 
8 & $6.2$ & $42$ & $41$ & $79$ & $0.0$ & $0.0$ & $51.2$ & $50.6$ \\ 
9 & $3.4$ & $85$ & $31$ & $102$ & $0.0$ & $0.0$ & $53.2$ & $73.3$ \\
10 & $2.1$ & $37$ & $69$ & $68$ & $0.0$ & $0.0$ & $60.9$ & $34.9$ \\  \hline
\end{tabular}
\end{center}
\caption{Equilibrium quantities under MS in the deterministic model. \textbf{Turnout} column represents the percentage of votes that goes either to candidate $A$ or $B$, this is computed as $(x_i+\alpha_i+y_i+\beta_i)/(x_i+\alpha_i+y_i+\beta_i+\gamma_i)$ for each region $i$. \textbf{VFT} $A$ represents the expected \textbf{V}otes \textbf{F}rom \textbf{T}urnout that go to candidate $A$, this is computed as $(x_i+\alpha_i)/(x_i+\alpha_i+y_i+\beta_i)$ for each region $i$.}
\label{tab:eqPSdet}
\end{table}

Candidates' investments in equilibrium shown in Table~\ref{tab:eqPSdet} is not just the result of the size and biases of regions, but also of their abstention. In order to understand the impact of the latter, we analyze the same instance shown in Table~\ref{tab:eqPSdet} with no abstention, i.e., $\gamma_i=0$ for every region $i$. 
%We analyze the case with no abstention. For this, consider the same instance given in Table~\ref{tab:eqPSdet} but with $\gamma_i=0$ for every region $i$. 
The equilibrium is given in Table~\ref{tab:eqPSdet_G0} from which we can make the following two observations: First, note that compared to the equilibrium obtained in the case with abstention (see Table~\ref{tab:eqPSdet}), both candidates shift some of their efforts from the first two regions onto the third one. This is for two reasons: (a) when setting the abstention parameters to zero, the third region becomes more attractive to invest in since it has one of the biggest abstention parameters in the original instance, and (b) the bias parameters ($\alpha_3$ and $\beta_3$) are the lowest and one of the closest among all the regions, and therefore, it is easier to influence the voters of that region. Second, since both candidates are investing in the first three regions under a no abstention setting, the result of Corollary~\ref{cor:UBG0} holds. Namely, the fraction of votes obtained by candidate $A$ is the same in all these regions (see last column of Table~\ref{tab:eqPSdet_G0}). %Indeed, we can see from the last column of Table~\ref{tab:eqPSdet_G0} that candidate $A$ gets the same fraction of votes (from the turnout) in the three first regions.}

\begin{table}[h]
\begin{center}
\begin{tabular}{|c|c|c|c|c|c|c|c|c|c|}
\hline
\textbf{Region} & \textbf{v {[}\%{]}} & \textbf{$\boldsymbol{\alpha}$ {[}\%{]}} & \textbf{$\boldsymbol{\beta}$ {[}\%{]}} & \textbf{$\boldsymbol{\gamma}$ {[}\%{]}} & \textbf{$\boldsymbol{x}$ {[}\%{]}} & \textbf{$\boldsymbol{y}$ {[}\%{]}} & \textbf{Turnout {[}\%{]}} & \textbf{VFT $A$ {[}\%{]}} \\ \hline
1 & $23.3$ & $45$ & $71$ & $0$ & $56.6$ & $25.2$ & $100.0$ & $51.4$ \\ 
2 & $18.5$ & $68$ & $37$ & $0$ & $12.7$ & $39.4$ & $100.0$ & $51.4$ \\ 
3 & $14.4$ & $32$ & $24$ & $0$ & $30.7$ & $35.4$ & $100.0$ & $51.4$ \\ 
4 & $8.9$ & $43$ & $39$ & $0$ & $0.0$ & $0.0$ & $100.0$ & $52.4$ \\ 
5 & $8.2$ & $76$ & $65$ & $0$ & $0.0$ & $0.0$ & $100.0$ & $53.9$ \\ 
6 & $8.2$ & $36$ & $61$ & $0$ & $0.0$ & $0.0$ & $100.0$ & $37.1$ \\ 
7 & $6.8$ & $51$ & $54$ & $0$ & $0.0$ & $0.0$ & $100.0$ & $48.6$ \\ 
8 & $6.2$ & $42$ & $41$ & $0$ & $0.0$ & $0.0$ & $100.0$ & $50.6$ \\ 
9 & $3.4$ & $85$ & $31$ & $0$ & $0.0$ & $0.0$ & $100.0$ & $73.3$ \\ 
10 & $2.1$ & $37$ & $69$ & $0$ & $0.0$ & $0.0$ & $100.0$ & $34.9$ \\  \hline
\end{tabular}
\end{center}
\caption{Equilibrium quantities under MS with no abstention in the deterministic model. \textbf{Turnout} column represents the percentage of votes that goes either to candidate $A$ or $B$, this is computed as $(x_i+\alpha_i+y_i+\beta_i)/(x_i+\alpha_i+y_i+\beta_i+\gamma_i)$ for each region $i$. \textbf{VFT} $A$ represents the expected \textbf{V}otes \textbf{F}rom \textbf{T}urnout that go to candidate $A$, this is computed as $(x_i+\alpha_i)/(x_i+\alpha_i+y_i+\beta_i)$ for each region $i$.}
\label{tab:eqPSdet_G0}
\end{table}

\subsubsection{Stochastic case:}
We now proceed to solve the MS game under a stochastic model. Table~\ref{tab:eqPSsto} shows the results with $k=10$. It is interesting to observe that the candidates' efforts are similar to those in the deterministic case (see Table~\ref{tab:eqPSdet}). As a result, the previously bias disadvantage effect in which candidates invest more in regions with a smaller bias parameter relative to the contender also holds. The probability that candidate $A$ wins is $57.4\%$.

\begin{table}[h]
\begin{center}
\begin{tabular}{|c|c|c|c|c|c|c|c|c|c|}
\hline
\textbf{Region} & \textbf{v {[}\%{]}} & \textbf{$\boldsymbol{\alpha}$ {[}\%{]}} & \textbf{$\boldsymbol{\beta}$ {[}\%{]}} & \textbf{$\boldsymbol{\gamma}$ {[}\%{]}} & \textbf{$\boldsymbol{x}$ {[}\%{]}} & \textbf{$\boldsymbol{y}$ {[}\%{]}} & \textbf{Turnout {[}\%{]}} & \textbf{VFT $A$ {[}\%{]}} \\ \hline
%1 & $23.3$ & $45$ & $71$ & $94$ & $68.0$ & $35.4$ & $70.0$ & $51.5$ \\ 
%2 & $18.5$ & $68$ & $37$ & $67$ & $26.2$ & $53.9$ & $73.4$ & $50.9$ \\ 
%3 & $14.4$ & $32$ & $24$ & $121$ & $5.9$ & $10.7$ & $37.5$ & $52.2$ \\ 
%4 & $8.9$ & $43$ & $39$ & $89$ & $0.0$ & $0.0$ & $48.0$ & $52.4$ \\ 
%5 & $8.2$ & $76$ & $65$ & $92$ & $0.0$ & $0.0$ & $60.5$ & $53.9$ \\ 
%6 & $8.2$ & $36$ & $61$ & $143$ & $0.0$ & $0.0$ & $40.4$ & $37.1$ \\ 
%7 & $6.8$ & $51$ & $54$ & $45$ & $0.0$ & $0.0$ & $70.0$ & $48.6$ \\ 
%8 & $6.2$ & $42$ & $41$ & $79$ & $0.0$ & $0.0$ & $51.2$ & $50.6$ \\ 
%9 & $3.4$ & $85$ & $31$ & $102$ & $0.0$ & $0.0$ & $53.2$ & $73.3$ \\ 
%10 & $2.1$ & $37$ & $69$ & $68$ & $0.0$ & $0.0$ & $60.9$ & $34.9$ \\ 
1 & $23.3$ & $45$ & $71$ & $94$ & $68.3$ & $36.4$ & $70.1$ & $51.3$ \\ 
2 & $18.5$ & $68$ & $37$ & $67$ & $25.8$ & $52.1$ & $73.2$ & $51.3$ \\ 
3 & $14.4$ & $32$ & $24$ & $121$ & $5.9$ & $11.5$ & $37.7$ & $51.7$ \\ 
4 & $8.9$ & $43$ & $39$ & $89$ & $0.0$ & $0.0$ & $48.0$ & $52.4$ \\ 
5 & $8.2$ & $76$ & $65$ & $92$ & $0.0$ & $0.0$ & $60.5$ & $53.9$ \\ 
6 & $8.2$ & $36$ & $61$ & $143$ & $0.0$ & $0.0$ & $40.4$ & $37.1$ \\ 
7 & $6.8$ & $51$ & $54$ & $45$ & $0.0$ & $0.0$ & $70.0$ & $48.6$ \\ 
8 & $6.2$ & $42$ & $41$ & $79$ & $0.0$ & $0.0$ & $51.2$ & $50.6$ \\ 
9 & $3.4$ & $85$ & $31$ & $102$ & $0.0$ & $0.0$ & $53.2$ & $73.3$ \\ 
10 & $2.1$ & $37$ & $69$ & $68$ & $0.0$ & $0.0$ & $60.9$ & $34.9$ \\  \hline
\end{tabular}
\end{center}
\caption{Equilibrium quantities under MS in stochastic model with $k=10$. \textbf{Turnout} column represents the percentage of votes that goes either to candidate $A$ or $B$, this is computed as $(x_i+\alpha_i+y_i+\beta_i)/(x_i+\alpha_i+y_i+\beta_i+\gamma_i)$ for each region $i$. \textbf{VFT} $A$ represents the expected \textbf{V}otes \textbf{F}rom \textbf{T}urnout that go to candidate $A$, this is computed as $(x_i+\alpha_i)/(x_i+\alpha_i+y_i+\beta_i)$ for each region $i$.}
\label{tab:eqPSsto}
\end{table}

\begin{figure}
    \centering
    \begin{subfigure}{.5\textwidth}
        \centering
        \includegraphics[width=.95\linewidth]{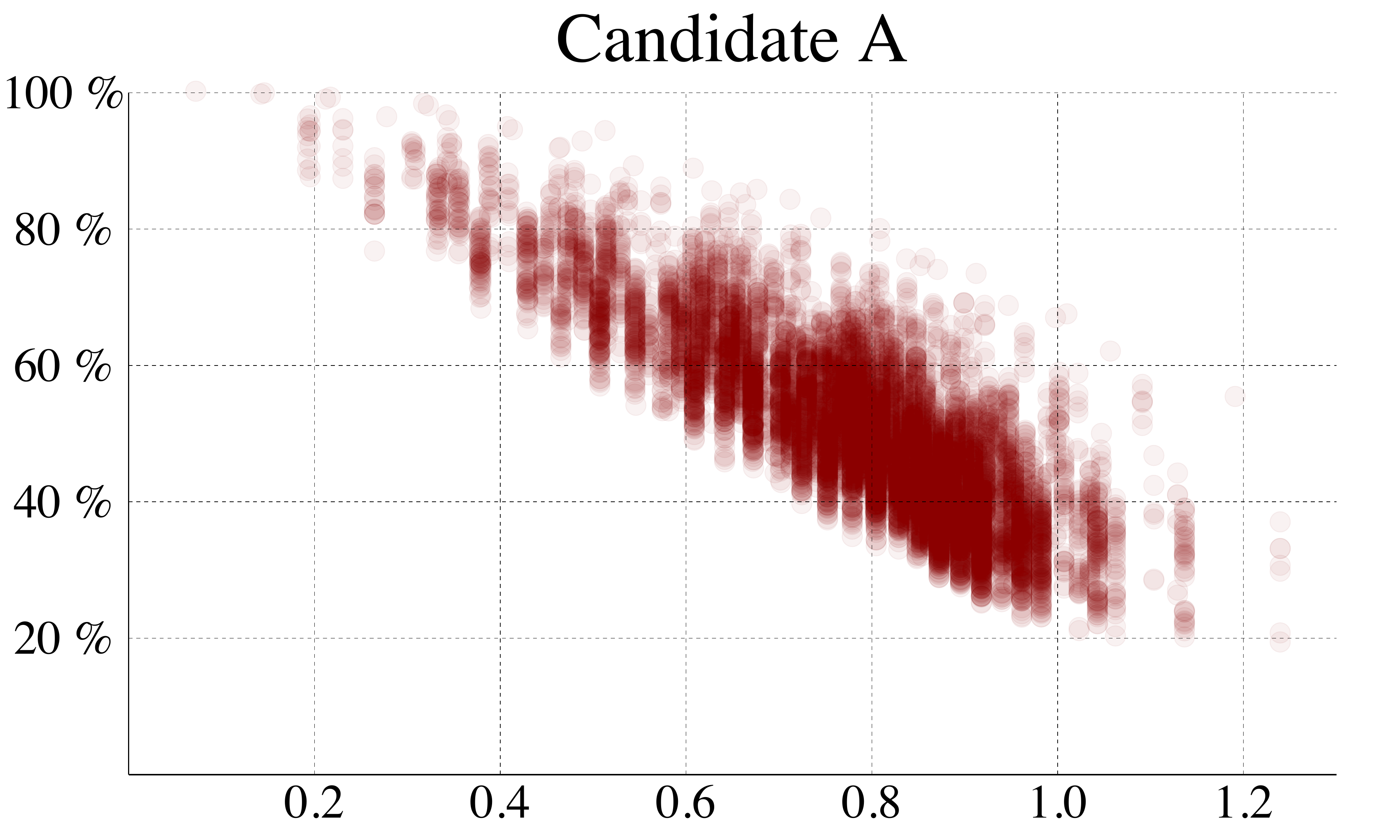}
        %\label{fig:sub1}
    \end{subfigure}%
    \begin{subfigure}{.5\textwidth}
        \centering
        \includegraphics[width=.95\linewidth]{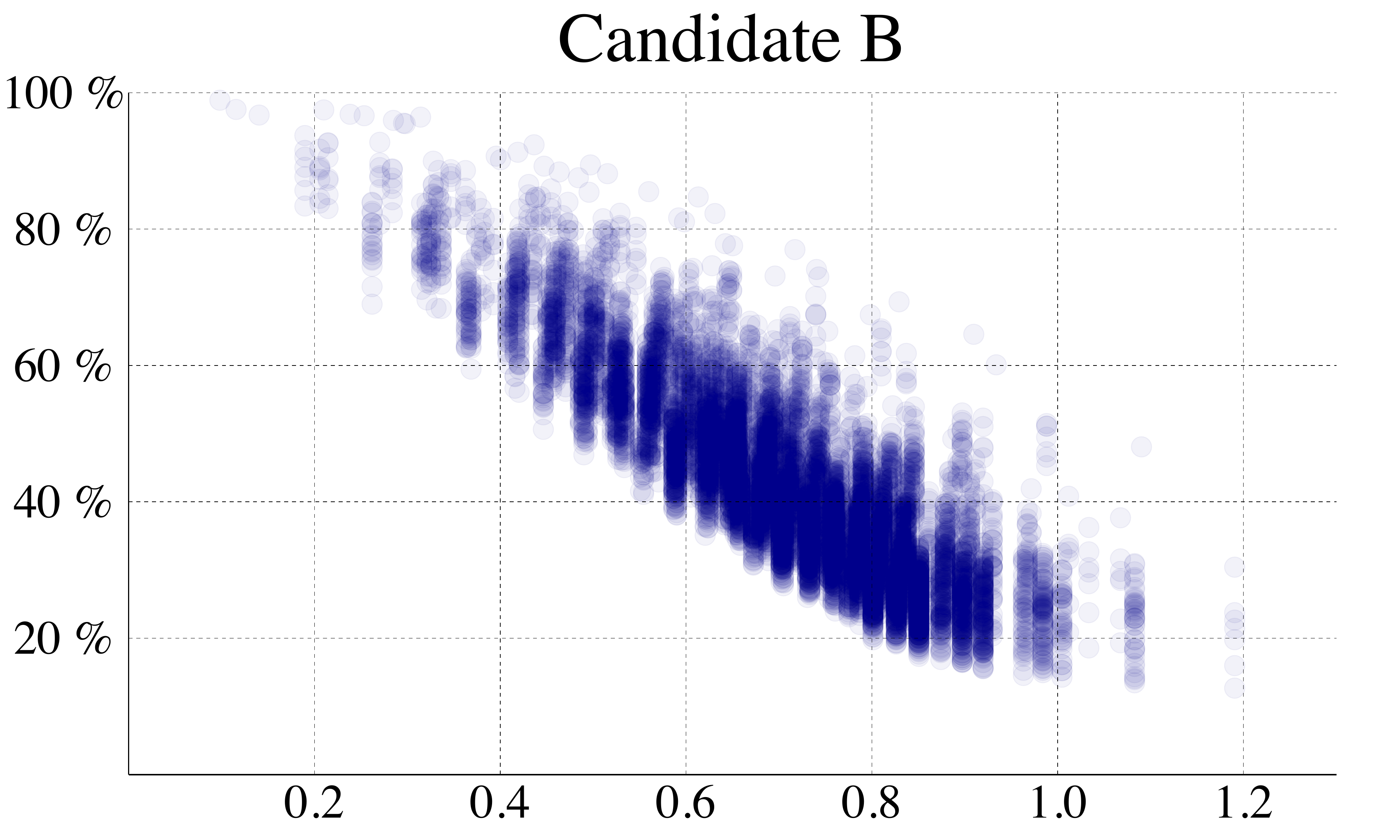}
        %\label{fig:sub2}
    \end{subfigure}
    \caption{Payoff ratio for unilateral deviations for different strategies in $\Delta_n$ for candidate $A$ in the left panel, and candidate $B$ in the right panel.}
    \label{fig:Corroboration_PS}
\end{figure}

The equilibrium quantities given in Table~\ref{tab:eqPSsto} are obtained by running Algorithm~\ref{alg:1}. Recall that we do not have a formal proof of existence and uniqueness of the equilibrium for the stochastic game. Because of the latter, an empirical analysis is performed to test whether or not the strategies obtained are indeed an equilibrium. Figure~\ref{fig:Corroboration_PS} shows the payoff ratio for different unilateral deviations for each player. For example, if the equilibrium obtained is $({\bf x},{\bf y})\in\Delta_n^2$, then the ratio for player $A$ at a strategy ${\bf x'}\in\Delta_n$ is computed as $\frac{\mathbb{P}(R^{A}>R^{B}|{\bf x'},{\bf y})}{\mathbb{P}(R^{A}>R^{B}|{\bf x},{\bf y})}$. Similarly for player $B$. The y-axis of Figure~\ref{fig:Corroboration_PS} corresponds to the ratios for unilateral deviations of both players, while the x-axis represents the Euclidean distance between the equilibrium point and the respective unilateral deviated strategy. Note that every unilateral deviation computed resulted in a ratio below 1. Therefore, it seems that neither player has an incentive to switch its strategies, at least from the testes unilateral deviations.%, none has ration above $100\%$, which means that, at least for the tested unilateral deviations, neither player has an incentive to switch her strategy.

\begin{figure}[H]
\begin{center}
\addtolength{\leftskip}{-1cm}
\includegraphics[width=.80\linewidth]{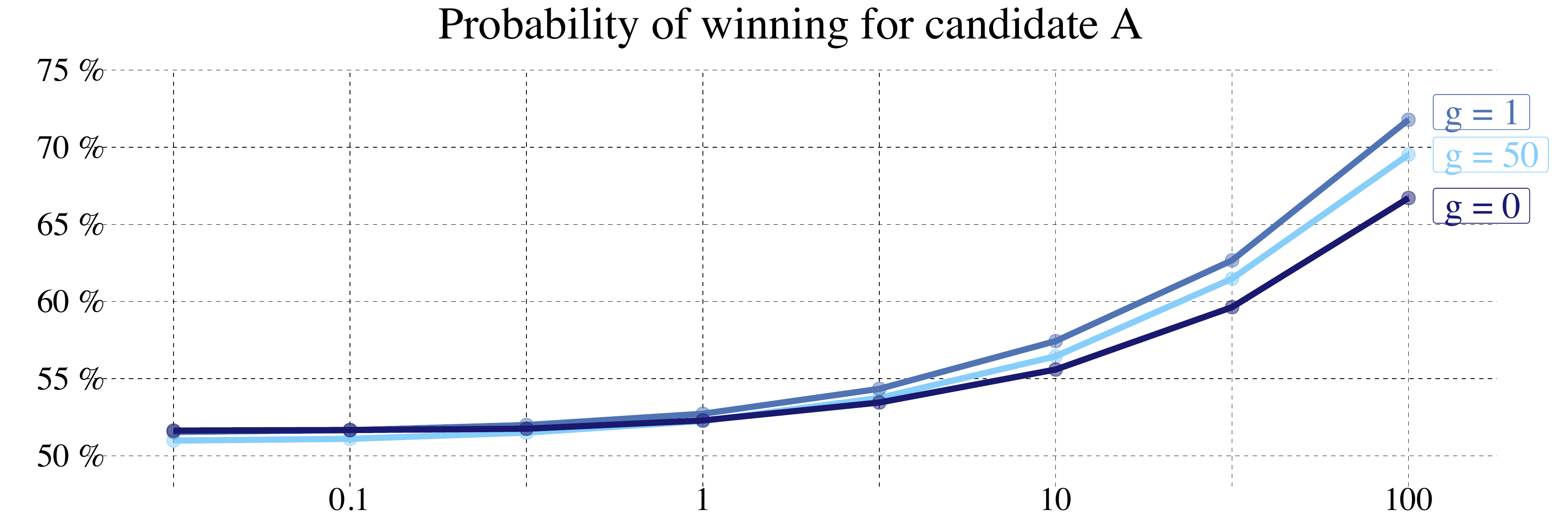}
\end{center}
\caption{Probability of winning for candidate $A$ (y-axis) for different levels of the parameter $k$ (x-axis in log-scale). %Recall that when $k \rightarrow \infty$, the system behaves as a deterministic model, converging to a completely certain triumph for either candidate (in this case, for $A$). 
}
\label{fig:Eq_var_k}
\end{figure}

We now analyze the outcome of the election for different levels of uncertainty. Recall that for the stochastic case, $k$ is the parameter that controls for uncertainty (see Equation~\eqref{eq:dir_distribution}). On the one hand, as $k\rightarrow\infty$ the variance approaches zero, and therefore the game resembles its deterministic version. On the other hand, at the limit where $k\rightarrow0$, the result in each region $i$ follows a discrete random variable where all votes go to either one candidate or to abstention. 
Figure~\ref{fig:Eq_var_k} plots the probability that candidate $A$ wins, in equilibrium, for different values of $k$ using the same instance parameters as before (see Table~\ref{tab:eqPSsto}) except for the abstention parameter {\boldmath$\gamma$}. More precisely, we look at different levels of abstention by scaling the original abstention vector {\boldmath$\gamma$} (from Table~\ref{tab:eqPSsto}) by a scalar factor $g\ge 0$. It can be seen on Figure~\ref{fig:Eq_var_k} that as the game becomes more deterministic ($k \rightarrow \infty$), the result becomes more predictable, and therefore the probability of winning for one of the candidates (candidate $A$ in this case) approaches $100\%$. Also, for a fixed variability level $k$, there is no clear trend on the probability of winning for candidate $A$ under the different abstention cases.

\vspace{0cm} 

%\subsubsection{The effect of polarization under MS}

%Note that in all the examples analyzed so far, we have fixed the values of the bias parameters, and so its magnitude relative to the sum of candidates effort (which add up to $100$ percentage points). Nevertheless, in reality it is not clear how big the effect of campaigning actually is when compared to the effect of the existing biases. Here we show what would happen when parameters 
%$\alpha$ and $\beta$ are proportionally scaled for different factors. Let $f>0$ denote the value of this factor so that the new bias parameters are ($f\alpha$, $f\beta$). Since the candidates' budget remains unchanged, different factors will represent different levels of power of campaigns. On the one hand, $f\rightarrow 0$ represents the case where there are virtually no bias parameters at all, so voters decision will rely mostly on candidates campaigns rather than previous biases. %In such cases, most states will behave as a swing state.
%On the other hand, the case when $f \rightarrow \infty$ makes the effect of campaigning negligible.%, except for those states in which the difference between candidates' bias parameters is small, i.e. swing states.

\subsection{Electoral College}\label{subse:EC}
\begin{figure}[h]
    \centering
    \begin{subfigure}{.5\textwidth}
        \centering
        \includegraphics[width=.95\linewidth]{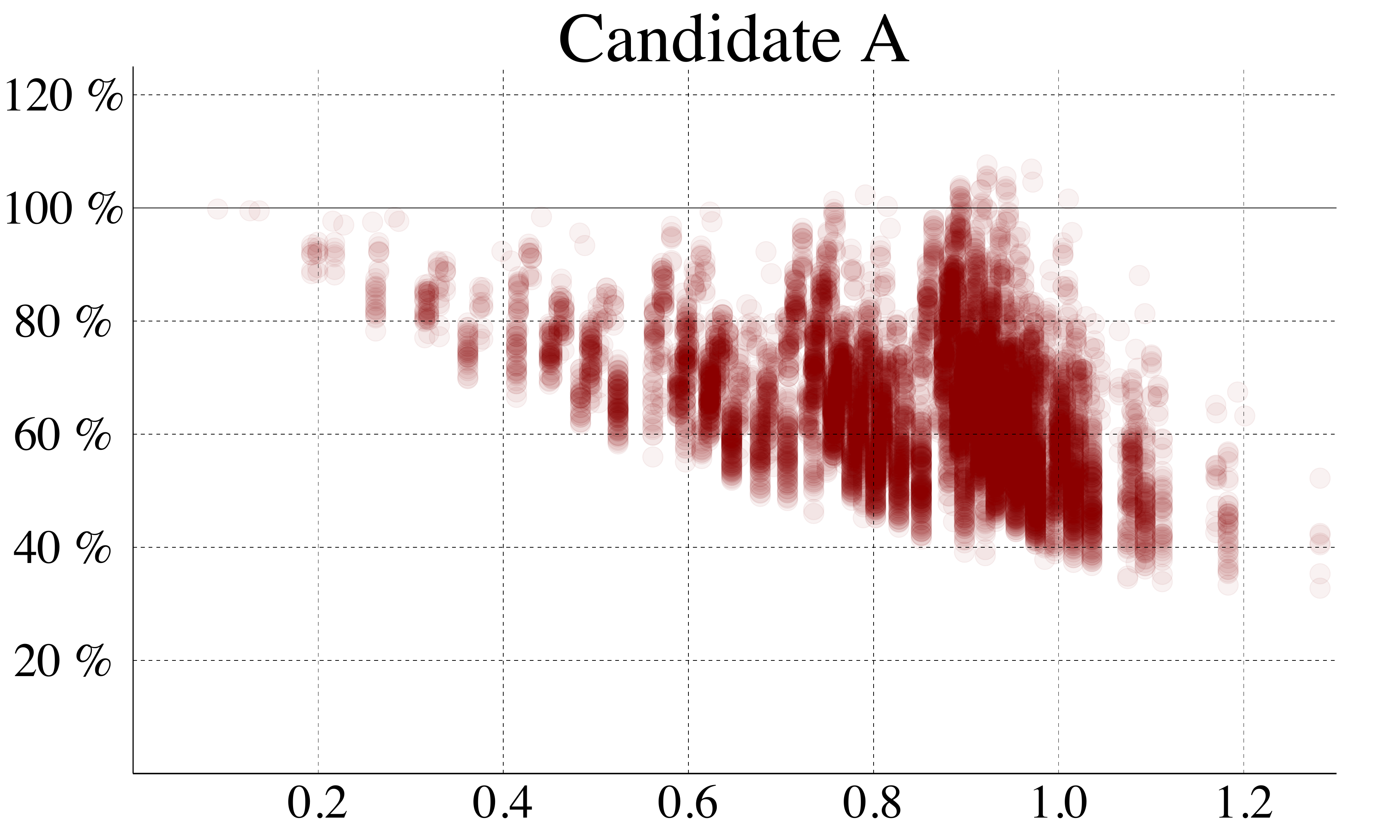}
        %\label{fig:sub1}
    \end{subfigure}%
    \begin{subfigure}{.5\textwidth}
        \centering
        \includegraphics[width=.95\linewidth]{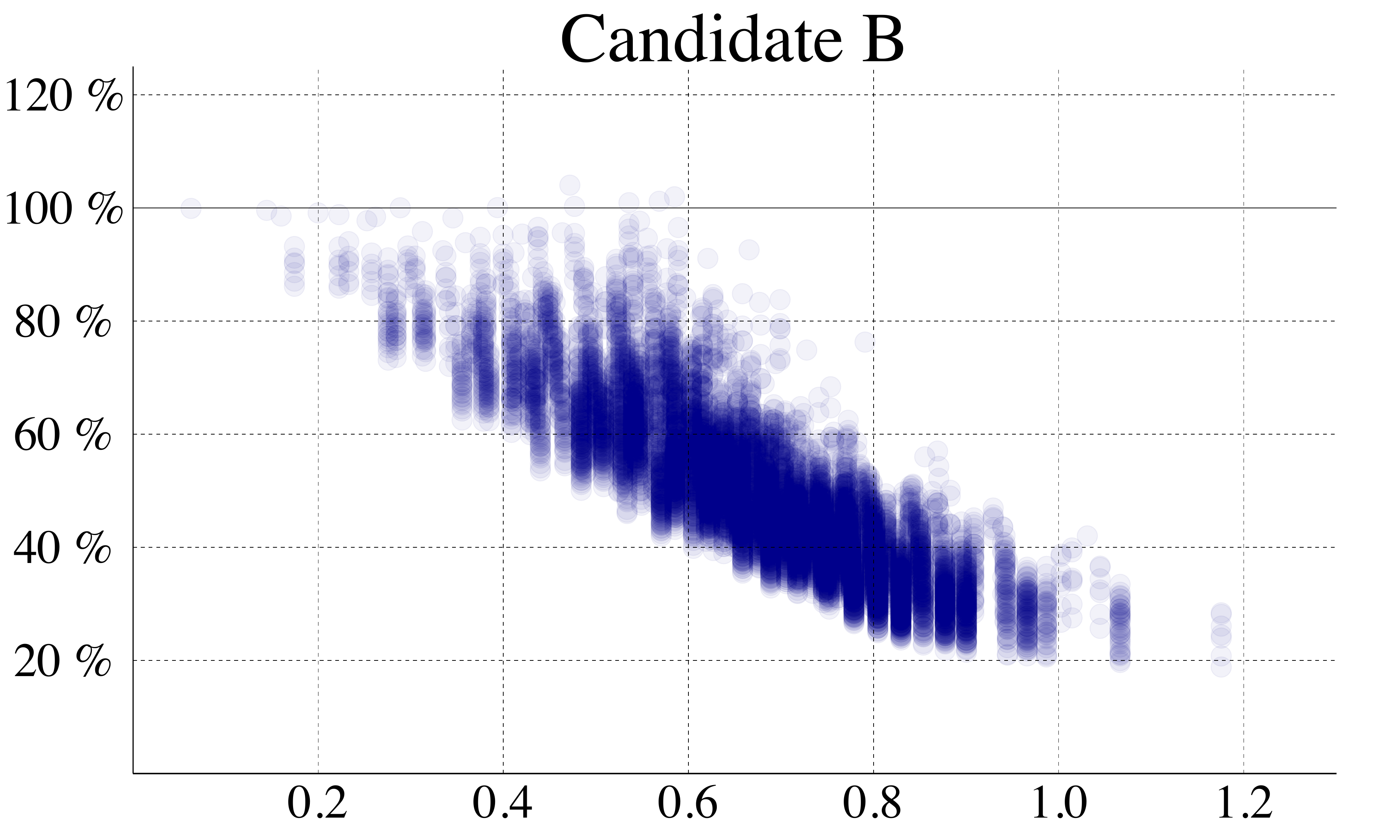}
        %\label{fig:sub2}
    \end{subfigure}
    \caption{Payoff ratio for unilateral deviations.}
    \label{fig:Corroboration_Local_Nash_EC_2}
\end{figure}

For the Electoral College case, we consider the same instance as in the Majority System, with the same bias and abstention parameters, except that the states have electoral votes. With the aim of obtaining an Equilibrium in fixed strategies, if there is any, we apply a Gradient Descent Ascent method as described in Section~\ref{sec:comp_eq_MS_sto} using the equations for the derivative values described in Section~\ref{sec:EC_pure}. Despite obtaining a pair of strategies for both candidates when doing the latter procedure, this pair of strategies is not an Equilibrium. Figure~\ref{fig:Corroboration_Local_Nash_EC_2} shows the payoff ratio for both candidates for unilateral deviations from the pair of strategies obtained. We can see that both candidates have an incentive to change their strategies to different ones. Nonetheless, it seems that, at least locally, there is no such incentive. Thus, the pair of strategies obtained might be a Local Nash Equilibrium.

\begin{comment}

    \begin{figure}[h]
        \begin{center}
            \addtolength{\leftskip}{-1cm}
            \includegraphics[scale = 0.45]{NubeUSANash3.pdf}
        \end{center}
    \caption{Payoff ratio for unilateral deviations.}
    \label{fig:Corroboration_Local_Nash_EC}
    \end{figure}
    
\end{comment}

Consequently, we run Algorithm~\ref{Alg_CE} in order to find an equilibrium under mixed strategies. Table~\ref{tab:Eq_CE} shows the instance parameters and the equilibrium obtained after running Algorithm~\ref{Alg_CE}. The vector efforts shown are the strategies obtained with positive probability. These probabilities are given in the last row of Table~\ref{tab:Eq_CE}. It can be seen that (i) both candidates randomized their strategies, and (ii) their efforts are mostly invested in the first three states, which are the ones with more electoral votes. Candidate $A$'s equilibrium strategies are very similar to each other, focusing most efforts in the first state, and less on the second and third states. As for candidate $B$'s efforts in equilibrium, these more evenly distributed between the three first states compared to candidate $A$, with emphasis on the first two regions. Also, it can be seen that  overall, the expected probability that candidate $A$ wins the election is $55.2\%$. %Note that candidate $A$ equilibrium strategies are such that with more than $80\%$ she invests more than $56\%$ of her efforts in the first region. In addition, she invests $100\%$ of her effort in this region with around $1\%$ chance. As for candidate $B$, her efforts in equilibrium are more evenly distributed between the three first states compared to candidate $A$. Overall, the expected probability that candidate $A$ wins the election is $55.2\%$. 

%we can see from Table~\ref{tab:Eq_CE} that in equilibrium her efforts are more evenly distributed between 
%These quantities are shown in the results provided below. The results after running Algorithm~\ref{Alg_CE} are shown in Table~\ref{tab:Eq_CE}. 

\begin{table}[h]
\begin{center}
\begin{tabular}{|c|c|c|c||c|c|c|c||c|c|c|c|}\hline
\textbf{Region} & $\boldsymbol{w}$ & \textbf{$\boldsymbol{\alpha}$} & \textbf{$\boldsymbol{\beta}$ }& $\boldsymbol{x^{(1)}}$ & $\boldsymbol{x^{(2)}}$ & $\boldsymbol{x^{(3)}}$ & $\boldsymbol{x^{(4)}}$ & $\boldsymbol{y^{(1)}}$ & $\boldsymbol{y^{(2)}}$ & $\boldsymbol{y^{(3)}}$ & $\boldsymbol{y^{(4)}}$ \\ \hline
1 & 34 & 45 & 71 & 76 & 75 & 75 & 0 & 64 & 52 & 0 & 0 \\
2 & 27 & 68 & 37 & 8 & 9 & 8 & 47 & 0 & 48 & 61 & 61 \\
3 & 21 & 32 & 24 & 16 & 16 & 17 & 46 & 36 & 0 & 39 & 38 \\
4 & 13 & 43 & 39 & 0 & 0 & 0 & 7 & 0 & 0 & 0 & 1 \\
5 & 12 & 76 & 65 & 0 & 0 & 0 & 0 & 0 & 0 & 0 & 0 \\
6 & 12 & 36 & 61 & 0 & 0 & 0 & 0 & 0 & 0 & 0 & 0 \\
7 & 10 & 51 & 54 & 0 & 0 & 0 & 0 & 0 & 0 & 0 & 0 \\
8 & 9 & 42 & 41 & 0 & 0 & 0 & 0 & 0 & 0 & 0 & 0 \\
9 & 5 & 85 & 31 & 0 & 0 & 0 & 0 & 0 & 0 & 0 & 0 \\
10 & 3 & 37 & 69 & 0 & 0 & 0 & 0 & 0 & 0 & 0 & 0 \\
\hline \hline
\multicolumn{4}{|c||}{\textbf{Probability [\%]}} & 11.8 & 1.1 & 82.9 & 4.2 & 28.4 & 35.9 & 24.0 & 11.7 \\
\hline
\end{tabular}
\end{center}
\caption{$\boldsymbol{x^{(i)}}$ and $\boldsymbol{y^{(i)}}$ with $i\in\{1,2,3,4\}$ correspond to the equilibrium effort vectors obtained with positive probability after running Algorithm~\ref{Alg_CE} with $q=100$.}
\label{tab:Eq_CE}
\end{table}

Table~\ref{tab:EQ_EC_prob} shows the probability that candidate $A$ wins the election for each combination of strategies of the mixed equilibrium given in Table~\ref{tab:Eq_CE}. Candidate $A$'s chances of winning are almost the same when playing any of the first three strategies regardless of what Candidate $B$ plays. On the contrary, if Candidate $A$ plays the fourth strategy, the winning odds will depend highly on the strategy of candidate $B$, taking values below $50\%$ for some cases, although the chances of the latter are below $2\%$ (see Table~\ref{tab:EQ_EC_prob_cases}).

\begin{center}
\begin{table}
\begin{minipage}{.45\textwidth}
   \centering
    \begin{tabular}{|c|c|c|c|c|}\hline
        \textbf{$\mathbb{P}$($A$ wins) [\%]}  & \textbf{$\boldsymbol{y^{(1)}}$} & \textbf{$\boldsymbol{y^{(2)}}$} & \textbf{$\boldsymbol{y^{(3)}}$} & \textbf{$\boldsymbol{y^{(4)}}$}\\ \hline
        \textbf{$\boldsymbol{x^{(1)}}$} & $55.1$ & $54.9$ & $55.1$ & $55.1$\\ 
        \textbf{$\boldsymbol{x^{(2)}}$} & $55.1$ & $54.6$ & $55.4$ & $55.4$\\ 
        \textbf{$\boldsymbol{x^{(3)}}$} & $54.8$ & $54.9$ & $55.4$ & $55.4$\\ 
        \textbf{$\boldsymbol{x^{(4)}}$} & $59.1$ & $58.4$ & $48.3$ & $48.4$\\ \hline
    \end{tabular}
    \caption{Probability that candidate $A$ wins for each pair of strategies.}
    
    \label{tab:EQ_EC_prob}
\end{minipage}\qquad
\begin{minipage}{.45\textwidth}
   \centering
    \begin{tabular}{|c|c|c|c|c|}\hline
    \textbf{$\mathbb{P}(\boldsymbol{x^{(i)},y^{(j)}})$ [\%]}  & \textbf{$\boldsymbol{y^{(1)}}$} & \textbf{$\boldsymbol{y^{(2)}}$} & \textbf{$\boldsymbol{y^{(3)}}$} & \textbf{$\boldsymbol{y^{(4)}}$}\\ \hline
    \textbf{$\boldsymbol{x^{(1)}}$} & $4.2$ & $3.3$ & $2.8$ & $1.4$\\ 
    \textbf{$\boldsymbol{x^{(2)}}$} & $0.4$ & $0.3$ & $0.3$ & $0.1$\\ 
    \textbf{$\boldsymbol{x^{(3)}}$} & $29.7$ & $23.5$ & $19.9$ & $9.7$\\ 
    \textbf{$\boldsymbol{x^{(4)}}$} & $1.5$ & $1.2$ & $1.0$ & $0.5$\\ 
    \hline
    \end{tabular}
   \caption{Probability of each case from the mixed equilibrium.}
\label{tab:EQ_EC_prob_cases}
\end{minipage}
\end{table}
\end{center}

\subsection{Equilibrium and States' Uncertainty}
The results shown above assume $k=10$. Recall that the parameter $k$ regulates the variability of the voters outcomes in each state; $k \rightarrow 0$ tends to the case where all electors choose one option, whereas $k \rightarrow \infty$ results in a more deterministic outcome (the variance, in fact, goes asymptotically to $0$). Before showing the equilibrium results for different levels of $k$, it is worth mentioning that Algorithm~\ref{Alg_CE} uses a set of starting points for both players' strategies. As a result, the equilibrium outcome might differ when running Algorithm~\ref{Alg_CE} with different starting points. 

For each $k$ we run Algorithm~\ref{Alg_CE} a total of $M\coloneqq 40$ times using a different random initial set of strategies for each candidate. Table~\ref{tab:EC_dif_k} shows for different levels of $k$: the average Earth Movers Distance using the Euclidean distance between all pairs of equilibrium strategies obtained for candidates $A$ and $B$ in the second and third columns respectively; the average probability that candidate $A$ wins among all $M$ runs in the fourth column; and the average equilibrium support cardinally obtained for candidates $A$ and $B$ in the fifth and sixth columns respectively.

Table~\ref{tab:EC_dif_k} shows that for higher values of $k$, Algorithm~\ref{Alg_CE} might lead to different outcomes, whereas for low values of $k$ the outcome always results in a single pure strategy for both candidates. The intuition behind this is that for a more deterministic outcome of the game (high value of $k$), players will tend to randomize their strategies since, otherwise, the opponent could take advantage of this deterministic outcome, similarly as in Matching Pennies game. On the contrary, a more stochastic game (low $k$) will induce pure strategies. Table~\ref{tab:EC_dif_k} also shows that for higher values $k$, the winning probability of candidate $A$ tends to a slight increase. %The latter makes sense, since a more deterministic version of the game might induce a more predictable outcome, and so more likely that one of the candidates wins compare to a more stochastic game.
However, it is worth noticing that the latter effect is much more pronounced in the MS rather than the EC (see Figure~\ref{fig:Eq_var_k} and Table~\ref{tab:EC_dif_k}). As a result, reducing voters' uncertainty (i.e., increasing $k$) does not translate in reducing the uncertainty of the election winner under EC, unlike the case of the MS.

\begin{table}[h]
\begin{center}
\begin{tabular}{|c|c|c|c|c|c|}
\hline
\textbf{$k$} & \textbf{$\mathbb{D}$ (on $x$)} & \textbf{$\mathbb{D}$ (on $y$)} & \textbf{$\mathbb{P}$($A$ wins) [\%]} & \textbf{$|\textbf{Sup}(x)|$} & \textbf{$|\textbf{Sup}(y)|$} \\ \hline
$1$ & $0.00$ & $0.00$ & $52.3$ & $1.0$ & $1.0$ \\
$2$ & $0.00$ & $0.00$ & $53.0$ & $1.0$ & $1.0$ \\
$5$ & $0.02$ & $0.00$ & $54.2$ & $1.0$ & $1.0$ \\
$10$ & $0.07$ & $0.07$ & $55.2$ & $4.125$ & $4.125$ \\
$20$ & $0.07$ & $0.08$ & $55.4$ & $4.85$ & $4.85$ \\
$50$ & $0.16$ & $0.15$ & $54.8$ & $8.5$ & $8.5$ \\
$100$ & $0.16$ & $0.15$ & $55.8$ & $13.35$ & $13.35$ \\
 \hline
\end{tabular}
\end{center}
\caption{The effects of the value of $k$ on the behavior of the equilibrium. }
\label{tab:EC_dif_k}
\end{table}

\subsection{Performance of Algorithm~\ref{Alg_CE}}

In this section we study the performance of Algorithm~\ref{Alg_CE} by solving different numerical instances. More precisely, we control for: (i) the number of states and (ii) the level of concentration of the electoral votes among the states. With respect to the former, we consider $n\in\{5,10,20,50\}$; while for the latter, states' electoral votes are sampled from a multinomial distribution where the number of electoral votes of the \textit{i}\textsuperscript{th} state is (in expectation) proportional to $\nu^{i}$, where $\nu\in(0,1]$ is a parameter that controls for the concentration. If $\nu=1$, states will have a similar number of electoral votes, whereas smaller values of $\nu$ will induce a more skewed distribution of electoral votes. We consider $\nu\in\{0.8,0.9,1\}$. In addition, the total number of electoral votes is set to 538 and each state is endorsed with 3 additional electoral votes besides the sampled ones (thus the parameter of the \textit{number of trials} of the multinomial distribution is $538-3n$). For each pair $(n,\nu)$ we sample and solve a total of $100$ instances.

\begin{table}[h]
\centering
\begin{tabular}{|C{0.5cm}C{0.5cm}|C{0.8cm}C{0.8cm}C{0.8cm}C{0.8cm}C{0.8cm}C{0.8cm}C{0.8cm}C{0.8cm}C{0.8cm}C{0.8cm}C{0.8cm}C{0.8cm}|}\hline
& & \multicolumn{2}{c}{Time [s]} & \multicolumn{2}{c}{Iterations} & \multicolumn{2}{c}{$|\textbf{Sup}(x)|$} & \multicolumn{2}{c}{$|\textbf{Sup}(y)|$} & \multicolumn{2}{c}{$|\textbf{Sup}_{+}(x)|$} & \multicolumn{2}{c|}{$|\textbf{Sup}_{+}(y)|$}\\
$\nu$ & $n$ & {Avg.} & {Max.} & {Avg.} & {Max.} & {Avg.} & {Max.} & {Avg.} & {Max.} & {Avg.} & {Max.} & {Avg.} & {Max.} \\
\hline
 $1.0$ & 5 & 23 & 237 & 13.8 & 33 & 36.8 & 127 & 36.2 & 86 & 3.9 & 9 & 3.9 & 9 \\
 $1.0$ & 10 & 347 & 1391 & 15.0 & 30 & 56.3 & 135 & 57.1 & 154 & 4.0 & 9 & 4.0 & 9 \\
 $1.0$ & 20 & 792 & 3087 & 14.2 & 33 & 86.4 & 198 & 89.6 & 270 & 3.5 & 7 & 3.5 & 7 \\
 $1.0$ & 50 & 1312 & 7556 & 12.2 & 30 & 136.9 & 354 & 142.0 & 438 & 2.9 & 6 & 2.9 & 6 \\ \hline
 $0.9$ & 5 & 23 & 245 & 13.8 & 33 & 36.8 & 127 & 36.1 & 86 & 3.9 & 9 & 3.9 & 9 \\
 $0.9$ & 10 & 303 & 1519 & 14.3 & 30 & 49.8 & 155 & 50.9 & 146 & 3.8 & 8 & 3.8 & 8 \\
 $0.9$ & 20 & 662 & 2577 & 14.0 & 28 & 61.0 & 191 & 67.5 & 157 & 3.6 & 7 & 3.6 & 7 \\
 $0.9$ & 50 & 786 & 4729 & 10.1 & 27 & 83.3 & 196 & 81.0 & 199 & 2.8 & 7 & 2.8 & 7 \\ \hline
 $0.8$ & 5 & 17 & 173 & 12.3 & 30 & 28.0 & 89 & 28.2 & 73 & 3.5 & 8 & 3.6 & 8 \\
 $0.8$ & 10 & 228 & 1110 & 12.7 & 29 & 37.7 & 92 & 39.8 & 102 & 3.7 & 8 & 3.7 & 8 \\
 $0.8$ & 20 & 417 & 2457 & 12.5 & 28 & 46.1 & 175 & 49.4 & 108 & 3.6 & 7 & 3.6 & 7 \\
 $0.8$ & 50 & 616 & 4239 & 9.9 & 28 & 73.3 & 146 & 70.1 & 131 & 3.1 & 6 & 3.0 & 6 \\ \hline
\end{tabular}
\caption{Average and maximum: solving times, iterations, and cardinality of the players' strategy sets when finishing Algorithm~\ref{Alg_CE}. $\textbf{Sup}_{+}(x)$ ($\textbf{Sup}_{+}(y)$) denotes the number of strategies that have a positive probability for candidate $A$ ($B$).}\label{tab:alg_outcomes}
\end{table}
With respect to Algorithm~\ref{Alg_CE}, Table~\ref{tab:alg_outcomes} shows the average and maximum: solving time, iterations, cardinality of the players' strategy sets when finishing Algorithm~\ref{Alg_CE}, and the number strategies in the support with positive probability. It can be seen that instances with lower concentration of electoral votes (i.e., higher $\nu$) have a higher solving time and a larger strategy sets. This is because when all states have a similar magnitude of electoral votes, candidates' strategies will have to consider investing in several states. This translates in higher solving times for the best responses while also inducing more strategies to be added in the players' strategy sets. It is worth observing from Table~\ref{tab:alg_outcomes} that the actual number of strategies with positive probability appears to be independent of the concentration of electoral votes. Also, it is interesting to note that the amount of iterations required do not seem to be particularly affected by the size of the instance (i.e., $n$), nor the concentration level. Although, there is a slight negative relation between the number of iterations and the size of the instance and the concentration of electoral votes. Despite the exponential cardinality of the simplex lattice, the running time of Algorithm~\ref{Alg_CE} does not show an exponential relation with the size of the instance. Furthermore,  the number of iterations is in average between 10 and 15, see fifth column of Table~\ref{tab:alg_outcomes}.

\subsection{Effect of Polarization under MS and EC}
In all the examples analyzed so far, we have fixed the values of the bias parameters, and so their magnitude relative to the candidates' budget. Nevertheless, it actually is not clear how big the effect of campaigning is relative to the effect of existing biases. The term \textit{polarization} is used to characterize the case when voters' position is inelastic with respect to candidates' campaign. The latter occurs when existing biases are large enough compared to candidates' budget. In this section we analyze the effect of polarization on candidates' equilibrium strategies. More precisely, the same instances given in Tables~\ref{tab:eqPSdet} and \ref{tab:Eq_CE} are solved while scaling the bias parameters {\boldmath$\alpha$} and {\boldmath$\beta$} for different factors. Let $f>0$ denote the value of this factor so that the new bias parameters are ($f${\boldmath$\alpha$}, $f${\boldmath$\beta$}). Since the candidates' budgets remain fixed, different factors will represent different levels of power of campaigns. On the one hand, $f\rightarrow 0$ represents a low polarization case since there are virtually no biases, and therefore the voters' decisions are triggered mostly by the candidates' campaigns. On the other hand, in the case when $f \rightarrow \infty$, the effect of campaigning becomes negligible, except for those states in which the difference between the candidates' bias parameters is still within the reach of what the campaign can affect.

\begin{table}[ht]
\centering
\captionsetup{width=0.1\linewidth}
%\subfloat{
% TABLA DE LA IZQUIERDA CON INFO DE F
\begin{tabular}[t]{|C{0.4cm}|L{0.3cm}}
\cline{1-1}
\multirow{2}{*}{$f$} & \\ \\ \cline{1-1}
\multirow{1}{*}{$0.1$} &  \\ 
\cline{1-1}
\multirow{1}{*}{$1$} & \\
\cline{1-1}
\multirow{1}{*}{$5$} & \\
\cline{1-1}
 \multirow{1}{*}{$10$} & \\ 
 \cline{1-1}
 \multirow{1}{*}{$50$} & \\
 \cline{1-1}
\end{tabular}%}
%\subfloat{
% TABLA DEL CENTRO CON INFO SOBRE ESTRATEGIAS PARA A
\begin{tabular}[t]{|C{0.4cm}C{0.4cm}C{0.4cm}C{0.4cm}C{0.4cm}C{0.4cm}C{0.4cm}C{0.4cm}|L{0.8cm}}
 \cline{1-8}
\multicolumn{8}{|c|}{\textbf{Strategies for $A$ [\%]}} \\  \cline{1-8}
$\textbf{1}$&$\textbf{2}$&$\textbf{3}$&$\textbf{4}$&$\textbf{5}$&$\textbf{6}$&$\textbf{7}$&$\textbf{8}$&\\
 \cline{1-8}
34 & 24 & 18 & 8 & 3 & 5 & 5 & 3 & \\ 
 \cline{1-8}
68 & 26 & 6 & &  &  &  & & \\
 \cline{1-8}
 100 &  & & &  &  &  &  & \\
 \cline{1-8}
 100 &  & & &  &  &  &  & \\
 \cline{1-8}
 100 &  & & &  &  &  &  & \\
  \cline{1-8}
 45 & 68 & 32 & 43 & 76 & 36 & 51 & 42 & $\alpha^{(0)}$ \\ 
  \cline{1-8}
\end{tabular}%}
%\subfloat{
% TABLA DE LA DERECHA CON INFO SOBRE ESTRATEGIAS PARA B
\begin{tabular}[t]{|C{0.4cm}C{0.4cm}C{0.4cm}C{0.4cm}C{0.4cm}C{0.4cm}C{0.4cm}C{0.4cm}|L{0.8cm}}
 \cline{1-8}
\multicolumn{8}{|c|}{\textbf{Strategies for $B$ [\%]}} & \\  \cline{1-8}
$\textbf{1}$&$\textbf{2}$&$\textbf{3}$&$\textbf{4}$&$\textbf{5}$&$\textbf{6}$&$\textbf{7}$&$\textbf{8}$& \\
 \cline{1-8}
32 & 27 & 18 & 8 & 4 & 2 & 5 & 3 &\\ 
 \cline{1-8}
36 & 52 & 12 & & & & & &\\ 
 \cline{1-8}
& 100 & & & & & & & \\
 \cline{1-8}
& 100 & & & & & & & \\
 \cline{1-8}
& 100 & & & & & & & \\
  \cline{1-8}%\morecmidrules\hline
 71 & 37 & 24 & 39 & 65 & 61 & 54 & 41 & $\beta^{(0)}$ \\ %\hline
 \cline{1-8}
\end{tabular}%}
\caption{Equilibrium for different bias parameters $(\alpha,\beta)=(f\alpha^{(0)}, f\beta^{(0)})$ under MS. 
For each $f=0.1,\; 1,\; 5,\; 10,\; 50$, candidate $A$ wins with probabilities $51.47\%$, $57.41\%$, $67.31\%$, $73.12\%$, and $91.75\%$ respectively. All strategies have null effort in states 9 and 10. The last row shows values of $\alpha^{(0)}$ and $\beta^{(0)}$. }
\label{EQ_MS_F}
\end{table}

% poner texto a talba

% hablar de cque cosas podemos concluir de la tabla

%Note that the budget constraint remains unchanged, so that the only difference here is how powerful the campaign actually is in each case. When $f\rightarrow 0$, we are approaching the case were there are virtually no bias parameters at all, so the campaign is everything the voters consider when it comes to decide for whom they are voting for. On the other hand, the case when $f \rightarrow \infty$ makes the effect of campaigning negligible, 

%We show above the results for a specific instance with an unbalance between the candidates $A$ and $B$ within each region. Nevertheless, in reality it is not clear how big the effect of campaigning actually is when compared to the effect of some previously existing biases. Here we show what would happen when parameters $\alpha$ and $\beta$ are proportionally scaled but different factors $f$ (so that the new bias parameters are going to be $f \alpha$ and $f\beta$). Note that the budget constraint will not change, so that the only difference here is how powerful the campaign actually is in each case. When $f\rightarrow 0$, we are approaching the case were there are virtually no bias parameters at all, so the campaign is everything the voters consider when it comes to decide for whom they are voting for. On the other hand, the case when $f \rightarrow \infty$ makes the effect of campaigning negligible, 

Results for the MS are shown in Table~\ref{EQ_MS_F}. We can see that for low levels of biases ($f=0.1$), candidates' efforts on average are directly related to the weight of the region size (in terms of number of votes). In addition, all states get some level of investment (except for states 9 and 10). The intuition behind the latter can be easily observed in the extreme case where $f\rightarrow0$ (i.e. there are almost no previous biases). If a candidate ignores a region, it takes the opponent just any positive effort to win most of its votes. For medium biases ($f=1$), candidates prioritize only those regions with more votes. Unlike the low bias case, investing in smaller regions no longer pays off. For higher biases ($f=5,10,50$), candidates concentrate all their efforts in a single region compensating the initial bias disadvantage (region 1 for candidate $A$; 2 for $B$). All in all, candidates allocate their efforts where they have the maximum marginal return%s in terms of votes gain
. The effect of polarization in candidate equilibrium strategies can be summarized as the interplay of the two following factors: (i) regions with a large number of votes, and (ii) the disadvantage bias.

%Generally speaking, candidates allocate efforts where they have the maximum marginal returns in terms of votes gain. As the bias grows, the more polarized regions decrease their marginal returns, and therefore, are less attractive to invest in. This explains why there is threshold for $f$ (between $f=1$ and $f=5$) from which their strategies are just an extreme point; Candidate $A$ is only focused on the largest region where she has an initial disadvantage (region 1), similar with candidate $B$ in region 2. 

%, which marginal return offsets all other regions marginal gain. The state 

%the state which has higher marginal vot

%in a single region the largest regions (in terms of votes)

%several states are not worth to invest in despite their weight (as for states 1 and 2); when the difference between candidates' biases are large enough, candidates refuse investing on these states since their electoral outcome will remain mostly unchanged. This is clear in the extreme case $f=50$, where both candidates invest only in state $4$, as it is one of the few where the difference between its bias parameters is within the reach of their campaign budget to offset its outcome. Note that the latter also applies to state 8, however its electoral vote weight is much lower than state 4 and therefore is not worth to invest in. In summary, as $f$ increases, candidates focus their efforts in those states with similar biases and high electoral votes, i.e. swing states.

% TABLA DE SUBTABLAS (principal diferencia es la estetica del caption)
\begin{table}[ht]
\centering
\captionsetup{width=0.1\linewidth}
%\subfloat{
\begin{tabular}[t]{|C{0.4cm}|C{0.05cm}}
\cline{1-1}
\multirow{2}{*}{$f$} \\ \\ \cline{1-1}\cline{1-1}
\multirow{6}{*}{$0.1$} & \\ \\ \\ \\ \\ \\
\cline{1-1}\cline{1-1}
\multirow{4}{*}{$1$} & \\  \\ \\ \\
\cline{1-1}\cline{1-1}
\multirow{2}{*}{$5$} & \\ \\
\cline{1-1}\cline{1-1}
 \multirow{2}{*}{$10$} & \\ \\ 
 \cline{1-1}\cline{1-1}
 \multirow{1}{*}{$50$} & \\
 \cline{1-1}
\end{tabular}%}
%\subfloat{
% TABLA DEL CENTRO CON INFO SOBRE ESTRATEGIAS PARA A
\begin{tabular}[t]{|C{0.4cm}C{0.4cm}C{0.4cm}C{0.4cm}C{0.4cm}C{0.4cm}C{0.4cm}C{0.4cm}|R{1.4cm}|L{0.05cm}}
\cline{1-9}
\multicolumn{8}{|c|}{\textbf{Strategies for $A$ [\%]}} & \multirow{2}{*}{\textbf{
$\sigma_{A}$ [\%]
}} & \\ \cline{1-8}
$\textbf{1}$&$\textbf{2}$&$\textbf{3}$&$\textbf{4}$&$\textbf{5}$&$\textbf{6}$&$\textbf{7}$&$\textbf{8}$&\\
\cline{1-9}\cline{1-9}
 51 &  & 24 & 9 & 3 & 7 & 4 & 2 & $10.8\;\;\;\;$ \\ 
40 & 27 & 18 & 6 & 1 & 5 & 2 & 1 & $26.8\;\;\;\;$ \\
40 & 26 & 19 & 6 & 1 & 5 & 2 & 1 & $46.0\;\;\;\;$ \\
2 & 40 & 30 & 9 & 4 & 8 & 4 & 3 & $1.7\;\;\;\;$ \\
1 & 40 & 30 & 9 & 4 & 9 & 4 & 3 & $5.8\;\;\;\;$ \\
1 & 40 & 30 & 9 & 4 & 8 & 4 & 4 & $8.8\;\;\;\;$ \\
\cline{1-9}\cline{1-9}
76 & 8 & 16 & &  &  &  &  & $11.8\;\;\;\;$ \\
75 & 9 & 16 &  &  &  &  &  & $1.1\;\;\;\;$ \\ 
75 & 8 & 17 &  &  &  &  &  & $82.9\;\;\;\;$ \\ 
 & 47 & 46 & 7 &  &  &  &  & $4.2\;\;\;\;$ \\ 
\cline{1-9}\cline{1-9}
 &  & 88 & 12 &  &  &  &  &  $19.6\;\;\;\;$\\
 &  & 87 & 13 &  &  &  &  &  $80.4\;\;\;\;$ \\ 
 \cline{1-9}\cline{1-9}
 &  & 86 & 14 &  &  &  &  & $93.3\;\;\;\;$ \\
 &  & 85 & 15 &  &  &  &  & $6.7\;\;\;\;$ \\ 
 \cline{1-9}\cline{1-9}
 &  &  & 100 &  &  &  &  & $100.0\;\;\;\;$ \\
 \cline{1-9}\cline{1-9}
 45 & 68 & 32 & 43 & 76 & 36 & 51 & 42 & $\alpha^{(0)}\;\;\;\;\;$ \\ 
 \cline{1-9}
\end{tabular}%}
%\subfloat{
% TABLA DE LA DERECHA CON INFO SOBRE ESTRATEGIAS PARA B
\begin{tabular}[t]{|C{0.4cm}C{0.4cm}C{0.4cm}C{0.4cm}C{0.4cm}C{0.4cm}C{0.4cm}C{0.4cm}|R{1.4cm}|L{0.05cm}}
\cline{1-9}\cline{1-9}
\multicolumn{8}{|c|}{\textbf{Strategies for $B$ [\%]}} & \multirow{2}{*}{\textbf{
$\sigma_{B}$ [\%]
}} & \\ \cline{1-8}
$\textbf{1}$&$\textbf{2}$&$\textbf{3}$&$\textbf{4}$&$\textbf{5}$&$\textbf{6}$&$\textbf{7}$&$\textbf{8}$& \\
\cline{1-9}\cline{1-9}
49 & & 29 & 7 & 4 & 4 & 3 & 4 & $7.0\;\;\;\;$ \\ 
38 & 29 & 20 & 6 & 2 & 3 & 1 & 1 & $31.7\;\;\;\;$ \\ 
38 & 29 & 20 & 6 & 2 & 2 & 2 & 1 & $4.0\;\;\;\;$ \\ 
38 & 28 & 20 & 6 & 2 & 3 & 2 & 1 & $2.8\;\;\;\;$ \\ 
37 & 29 & 20 & 6 & 2 & 3 & 2 & 1 & $37.9\;\;\;\;$ \\ 
& 41 & 28 & 11 & 6 & 6 & 5 & 3 & $16.6\;\;\;\;$ \\ 
\cline{1-9}\cline{1-9}
64 & & 36 & & & & & & $28.4\;\;\;\;$ \\ 
52 & 48 & & & & & & & $35.9\;\;\;\;$ \\ 
& 61 & 39 & & & & & & $24.0\;\;\;\;$ \\ 
& 61 & 38 & 1 & & & & & $11.7\;\;\;\;$ \\ 
\cline{1-9}\cline{1-9}
& & 100 & & & & & & $53.3\;\;\;\;$ \\
& & & 55 & 45 & & & & $46.7\;\;\;\;$ \\ 
\cline{1-9}\cline{1-9}
& & 100 & & & & & & $48.7\;\;\;\;$ \\
& & & 83 & & & & 17 & $51.3\;\;\;\;$ \\
\cline{1-9}\cline{1-9}
& & & 100 & & & & & $100.0\;\;\;\;$ \\
 \cline{1-9}\cline{1-9}
 71 & 37 & 24 & 39 & 65 & 61 & 54 & 41 & $\beta^{(0)}\;\;\;\;\;$ \\ 
\cline{1-9}

\end{tabular}%}
\caption{Equilibrium for different bias parameters $(\alpha,\beta)=(f\alpha^{(0)}, f\beta^{(0)})$ under EC. 
For each $f=0.1,\; 1,\; 5,\; 10,\; 50$, candidate $A$ wins with probabilities $51.54\%$, $55.05\%$, $58.90\%$, $69.34\%$, and $93.62\%$ respectively. All strategies have null effort in states 9 and 10. The last row shows values for $\alpha^{(0)}$ and $\beta^{(0)}$. }
\label{EQ_CE_F}
\end{table}

Table~\ref{EQ_CE_F} shows the equilibrium obtained for EC when running Algorithm~\ref{Alg_CE} for different levels of $f$. We can observe that for low ($f=0.1$) and medium biases ($f=1$), the equilibria obtained under EC follow the same structure as the one shown under MS; low biases make almost every state worth to investing in, whereas for medium biases smaller states (in terms of electoral votes) become less profitable. 
%but as $f$ increases, the smallest states become less profitable (now not in terms of votes gains, but of the probability of winning the state' electoral votes). 
For higher biases ($f \ge 5$), a clear difference emerges between the two systems. Under EC, unlike what was shown under MS, several states are not worth investing in despite their large electoral weight (such as states 1 and 2). When the difference between the candidates' biases is large enough, candidates refuse to invest in these states since their electoral outcome will remain mostly unchanged. The intuition behind the latter is that the effect of a candidate campaign on the probability of winning in a such a state will be negligible. This is clear in the extreme case where $f=50$. Both candidates invest only in state $4$, as it is one of the few where the difference between its bias parameters is within the reach of their campaign budgets to offset its outcome. In summary, polarization under EC will induce the candidates to campaign according to the two following factors: (i) states with similar biases, and (ii) high electoral votes. States that combine both of these elements are usually called \textit{swing states}.

The previous analysis helps us to understand candidates' decisions under both election systems for different levels of polarization. Under MS, candidates put their efforts into seeking to get the greater number of votes. The bigger the biases are, the more they tend to invest only in those regions where there are more people to convince to vote for them; large regions with relative initial disadvantages. %Under EC, voters do not necessarily dictate their respective electoral vote and therefore the origin of a vote says a lot about how important this is
%on each state, people's vote will not 
Under EC, because of the winner-take-all policy, some votes do not translate into its respective electoral vote. 
%. Because of the winner-take-all policy,
As a result,
when a candidate faces a state with an initial disadvantage, such that it is virtually impossible to induce any substantial change in the probability of winning, it is simply not worth investing in, even though it might actually be the largest state. Similarly in states with a considerable initial advantage. Therefore, under a highly polarized scenario, the campaign is only relevant in the undecided states; the swing states. 

An interesting insight from the last result can be applied to the effect of polarization on political campaigns. In a polarized country, we would expect higher bias values, and therefore strategies should be more focused on a few states. In reality, the latter observation can have additional consequences regarding not just the candidates' resource allocation strategies, but also on the election promises made in the different states. For example, a candidate might be more tempted to offer higher infrastructure expenditures in a swing state (under a polarized EC), despite the fact that that state might have only a small fraction of the national population, but it plays a key role in winning the election.

\section{Conclusions}\label{sec:conclusions}

The presented models and results presented show how different electoral systems and political realities might affect the optimal solution to the resources allocation problem of an election campaign. 

Under a deterministic model for the Majority System (MS), it is possible to show the existence and uniqueness of the equilibrium. In addition, a closed form solution is provided for some particular cases of the problem. For general cases, the equilibrium can be obtained by using a gradient descent ascent method. The latter is performed by using a simulation procedure for the stochastic version of the game, enabling the computation of an estimate of the candidate's objective function and gradient as well, which reuses simulated values of previous iterations. Under the Electoral College (EC), unlike MS, numerical computations indicate that there is no equilibrium in pure strategies. In order to explore mixed strategies, we propose an algorithm that returns a mixed equilibrium in a subset of the simplex lattice by augmenting candidates strategy sets in an iterative method.

When facing MS, candidates tend to focus mainly on the largest regions, with special attention to those where they are less popular than their rivals. In addition, it is observed that in equilibrium, the votes from turnout for each candidate happen to be almost the same among the regions chosen to invest in. Moreover these quantities happen to be exactly the same in the deterministic model, and in the stochastic with no abstention.

For the EC, we observed mixed strategies in the game equilibrium. In particular, we detected a connection between the uncertainty within the election and the number of strategies (support) of the obtained equilibrium. The more randomness, the more pure are the candidates' equilibrium strategies. As a result, an election in which there is little uncertainty of the outcome will induce a less predictable behaviour (due to mixing). On the contrary, under a lot of noise, candidates' campaigns narrow to a single pure strategy. 

It is interesting to note the impact of uncertainty in the winning probability under both election systems. In the case of MS, the stochastic model will resemble its deterministic version when reducing the uncertainty, and therefore candidates' winning probability will approach one or zero. In the case of EC, the probability of winning will be mildly affected by the noise reduction. %Due to the mixing of strategies, in equilibrium both candidates will have several combinations (among the strategy profiles in equilibrium) where they are likely to win, and 
Therefore, both candidates will continue to have a significant chance of winning the election.

Another important element analyzed is the effect of polarization on candidates' strategies under both election systems. In a scenario with low polarization, the most relevant information for the strategies is the size of the state. Furthermore, in such a case, one might expect to see candidates investing in almost every region. As polarization increases, candidates will focus on only a few regions. Indeed, in MS, candidates' strategies are centered on regions with more potential votes while emphazising those in which they have an initial bias disadvantage. On the other hand, under EC, candidate investments are concentrated in swing states. Namely, states with no clear tendency towards any candidate, while having a non-negligible number of electoral votes. As observed in some instances, even though there might be larger states than others, as polarization increases, the efforts are more concentrated on those swing states alone. This is something that actually happens: In the US, California and Texas are the two largest states. Yet, only one election campaign event was held in each of them during the 2016 election, while Florida (the third largest state) had 71 campaign events in total.

It is interesting to note that here it is assumed that the resource being allocated (the \textit{strategy}) is the time that the candidate invests in each state. This leads to a symmetrical budget constraint.%, since usually both candidates have the same amount of time to campaign. 
However, the same model can be applied for studying the campaign resource allocation strategy in terms any other resources rather than time, such as: advertisement budgets, election promises, etc. 
%\subsection{Ongoing and Future Work}
%In the current work, we have emphasize only on some aspects of candidates' electoral campaign game. Nonetheless, there are several other interesting aspects of the problem which can be studied. Some of the future research directions that can be extended are: (i) incorporate cross effects of campaigns among different states, (ii) consider the time dimension aspect into the problem in which candidates decide over time how to allocate their campaign strategy at a given time of the campaign time-frame, (iii) consider an electoral setting with more than two candidates, (iv) consider the multiple dimension setting in which candidates have multiple campaign resources which need to be allocated among the states, and (v) use alternative probabilistic models which allow more general behaviour of voters.

% Acknowledgments here
\ACKNOWLEDGMENT{The authors gratefully acknowledge financial support from CONICYT PIA/BASAL AFB180003. In addition, we thank Jose Correa for his useful comments that used to improve this work.}

% References here (outcomment the appropriate case)

% CASE 1: BiBTeX used to constantly update the references
%   (while the paper is being written).
\bibliographystyle{ormsv080} % outcomment this and next line in Case 1
\bibliography{litcompet} % if more than one, comma separated

\begin{thebibliography}{33}
\expandafter\ifx\csname natexlab\endcsname\relax\def\natexlab#1{#1}\fi
\expandafter\ifx\csname url\endcsname\relax
  \def\url#1{{\tt #1}}\fi
\expandafter\ifx\csname urlprefix\endcsname\relax\def\urlprefix{URL }\fi
\expandafter\ifx\csname urlstyle\endcsname\relax
  \expandafter\ifx\csname doi\endcsname\relax
  \def\doi#1{doi:\discretionary{}{}{}#1}\fi \else
  \expandafter\ifx\csname doi\endcsname\relax
  \def\doi{doi:\discretionary{}{}{}\begingroup \urlstyle{rm}\Url}\fi \fi

\bibitem[{Banzhaf~III(1964)}]{banzhaf1964weighted}
Banzhaf~III, John~F. 1964.
\newblock Weighted voting doesn't work: A mathematical analysis.
\newblock {\it Rutgers L. Rev.\/} {\bf 19} 317.

\bibitem[{Barnett(1976)}]{barnett1976more}
Barnett, Arnold~I. 1976.
\newblock More on a market share theorem.
\newblock {\it Journal of Marketing Research\/} {\bf 13}(1) 104--109.

\bibitem[{Bell et~al.(1975)Bell, Keeney, and Little}]{bell1975market}
Bell, David~E, Ralph~L Keeney, John~DC Little. 1975.
\newblock A market share theorem.
\newblock {\it Journal of Marketing Research\/} {\bf 12}(2) 136--141.

\bibitem[{Borel(1921)}]{borel1921theorie}
Borel, Emile. 1921.
\newblock La th{\'e}orie du jeu et les {\'e}quations int{\'e}gralesa noyau
  sym{\'e}trique.
\newblock {\it Comptes rendus de l’Acad{\'e}mie des Sciences\/} {\bf
  173}(1304-1308) 58.

\bibitem[{Brams and Davis(1974)}]{brams19743}
Brams, Steven~J, Morton~D Davis. 1974.
\newblock The 3/2's rule in presidential campaigning.
\newblock {\it American Political Science Review\/} {\bf 68}(1) 113--134.

\bibitem[{Buchanan et~al.(1980)Buchanan, Tollison, and
  Tullock}]{buchanan1980efficient}
Buchanan, James~M, Robert~D Tollison, Gordon Tullock. 1980.
\newblock Efficient rent seeking.
\newblock {\it Toward a Theory of the Rent Seeking Society\/}  97--121.

\bibitem[{Duffy and Matros(2015)}]{duffy2015stochastic}
Duffy, John, Alexander Matros. 2015.
\newblock Stochastic asymmetric blotto games: Some new results.
\newblock {\it Economics Letters\/} {\bf 134} 4--8.

\bibitem[{Duffy and Matros(2017)}]{duffy2017stochastic}
Duffy, John, Alexander Matros. 2017.
\newblock Stochastic asymmetric blotto games: An experimental study.
\newblock {\it Journal of Economic Behavior \& Organization\/} {\bf 139}
  88--105.

\bibitem[{Friedman(1958)}]{friedman1958game}
Friedman, Lawrence. 1958.
\newblock Game-theory models in the allocation of advertising expenditures.
\newblock {\it Operations research\/} {\bf 6}(5) 699--709.

\bibitem[{Gross and Wagner(1950)}]{gross1950continuous}
Gross, Oliver, Robert Wagner. 1950.
\newblock A continuous colonel blotto game.
\newblock Tech. rep., RAND PROJECT AIR FORCE SANTA MONICA CA.

\bibitem[{Kaplan and Barnett(2003)}]{kaplan2003new}
Kaplan, Edward~H, Arnold Barnett. 2003.
\newblock A new approach to estimating the probability of winning the
  presidency.
\newblock {\it Operations Research\/} {\bf 51}(1) 32--40.

\bibitem[{Klumpp and Polborn(2006)}]{klumpp2006primaries}
Klumpp, Tilman, Mattias~K Polborn. 2006.
\newblock Primaries and the new hampshire effect.
\newblock {\it Journal of Public Economics\/} {\bf 90}(6-7) 1073--1114.

\bibitem[{Kovenock and Roberson(2012)}]{kovenockconflicts}
Kovenock, Dan, Brian Roberson. 2012.
\newblock Conflicts with multiple battlefields.
\newblock {\it The Oxford Handbook of the Economics of Peace and Conflict\/}.
  Oxford University Press.

\bibitem[{Kovenock and Roberson(2020)}]{kovenock2020generalizations}
Kovenock, Dan, Brian Roberson. 2020.
\newblock Generalizations of the general lotto and colonel blotto games.
\newblock {\it Economic Theory\/}  1--36.

\bibitem[{Lake(1979)}]{lake1979new}
Lake, Mark. 1979.
\newblock A new campaign resource allocation model.
\newblock {\it Applied game theory\/}. Springer, 118--132.

\bibitem[{Laslier and Picard(2002)}]{laslier2002distributive}
Laslier, Jean-Francois, Nathalie Picard. 2002.
\newblock Distributive politics and electoral competition.
\newblock {\it Journal of Economic Theory\/} {\bf 103}(1) 106--130.

\bibitem[{Monahan(1987)}]{monahan1987structure}
Monahan, George~E. 1987.
\newblock The structure of equilibria in market share attraction models.
\newblock {\it Management Science\/} {\bf 33}(2) 228--243.

\bibitem[{Nagler and Leighley(1992)}]{nagler1992presidential}
Nagler, Jonathan, Jan Leighley. 1992.
\newblock Presidential campaign expenditures: Evidence on allocations and
  effects.
\newblock {\it Public Choice\/} {\bf 73}(3) 319--333.

\bibitem[{{National Popular Vote Inc.}(2019)}]{Eventos_US_2016}
{National Popular Vote Inc.} 2019.
\newblock Two thirds of the presidential campaign is in just 6 states.
\newblock \url{https://www.nationalpopularvote.com/campaign-events-2016}.
\newblock (accessed June, 2020).

\bibitem[{Osorio(2013)}]{osorio2013lottery}
Osorio, Antonio. 2013.
\newblock The lottery blotto game.
\newblock {\it Economics Letters\/} {\bf 120}(2) 164--166.

\bibitem[{{Our World in Data}(2019)}]{owid}
{Our World in Data}. 2019.
\newblock Democracy.
\newblock \url{https://ourworldindata.org/democracy}.
\newblock (accesed June, 2020).

\bibitem[{Rigdon et~al.(2009)Rigdon, Jacobson, Tam~Cho, Sewell, and
  Rigdon}]{rigdon2009bayesian}
Rigdon, Steven~E, Sheldon~H Jacobson, Wendy~K Tam~Cho, Edward~C Sewell,
  Christopher~J Rigdon. 2009.
\newblock A bayesian prediction model for the us presidential election.
\newblock {\it American Politics Research\/} {\bf 37}(4) 700--724.

\bibitem[{Rigdon et~al.(2015)Rigdon, Sauppe, and
  Jacobson}]{rigdon2015forecasting}
Rigdon, Steven~E, Jason~J Sauppe, Sheldon~H Jacobson. 2015.
\newblock Forecasting the 2012 and 2014 elections using bayesian prediction and
  optimization.
\newblock {\it SAGE Open\/} {\bf 5}(2) 2158244015579724.

\bibitem[{Roberson(2006)}]{roberson2006colonel}
Roberson, Brian. 2006.
\newblock The colonel blotto game.
\newblock {\it Economic Theory\/} {\bf 29}(1) 1--24.

\bibitem[{Robson et~al.(2005)}]{robson2005multi}
Robson, Alexander~RW, et~al. 2005.
\newblock Multi-item contests.
\newblock {\it Working paper\/} .

\bibitem[{Rosen(1965)}]{rosen1965existence}
Rosen, J~Ben. 1965.
\newblock Existence and uniqueness of equilibrium points for concave n-person
  games.
\newblock {\it Econometrica: Journal of the Econometric Society\/}  520--534.

\bibitem[{Scheff{\'e}(1958)}]{scheffe1958experiments}
Scheff{\'e}, Henry. 1958.
\newblock Experiments with mixtures.
\newblock {\it Journal of the Royal Statistical Society: Series B
  (Methodological)\/} {\bf 20}(2) 344--360.

\bibitem[{Schwartz et~al.(2014)Schwartz, Loiseau, and
  Sastry}]{schwartz2014heterogeneous}
Schwartz, Galina, Patrick Loiseau, Shankar~S Sastry. 2014.
\newblock The heterogeneous colonel blotto game.
\newblock {\it 2014 7th International Conference on NETwork Games, COntrol and
  OPtimization (NetGCoop)\/}. IEEE, 232--238.

\bibitem[{Shaw(1999)}]{shaw1999methods}
Shaw, Daron~R. 1999.
\newblock The methods behind the madness: Presidential electoral college
  strategies, 1988-1996.
\newblock {\it The Journal of Politics\/} {\bf 61}(4) 893--913.

\bibitem[{Snyder(1989)}]{snyder1989election}
Snyder, James~M. 1989.
\newblock Election goals and the allocation of campaign resources.
\newblock {\it Econometrica: Journal of the Econometric Society\/}  637--660.

\bibitem[{Stromberg(2008)}]{stromberg2008electoral}
Stromberg, David. 2008.
\newblock How the electoral college influences campaigns and policy: the
  probability of being florida.
\newblock {\it American Economic Review\/} {\bf 98}(3) 769--807.

\bibitem[{Thomas(2018)}]{thomas2018n}
Thomas, Caroline. 2018.
\newblock N-dimensional blotto game with heterogeneous battlefield values.
\newblock {\it Economic Theory\/} {\bf 65}(3) 509--544.

\bibitem[{Wang et~al.(2015)Wang, Rothschild, Goel, and
  Gelman}]{wang2015forecasting}
Wang, Wei, David Rothschild, Sharad Goel, Andrew Gelman. 2015.
\newblock Forecasting elections with non-representative polls.
\newblock {\it International Journal of Forecasting\/} {\bf 31}(3) 980--991.

\end{thebibliography}

% CASE 2: BiBTeX used to generate mypaper.bbl (to be further fine tuned)
%\input{mypaper.bbl} % outcomment this line in Case 2

%If you don't use BiBTex, you can manually itemize references as shown below.

%\bibliographystyle{nonumber}

%\begin{thebibliography}{}

%\bibitem[{American Butter Institute(2005)}]{abi}
%American Butter Institute. 2005. Dairy market report. Retrieved June
%14, 2005, www.butterinstitute.org.
%\bibitem[{Rosen 1965}]{R65}
%Rosen, J. Ben. "Existence and uniqueness of equilibrium points for concave n-person games." Econometrica: Journal of the Econometric Society (1965): 520-534.

%Kaplan, E. H., & Barnett, A. (2003). A new approach to estimating the probability of winning the presidency. Operations Research, 51(1), 32-40.
%\end{thebibliography}

\newpage

\begin{APPENDICES}

\section{Proof of Theorem~\ref{the:det_ex}. }\label{proof:the:ex}
It suffices to note that (i) the players' strategy sets are convex, closed, and bounded; and (ii) the players' utilities are concave. We will focus the analysis on the first player. Because of the zero-sum nature of the game, (i) is direct. To show (ii), recall that the utility of the first player is given by $Q^A$ in the objective function of Equation~\eqref{eq:det_max_A}. In order to compute the hessian, let us first compute the gradient. We get $\frac{\partial Q^A}{\partial x_i}=\frac{v_im_i}{\sigma_i p}$, where $p\coloneqq\sum_{i\in\mathcal{I}}v_{i}(s_{i}^A+s_{i}^B)$, $\sigma_i \coloneqq x_i + \alpha_i + y_i + \beta_i + \gamma_i$, $m_i \coloneqq b_{i} p + c_i \sum_{k}v_{k}b_{k}$, $c_i \coloneqq \frac{\gamma_i}{\sigma_i}$, and $b_{i}\coloneqq \frac{y_i+\beta_i}{\sigma_i}$. Then, the hessian matrix of the first player utility is given by  $\left(\nabla_{xx}  Q^A \right)_{ij} = \frac{\partial^2 R^A}{\partial x_i x_j} = - \left( \frac{v_i v_j}{p^3 \sigma_i \sigma_j} (c_i m_j + c_j m_i) + \mathbb{1}_{\{i=j\}} \frac{2 v_i m_i}{p^2 \sigma_i^2} \right)$ for all $i,j\in \mathcal{I}$. To demonstrate the concavity of the utility function, we will show that for any ${\bf z}\in\mathbb{R}^{N}$, it holds that $({\bf z^{T}} \nabla_{xx} Q^A {\bf z} \leq 0$. Indeed, we have
\begin{eqnarray*}
{\bf z^{T}} \nabla_{xx} Q^A  {\bf z} &=& -\sum_{i}\sum_{j} \left(z_iz_j\frac{v_i v_j}{p^3 \sigma_i \sigma_j} (c_i m_j + c_j m_i) + \mathbb{1}_{\{i=j\}} z_i^2\frac{2 v_i m_i}{p^2 \sigma_i^2}\right)\\
&=& -\sum_{i}\sum_{j} \left(\frac{1}{2}w_iw_j(c_i m_j + c_j m_i) + \mathbb{1}_{\{i=j\}} w_i^2pv_{i}^{-1}\right)
\end{eqnarray*}
where $w_{i}\coloneqq z_iv_i\sigma_i^{-1}p^{-3/2}\sqrt{2}$. Then:
\begin{eqnarray} \nonumber
{\bf z^{T}} \nabla_{xx} R^A {\bf z} &=&-\sum_{i}\sum_{j} \left(w_iw_jc_i m_j + \mathbb{1}_{\{i=j\}} w_i^2pm_{i}v_{i}^{-1}\right) \\ \nonumber &=& -\sum_{i}\sum_{j} \left(w_iw_jc_i\left(b_jp+c_j\sum_{k}v_kb_k\right) + \mathbb{1}_{\{i=j\}} \frac{w_i^2p}{v_{i}}\left(b_ip+c_i\sum_{k}v_kb_k\right)\right) \\ \nonumber &\leq& -\sum_{i}\sum_{j} \left(w_iw_jc_i\left(b_jp+c_j\sum_{k}v_kb_k\right) + \mathbb{1}_{\{i=j\}} \frac{w_i^2p^2b_i}{4v_{i}}\right) \\ \label{eq:polP} &=& -\left(\sum_{i}w_ic_i\right)\left(\sum_{i}w_ib_i\right)p-\left(\sum_{i}w_ic_i\right)^2\left(\sum_{i}v_ib_i\right) - \sum_{i}\frac{w_i^2b_i}{4v_{i}}p^2\\ \nonumber
&\le&-\left(\sum_{i}w_ic_i\right)^2\left(\sum_{i}v_ib_i\right)+\frac{\left(\sum_{i}w_ic_i\right)^2\left(\sum_{i}w_ib_i\right)^2}{\sum_{i}\frac{w_i^2b_i}{v_{i}}}\\ \nonumber
&=&-\frac{\left(\sum_{i}w_ic_i\right)^2}{\sum_{i}\frac{w_i^2b_i}{v_{i}}}\left(\left(\sum_{i}v_ib_i\right)\left(\sum_{i}\frac{w_i^2b_i}{v_{i}}\right)-\left(\sum_{i}w_ib_i\right)^2\right)\\ \nonumber
&\le&0.
\end{eqnarray}
The first inequality is because $p>0$ and $c_i,b_i,v_i\ge0$. The second inequality is because Expression~\eqref{eq:polP} is a second degree polynomial of $p$ which is maximized at $p$ equal to $-a_{1}/(2a_{2})$, where $a_i$ is the $i^{th}$ coefficient (i.e. of the variable $p^i$) for $i=1,2$. The last inequality, follows from Lemma~\ref{lem:1}, by replacing ${\bf w}$ with ${\bf x}$, ${\bf b}$ with ${\bf y}$, and the diagonal matrix with $b_{i}/v_{i}$ in the row-column $i$ with ${\bf D}$.
\begin{lemma}\label{lem:1}
For any ${\bf x,y}\in\mathbb{R}^n$, and positive diagonal matrix ${\bf D}\in\mathbb{R}^{n\times n}$, it holds that $\left({\bf x^T}{\bf y}\right)^2\le{\bf x^T}{\bf D}{\bf y}{\bf x^T}{\bf D^{-1}}{\bf y}$.
\proof{Proof.} 
\begin{eqnarray*}
\left({\bf x^T}{\bf y}\right)^2&=&\|{\bf x^T}{\bf y}\|^2_2 \\
&=&\|{\bf x^T}{\bf D^{1/2}}{\bf D^{-1/2}}{\bf y}\|^2_2 \\
&\le&\|{\bf x^T}{\bf D^{1/2}}\|^2_2\|{\bf D^{-1/2}}{\bf y}\|^2_2 \\
&=&{\bf x^T}{\bf D}{\bf y}{\bf x^T}{\bf D^{-1}}{\bf y},
\end{eqnarray*}
where the inequality follows from the Cauchy-Schwartz inequality.
\Halmos
\endproof 
\end{lemma}

\section{Proof of Theorem~\ref{the:det_un}. }\label{proof:the:un}

Let 
\begin{eqnarray*}
G({\bf x},{\bf y})\coloneqq\left[\begin{matrix} \nabla_{xx}Q^A & \nabla_{yx}Q^A \\ \nabla_{xy}Q^B & \nabla_{yy}Q^B\end{matrix} \right].
\end{eqnarray*}
Using Theorem 2 of~\cite{rosen1965existence}, we need to show that $G({\bf x},{\bf y})+G^{T}({\bf x},{\bf y})$ is negative definite. Note that  $\nabla_{yx}Q^A$ and $\nabla_{xy}Q^B$ are symmetric matrices, since the utility functions, $R^A$ and $R^B$, and have both continuous second derivatives. Then 
\begin{eqnarray*}
\left(\nabla_{xy}Q^B\right)^{T}&=&\left(\nabla_{xy}\left(1-Q^A\right)\right)^{T}\\
&=&-\left(\nabla_{xy}Q^A\right)^{T}\\
&=&-\nabla_{xy}Q^A\\
&=&-\nabla_{yx}Q^A.
\end{eqnarray*}
Then 
%Then, to show that $G({\bf x},{\bf y})+G^{T}({\bf x},{\bf y})$ is negative definite. But
$$G({\bf x},{\bf y})+G^{T}({\bf x},{\bf y})=2 \times \left[ \begin{matrix} \nabla_{xx}Q^A & {\bf 0_{N\times N}} \\ {\bf 0_{N\times N}} & \nabla_{yy}Q^B\end{matrix} \right].$$
But we have already shown that $\nabla_{xx}Q^A$ and $\nabla_{yy}Q^B$ are negative definite in the proof of Theorem~\ref{the:det_ex}. Therefore, $G({\bf x},{\bf y})+G^{T}({\bf x},{\bf y})$ is also negative definite, which concludes the proof. 

\section{Proof of Proposition ~\ref{pro:CUB}}\label{proof:pro:CUB}
To do so, we solve the following double KKT equations system:
\begin{eqnarray}\label{pro:CUB_FOC_A}
    \frac{v_i(Q^A + Q^B)(s^B_i + s^C_i)}{\sigma_i} - \frac{Q^A v_i s^C_i}{\sigma_i} &=& \lambda (Q^A + Q^B)^2 \qquad \forall j \in \mathcal{I^*} \\
%\end{equation}
%\begin{equation}
    \label{pro:CUB_FOC_B} \frac{v_i(Q^A + Q^B)(s^A_i + s^C_i)}{\sigma_i} - \frac{Q^B v_i s^C_i}{\sigma_i} &=& \eta (Q^A + Q^B)^2 \qquad \forall j \in \mathcal{I^*} \\
%\end{equation}
%\begin{equation}
     \label{pro:CUB_FOC_sumx} \sum_{j \in \mathcal{I^*}} x_j &=& 1 \\
%\end{equation}
%\begin{equation}
    \label{pro:CUB_FOC_sumy} \sum_{j \in \mathcal{I^*}} y_j &=& 1
\end{eqnarray}
Adding equations~\eqref{pro:CUB_FOC_A} and~\eqref{pro:CUB_FOC_B}, plus using the fact that $s^A_j+  s^B_j + s^C_j = 1$, we get $\frac{v_j}{\sigma_j} = (\lambda + \eta)(Q^A + Q^B)$ for all $j \in \mathcal{I^*}$, equivalently
\begin{eqnarray}\label{eq:pro:CUB_1}
v_j = (\lambda + \eta)(Q^A + Q^B)\sigma_j
\end{eqnarray}
for all $j\in \mathcal{I^{*}}$. Adding up Equation~\eqref{eq:pro:CUB_1} over all $j\in\mathcal{I^{*}}$, and using Equation~\eqref{pro:CUB_FOC_sumx} and~\eqref{pro:CUB_FOC_sumy}, we get
\begin{eqnarray}\label{eq:pro:CUB_2}
v_{\mathcal{I^{*}}}=(\lambda+\eta)(Q^A+Q^B)(2+\alpha_{\mathcal{I^*}}+\beta_{\mathcal{I^*}}+\gamma_{\mathcal{I^*}})
\end{eqnarray}
where for any ${\bf z}\in\mathbb{R}^{n}$ and $\mathcal{I^*}\subseteq\mathcal{I}=\{1,\dots,n\}$, we define $z_{\mathcal{I^*}}\coloneqq \sum_{j\in\mathcal{I^*}}z_{j}$. Replacing the term $(\lambda+\eta)(Q^{A}+Q^{B})$ from Equation~\eqref{eq:pro:CUB_2} into Equation~\eqref{eq:pro:CUB_1} leads to the following identity
\begin{equation}\label{eq:pro:CUB_3}
    \frac{v_j}{\sigma_j} = \frac{v_{\mathcal{I^*}}}{2 + \alpha_{\mathcal{I^*}} + \beta_{\mathcal{I^*}} + \gamma_{\mathcal{I^*}}}
\end{equation}
for all $j\in\mathcal{I^*}$. Let $Q^A_{\mathcal{I^*}}$ be the number of votes obtained by candidate $A$ on the set of regions $\mathcal{I^{*}}$. Similarly define $Q^B_{\mathcal{I^*}}$ for candidate $B$, and $Q^C_{\mathcal{I^*}}$ for the sum of abstention votes. Then, using Equation~\eqref{eq:pro:CUB_3} on the definition of $Q^{A}_{\mathcal{I^*}}$, $Q^{B}_{\mathcal{I^*}}$, and $Q^{C}_{\mathcal{I^*}}$ results in
\begin{eqnarray}
\label{pro:CUB_QAI} Q^A_{\mathcal{I^*}} &=& \sum_{j \in \mathcal{I^*}} v_j  \frac{x_j +\alpha_j}{\sigma_j} =  v_{I^*} \frac{1 + \alpha_{\mathcal{I^*}}}{2 + \alpha_{\mathcal{I^*}} + \beta_{\mathcal{I^*}} + \gamma_{\mathcal{I^*}}} \\
\label{pro:CUB_QBI} Q^B_{\mathcal{I^*}} &=& \sum_{j \in \mathcal{I^*}} v_j  \frac{y_j +\beta_j}{\sigma_j} =  v_{I^*} \frac{1 + \beta_{\mathcal{I^*}}}{2 + \alpha_{\mathcal{I^*}} + \beta_{\mathcal{I^*}} + \gamma_{\mathcal{I^*}}} \\
\label{pro:CUB_QCI} Q^C_{\mathcal{I^*}} &=& \sum_{j \in \mathcal{I^*}} v_j  \frac{\gamma_j}{\sigma_j} =  v_{I^*} \frac{\gamma_{\mathcal{I^*}}}{2 + \alpha_{\mathcal{I^*}} + \beta_{\mathcal{I^*}} + \gamma_{\mathcal{I^*}}}.
\end{eqnarray}
%We can also define $Q^A_{(\mathcal{I^*})^c}$ (and $Q^B_{(\mathcal{I^*})^c}$) as the number of votes obtained by candidate $A$ ($B$) from regions in $(\mathcal{I^*})^c=\mathcal{I}\setminus \mathcal{I^*}$. Note these quantities are constant as they do not depend on the candidates' decision variables. 
Multiplying Equation~\eqref{pro:CUB_FOC_A} by $\sigma_j$ and adding up over all $j \in \mathcal{\mathcal{I^*}}$, we get 
\begin{equation}\label{eq:pro:CUB_4}
\lambda (Q^A + Q^B)^2 (2 + \alpha_{\mathcal{I^*}} + \beta_{\mathcal{I^*}} + \gamma_{\mathcal{I^*}}) = (Q^A+Q^B)(Q^C_{\mathcal{I^*}}+Q^B_{\mathcal{I^*}}) - Q^A Q^C_{\mathcal{I^*}}.
\end{equation} 
Using Equations~\eqref{pro:CUB_QAI}, \eqref{pro:CUB_QBI}, and \eqref{pro:CUB_QCI} on Equation~\eqref{eq:pro:CUB_4} leads to
\begin{equation}\label{eq:pro:CUB_5}
    \lambda = v_{\mathcal{I^*}} \frac{ (Q^A + Q^B)(1 + \beta_{\mathcal{I^*}} +\gamma_{\mathcal{I^*}}) - Q^A \gamma_{\mathcal{I^*}} }{(Q^A + Q^B)^2(2 + \alpha_{\mathcal{I^*}} + \beta_{\mathcal{I^*}} + \gamma_{\mathcal{I^*}})^2} 
\end{equation}
where all values are already known. Doing the analogous steps, we can conclude that
\begin{equation}\label{eq:pro:CUB_6}
    \eta = v_{\mathcal{I^*}} \frac{ (Q^A + Q^B)(1 + \alpha_{\mathcal{I^*}} +\gamma_{\mathcal{I^*}}) - Q^B \gamma_{\mathcal{I^*}} }{(Q^A + Q^B)^2(2+ \alpha_{\mathcal{I^*}} + \beta_{\mathcal{I^*}} + \gamma_{\mathcal{I^*}})^2} 
\end{equation}
Replacing Equations~\eqref{eq:pro:CUB_5} and~\eqref{eq:pro:CUB_6} in Equations~\eqref{pro:CUB_FOC_A} and~\eqref{pro:CUB_FOC_B} %(recall $s^A_j = \frac{x_j + \alpha_j}{\sigma_j}$) 
and after arranging some terms, we get
\begin{equation}\label{eq:x_CUB_app}
x_i^{UB(\mathcal{I^*})} = \frac{v_i}{v_{\mathcal{I^*}}} \left( 1 + \alpha_{\mathcal{I^*}} + \frac{Q^A}{Q^A + Q^B} \gamma_{\mathcal{I^*}} \right) - \frac{Q^A}{Q^A + Q^B} \gamma_i - \alpha_i
\end{equation}
\begin{equation}\label{eq:y_CUB_app}
y_i^{UB(\mathcal{I^*})} = \frac{v_i}{v_{\mathcal{I^*}}} \left( 1 + \beta_{\mathcal{I^*}} + \frac{Q^B}{Q^A + Q^B} \gamma_{\mathcal{I^*}} \right) - \frac{Q^B}{Q^A + Q^B} \gamma_i - \beta_i
\end{equation}
for all $i \in\mathcal{I^*}$. Finally, $Q^A$ and $Q^{B}$, the  number of votes obtained by candidates $A$ and $B$ respectively, can be obtained from Equations~\eqref{pro:CUB_QAI} and \eqref{pro:CUB_QBI} plus the votes obtained in the regions with no campaign. Namely
\begin{eqnarray*}
Q^{A} &=& \sum_{j\in\mathcal{I}}v_{j}\frac{x_i+\alpha_i}{x_i+\alpha_i+y_i+\beta_i+\gamma_i} \\
&=& Q^{A}_{\mathcal{I^*}} + \sum_{j\not\in\mathcal{I^*}}v_{j}\frac{\alpha_{i}}{\alpha_i+\beta_i+\gamma_i}
\end{eqnarray*}
where $Q^{A}_{\mathcal{I^*}}$ is given in Equation~\eqref{pro:CUB_QAI}. Similarly with $Q^B$, concluding the proof.

\section{MS Unbounded Equilibrium with $\mathcal{I}^{*}=\mathcal{I}$}\label{proof:cor:UBA}
In this case, $Q^A = Q^A_{\mathcal{I^*}} = v_{\mathcal{I^*}} \frac{1 + \alpha_{\mathcal{I^*}}}{2 + \alpha_{\mathcal{I^*}} + \beta_{\mathcal{I^*}} + \gamma_{\mathcal{I^*}}} = (\sum_j v_j) \frac{1 + \sum_j \alpha_j}{2 + \sum_j \alpha_j + \sum_j \beta_j + \sum_j \gamma_j}$. Analogous with $Q^B$. Therefore, we have $\frac{Q^A}{Q^A + Q^B} =  \frac{1 + \sum_j \alpha_j}{2 + \sum_j \alpha_j + \sum_j \beta_j}$. Replacing these into the values of $x_i^{UB(\mathcal{I^*})}$ and $y_i^{UB(\mathcal{I^*})}$ of Equations~\eqref{eq:x_CUB_app} and \eqref{eq:y_CUB_app} lead to: 
\begin{equation*}
    x_i^{UB} = \frac{1 + \sum_j \alpha_j}{2 + \sum_j (\alpha_j + \beta_j)} \left[ \frac{v_i}{ \sum_j v_j} \left( 2 + \sum_j (\alpha_j + \beta_j + \gamma_j) \right) - \gamma_i \right] -  \alpha_i
\end{equation*}
\begin{equation*}
    y_i^{UB} = \frac{1 + \sum_j \beta_j}{2 + \sum_j (\alpha_j + \beta_j)} \left[ \frac{v_i}{ \sum_j v_j} \left( 2 + \sum_j (\alpha_j + \beta_j + \gamma_j) \right) - \gamma_i \right] -  \beta_i
\end{equation*}
for all $i\in\mathcal{I}$.

\section{Proof of Corollary~\ref{cor:UBG0}}\label{proof:cor:UBG0}
From Proposition~\ref{pro:CUB}, we know that $x_i^{UB(\mathcal{I^*})} = \frac{v_i}{v_{\mathcal{I^*}}} \left( (1 + \alpha_{\mathcal{I^*}}) + \frac{Q^A}{Q^A + Q^B} \gamma_{\mathcal{I^*}} \right) - \frac{Q^A}{Q^A + Q^B} \gamma_i - \alpha_i$. But, if $\gamma_j = 0$ for all $j \in \mathcal{I}^*$ (and therefore, $\gamma_{\mathcal{I}^*} = 0$), then the last expressions is $x_i^{UB(\mathcal{I^*})} = \frac{v_i}{v_{\mathcal{I^*}}} \left(1 + \alpha_{\mathcal{I^*}} \right) - \alpha_i$ (analogous for $y$). By replacing this expression, we get
\begin{equation}
    \frac{x^{UB(\mathcal{I}^*)}_i + \alpha_i}{x^{UB(\mathcal{I}^*)}_i + \alpha_i + y^{UB(\mathcal{I}^*)}_i + \beta} = \frac{1 + \alpha_{\mathcal{I}^*}}{2 + \alpha_{\mathcal{I}^*} + \beta_{\mathcal{I}^*}}
\end{equation}
\begin{equation}
    \frac{y^{UB(\mathcal{I}^*)}_i + \beta_i}{x^{UB(\mathcal{I}^*)}_i + \alpha_i + y^{UB(\mathcal{I}^*)}_i + \beta} = \frac{1 + \beta_{\mathcal{I}^*}}{2 + \alpha_{\mathcal{I}^*} + \beta_{\mathcal{I}^*}}
\end{equation}

\section{Proof of Theorem~\ref{the:sto_ex}. }\label{proof:the:sto:ex}
The existence of equilibrium in mixed strategies follows from the compactness of the strategy spaces, and the continuity of the utility functions. The former statement is direct, whereas the latter occurs since the winning probability of candidate $A$ is an integral of continuous functions on $x_i$ (see Equation~\eqref{eq:probRA}).

\section{Proof of Proposition~\ref{pro:grad}. }\label{proof:pro:grad}
For ease of exposition, assume that ${\bf x}\in\text{Int}(\Delta_{n})$. Let us do the transformation ${\bf x}=h({\bf w})$ such that $h_{i}({\bf w})=e^{w_i}/\left(\sum_{j\in\mathcal{I}}e^{w_j}\right)$. Note that for any ${\bf x}\in\text{Int}(\Delta_{n})$, there exists a ${\bf w}\in\mathbb{R}^{n}$ such that ${\bf x}=h({\bf w})$; indeed $w_{i}=\ln(x_i/x_1)+w_1$ for $i>1$, while $w_1$ can take any arbitrary value. Then, we look for the directional derivative of $h({\bf w})$ in the direction of maximum growth of $f$, i.e. $\nabla^{w}f(h({\bf w}))$. Then, for every $i\in\mathcal{I}$ we have
\begin{eqnarray}\label{eq:d_i} \nonumber
d^{x}_{i} &=& \lim_{t\rightarrow0} \frac{h_{i}({\bf w}+t\nabla^{w}f(h({\bf w})))-h_{i}({\bf w})}{t} \\
&=& \nabla^{w}h_{i}({\bf w}) \boldsymbol{\cdot} \nabla^{w}f(h({\bf w}))
\end{eqnarray}
where $\boldsymbol{\cdot}$ is the dot product. Then
\begin{eqnarray}\label{eq:partial_h} \nonumber
\frac{\partial h_{i}}{\partial w_{j}} &=& \frac{\partial }{\partial w_{j}} \left(\frac{e^{w_i}}{\sum_{k\in\mathcal{I}}e^{w_k}}\right) \\ \nonumber
&=& \mathbbm{1}_{\left\{i=j\right\}}\frac{e^{w_i}}{\sum_{k\in\mathcal{I}}e^{w_k}}-\frac{-e^{w_i}e^{w_j}}{\left(\sum_{k\in\mathcal{I}}e^{w_k}\right)^2} \\ \nonumber
&=& h_{i}({\bf w})\left(\mathbbm{1}_{\left\{i=j\right\}}-h_{j}({\bf w})\right) \\
&=& x_{i}\left(\mathbbm{1}_{\left\{i=j\right\}}-x_{j}\right) 
\end{eqnarray}
and
\begin{eqnarray}\label{eq:partial_f} \nonumber
\frac{\partial f}{\partial w_{j}} &=&
\sum_{k\in\mathcal{I}} \frac{\partial f}{\partial x_{k}}\frac{\partial h_{k}}{\partial w_{j}} \\ \nonumber
&=& \sum_{k\in\mathcal{I}} \frac{\partial f}{\partial x_{k}}h_{k}({\bf w})\left(\mathbbm{1}_{\left\{k=j\right\}}-h_{j}({\bf w})\right) \\ \nonumber
&=& h_{j}({\bf w})\left(\frac{\partial f}{\partial x_{j}}-\sum_{k\in\mathcal{I}} \frac{\partial f}{\partial x_{k}}h_{k}({\bf w})\right) \\ \nonumber
&=& x_{j}\left(\frac{\partial f}{\partial x_{j}}-\sum_{k\in\mathcal{I}} \frac{\partial f}{\partial x_{k}}x_{k}\right) \\
&\eqqcolon& \tau^{x}_{i}.
\end{eqnarray}
Putting together Equations~\eqref{eq:partial_h} and~\eqref{eq:partial_f} into Equation~\eqref{eq:d_i}, we get
\begin{eqnarray*}\label{eq:d_i2}
d_{i}^{x} &=& \sum_{j\in\mathcal{I}} x_{i}\left(\mathbbm{1}_{\left\{i=j\right\}}-x_{j}\right) \tau^{x}_{j} \\
&=& x_{i}\left(\tau^{x}_{i}-\sum_{j\in\mathcal{I}} x_{j}\tau^{x}_{j}\right).
\end{eqnarray*}

\section{Gradient formulas for the majority system stochastic model}\label{app:f_gradf}
Let us denote the event $W\coloneqq \sum_{i\in\mathcal{I}}v_{i}s_{i}^{A}>\sum_{i\in\mathcal{I}}v_{i}s_{i}^{B}$. Then
\begin{eqnarray*}
\mathbb{P}\left(R^A>R^B\right)&=&\int_{\Delta_{3}}\dots \int_{\Delta_{3}}\mathbbm{1}_{\left\{W\right\}}\prod_{i\in\mathcal{I}}f_{i}ds_i.
\end{eqnarray*}
The derivatives with respect to each candidate investing component, $x_i$ and $y_i$, can be computed as
\begin{eqnarray*}
\frac{\partial}{\partial x_i}\mathbb{P}\left(R^A>R^B\right)&=&k\int_{\Delta_{3}}\dots \int_{\Delta_{3}}\mathbbm{1}_{\left\{W\right\}}\ln(s_{i}^{A})\prod_{i\in\mathcal{I}}f_{i}ds_i + k z_i^A \int_{\Delta_{3}}\dots \int_{\Delta_{3}}\mathbbm{1}_{\left\{W\right\}}\prod_{i\in\mathcal{I}}f_{i}ds_i \\
\frac{\partial}{\partial y_i}\mathbb{P}\left(R^A>R^B\right)&=&k\int_{\Delta_{3}}\dots \int_{\Delta_{3}}\mathbbm{1}_{\left\{W\right\}}\ln(s_{i}^{B})\prod_{i\in\mathcal{I}}f_{i}ds_i + k z_i^B \int_{\Delta_{3}}\dots \int_{\Delta_{3}}\mathbbm{1}_{\left\{W\right\}}\prod_{i\in\mathcal{I}}f_{i}ds_i
\end{eqnarray*}
where $z_{i}^A=\psi\left(k(\alpha_i+x_{i}+\beta_i+y_i+\gamma_i)\right) - \psi\left(k(\alpha_i+x_{i})\right)$ and $z_{i}^B=\psi\left(k(\alpha_i+x_{i}+\beta_i+y_i+\gamma_i)\right) - \psi\left(k(\beta_i+y_{i})\right)$, where $\psi(\cdot)$ denotes the di-gamma function.

\section{Demonstration of Proposition~\ref{pro:close}}\label{proof:pro:close}
Let $\textbf{s}=\{(s_{i}^A,s_{i}^{B},s_{i}^{C})\}_{i=1}^{n}$ be $n$ samples of the random variable $\textbf{S}$. The density at each point, given the parameters $({\bf x},{\bf y})$ will be denoted as $f_{i|{\bf x},{\bf y}}(s_{i}^{A},s_{i}^{B},s_{i}^C)$. Then, we can write the density of the Dirichlet random variable $(S_{i}^A,S_{i}^{B},S_{i}^{C})$ at a particular $({\bf x}+\Delta {\bf x},{\bf y}+{\bf \Delta y})$ as
\begin{eqnarray*}
f_{i|{\bf x}+{\bf \Delta x}, {\bf y}+{\bf \Delta y}}(s_{i}^{A},s_{i}^{B},s_{i}^C) &=& \frac{(s^A_j)^{k(x_j + \Delta x_j  + \alpha_j)-1}(s^B_j)^{k(y_j +  \Delta y_j + \beta_j)-1}(s^C_j)^{k\gamma_j-1}}{\text{B}(k(x_j+\Delta x_j+\alpha_j),k(y_j+\Delta y_j+\beta_j),k\gamma_j)} \\
&=& (s^A_j)^{k\Delta x_j}(s^B_j)^{k\Delta y_j} \frac{(s^A_j)^{k(x_j  + \alpha_j)-1}(s^B_j)^{k(y_j +  \beta_j)-1}(s^C_j)^{k\gamma_j-1}}{\text{B}(k(x_j+\Delta x_j+\alpha_j),k(y_j+\Delta y_j+\beta_j),k\gamma_j)} \\
&=& (s^A_j)^{k\Delta x_j}(s^B_j)^{k\Delta y_j} \frac{\text{B}(k(x_j+\alpha_j),k(y_j+\beta_j),k\gamma_j}{\text{B}(k(x_j+\Delta x_j+\alpha_j),k(y_j+\Delta y_j+\beta_j),k\gamma_j)} f_{i|{\bf x},{\bf y}}(s_{i}^{A},s_{i}^{B},s_{i}^C) \\
&=& (s^A_j)^{k\Delta x_j}(s^B_j)^{k\Delta y_j} \times K_{j} \times f_{i|{\bf x},{\bf y}}(s_{i}^{A},s_{i}^{B},s_{i}^C)
\end{eqnarray*}
where $K_j = \frac{\text{B}(k(x_j+\alpha_j),k(y_j+\beta_j),k\gamma_j)}{\text{B}(k(x_j+\Delta x_j+\alpha_j),k(y_j+\Delta y_j+\beta_j),k\gamma_j)}$. Let $K=\Pi_i K_i$. Then, we can write the expectation of $g(\textbf{S})$ given the efforts $({\bf x}+{\bf \Delta x},{\bf y}+{\bf \Delta y})$ as
\begin{eqnarray*}
\mathbb{E}(g(\textbf{S})|{\bf x}+{\bf \Delta x},{\bf y}+{\bf \Delta y}) &=& \int \dots \int _{(\Delta_3)^n} g(\textbf{s}) \times \prod_{j}f_{j|{\bf x}+{\bf \Delta x}, {\bf y}+{\bf \Delta y}}(s_{j}^{A},s_{j}^{B},s_{j}^C)ds_{j}^{A}ds_{j}^{B}ds_{j}^{C} \\
&=& K\times \int \dots \int _{(\Delta_3)^n} g(\textbf{s}) \times \prod_{j}(s^A_j)^{k\Delta x_j}(s^B_j)^{k\Delta y_j} f_{j|{\bf x},{\bf y}}(s_{j}^{A},s_{j}^{B},s_{j}^C)ds_{j}^{A}ds_{j}^{B}ds_{j}^{C} \\
&=&K \times \mathbb{E}\left(g(\textbf{S}) \times \prod_{j} (S^A_j)^{k\Delta x_j}(S^B_j)^{k\Delta y_j} | {\bf x},{\bf y}\right).
\end{eqnarray*}

\section{Demonstration of Proposition~\ref{pro:close_var}}\label{proof:pro:close_var}
Let $W$ and $V$ be the rv of the candidate who wins the election at point $({\bf x}+{\bf \Delta x},{\bf y}+{\bf \Delta y})$ and $({\bf x},{\bf y})$ respectively. So $W=A$ if candidate $A$ wins at strategies $({\bf x}+{\bf \Delta x},{\bf y}+{\bf \Delta y})$, and $W=B$ otherwise; similarly for $V$. Let us denote $p=\mathbb{P}(W=A)=\mathbb{P}(g(\textbf{S})|{\bf x}+{\bf \Delta x},{\bf y}+{\bf \Delta y})$, $q=\mathbb{P}(V=A)=\mathbb{P}(g(\textbf{S})|{\bf x},{\bf y})$. Note we are assuming $p>q$. Denote the function  $h(\textbf{S})=K\times\prod_{j}(S_{j}^{A})^{k\times\Delta x_j}(S_{j}^{B})^{k\times\Delta y_j}$. Using the law of total variance, we have
\begin{eqnarray}\label{eq:var1}
\text{Var}\left(g(\textbf{S})|{\bf x}+{\bf \Delta x},{\bf y}+{\bf \Delta y}\right)&=&\mathbb{E}\left[\text{Var}\left(g(\textbf{S})|{\bf x}+{\bf \Delta x},{\bf y}+{\bf \Delta y},W\right)\right]+\text{Var}\left(\mathbb{E}\left[g(\textbf{S})|{\bf x}+{\bf \Delta x},{\bf y}+{\bf \Delta y},W\right]\right).
\end{eqnarray}
With respect to the first term of the RHS of Equation~\eqref{eq:var1} we have
\begin{eqnarray}\label{eq:var2} \nonumber
&&\mathbb{E}\left[\text{Var}\left(g(\textbf{S})|{\bf x}+{\bf \Delta x},{\bf y}+{\bf \Delta y},W\right)\right]\\ \nonumber
&&=\text{Var}\left(g(\textbf{S})|{\bf x}+{\bf \Delta x},{\bf y}+{\bf \Delta y},W=A\right)\mathbb{P}(W=A)+\text{Var}\left(g(\textbf{S})|{\bf x}+{\bf \Delta x},{\bf y}+{\bf \Delta y},W=B\right)\mathbb{P}(W=B) \\ \nonumber
&&=\mathbb{E}\left[\left(g(\textbf{S})-\mathbb{E}\left[g(\mathbf{S})|{\bf x}+{\bf \Delta x},{\bf y}+{\bf \Delta y},W=A\right]\right)^2|{\bf x}+{\bf \Delta x},{\bf y}+{\bf \Delta y},W=A\right]\mathbb{P}(W=A) \\ \nonumber
&&\qquad+\mathbb{E}\left[\left(g(\textbf{S})-\mathbb{E}\left[g(\mathbf{S})|{\bf x}+{\bf \Delta x},{\bf y}+{\bf \Delta y},W=B\right]\right)^2|{\bf x}+{\bf \Delta x},{\bf y}+{\bf \Delta y},W=B\right]\mathbb{P}(W=B) \\ \nonumber
&&=\mathbb{E}\left[\left(1-1\right)^2\right]p+\mathbb{E}\left[\left(0-0\right)^2\right](1-p) \\ \nonumber
&&= 0\cdot p+0\cdot(1-p) \\ 
&&= 0.
\end{eqnarray}
With respect to the second term of the RHS of Equation~\eqref{eq:var1} we have
\begin{eqnarray}\label{eq:var3} \nonumber
&&\text{Var}\left(\mathbb{E}\left[g(\textbf{S})|{\bf x}+{\bf \Delta x},{\bf y}+{\bf \Delta y},W\right]\right)\\ \nonumber
&&= \left(1-\mathbb{E}\left[\mathbb{E}\left[g(\textbf{S})|{\bf x}+{\bf \Delta x},{\bf y}+{\bf \Delta y},W\right]\right]\right)^2\mathbb{P}(W=A) + \left(0-\mathbb{E}\left[\mathbb{E}\left[g(\textbf{S})|{\bf x}+{\bf \Delta x},{\bf y}+{\bf \Delta y},W\right]\right]\right)^2\mathbb{P}(W=B) \\ \nonumber
&&= \left(1-\mathbb{E}\left[g(\textbf{S})|{\bf x}+{\bf \Delta x},{\bf y}+{\bf \Delta y}\right]\right)^2p + \left(0-\mathbb{E}\left[g(\textbf{S})|{\bf x}+{\bf \Delta x},{\bf y}+{\bf \Delta y}\right]\right)^2(1-p) \\ \nonumber
&&= \left(1-p\right)^2p + \left(0-p\right)^2(1-p) \\
&&=p(1-p).
\end{eqnarray}
Putting together Equations~\eqref{eq:var2} and~\eqref{eq:var3} into Equation~\eqref{eq:var1}, we get
\begin{eqnarray}\label{eq:var4}
\text{Var}\left(g(\textbf{S})|{\bf x}+{\bf \Delta x},{\bf y}+{\bf \Delta y}\right)&=&p(1-p).
\end{eqnarray}

Again, using the law of total variance, we have
\begin{eqnarray}\label{eq:var5}
\text{Var}\left(h(\textbf{S})g(\textbf{S})|{\bf x},{\bf y}\right)&=&\mathbb{E}\left[\text{Var}\left(h(\textbf{S})g(\textbf{S})|{\bf x},{\bf y},V\right)\right]+\text{Var}\left(\mathbb{E}\left[h(\textbf{S})g(\textbf{S})|{\bf x},{\bf y},V\right]\right).
\end{eqnarray}
With respect to the first term of the RHS of Equation~\eqref{eq:var5} we have
\begin{eqnarray}\label{eq:var6} \nonumber
&&\mathbb{E}\left[\text{Var}\left(h(\textbf{S})g(\textbf{S})|{\bf x},{\bf y},V\right)\right]\\ \nonumber
&&=\text{Var}\left(h(\textbf{S})g(\textbf{S})|{\bf x},{\bf y},V=A\right)\mathbb{P}(V=A)+\text{Var}\left(h(\textbf{S})g(\textbf{S})|{\bf x},{\bf y},V=B\right)\mathbb{P}(V=B) \\ \nonumber
&&=\mathbb{E}\left[\left(h(\textbf{S})g(\textbf{S})-\mathbb{E}\left[h(\textbf{S})g(\mathbf{S})|{\bf x},{\bf y},V=A\right]\right)^2|{\bf x},{\bf y},V=A\right]\mathbb{P}(V=A) \\ \nonumber
&&\qquad+\mathbb{E}\left[\left(h(\textbf{S})g(\textbf{S})-\mathbb{E}\left[h(\textbf{S})g(\mathbf{S})|{\bf x},{\bf y},V=B\right]\right)^2|{\bf x},{\bf y},V=B\right]\mathbb{P}(V=B) \\ \nonumber
&&=\mathbb{E}\left[\left(h(\textbf{S})g(\textbf{S})-\frac{p}{q}\right)^2|{\bf x},{\bf y},V=A\right]\mathbb{P}(V=A) +\mathbb{E}\left[\left(h(\textbf{S})g(\textbf{S})-0\right)^2|{\bf x},{\bf y},V=B\right]\mathbb{P}(V=B) \\ \nonumber
&&=\mathbb{E}\left[\left(h(\textbf{S})-\frac{p}{q}\right)^2|{\bf x},{\bf y},V=A\right]q+\mathbb{E}\left[\left(0-0\right)^2\right](1-q) \\ \label{eq:var6b}
&&> 0
\end{eqnarray}
where in the third equality we use the fact that $\mathbb{E}\left[h(\textbf{S})g(\textbf{S})|{\bf x},{\bf y},V=A\right]=1$. This is true since we have that \begin{eqnarray*}
p&=&\mathbb{E}\left[h(\textbf{S})g(\textbf{S})|{\bf x},{\bf y}\right]\\
&=&\mathbb{E}\left[h(\textbf{S})g(\textbf{S})|{\bf x},{\bf y},V=A\right]\mathbb{P}(V=A)+\mathbb{E}\left[h(\textbf{S})g(\textbf{S})|{\bf x},{\bf y},V=B\right]\mathbb{P}(V=B)\\
&=&\mathbb{E}\left[h(\textbf{S})g(\textbf{S})|{\bf x},{\bf y},V=A\right]\cdot q+0\cdot(1-q)\\
&=&\mathbb{E}\left[h(\textbf{S})g(\textbf{S})|{\bf x},{\bf y},V=A\right]\cdot q.
\end{eqnarray*}
Then $\mathbb{E}\left[h(\textbf{S})g(\textbf{S})|{\bf x},{\bf y},V=A\right]=\frac{p}{q}$. With respect to the second term of the RHS of Equation~\eqref{eq:var5} we have
\begin{eqnarray}\label{eq:var7} \nonumber
&&\text{Var}\left(\mathbb{E}\left[h(\textbf{S})g(\textbf{S})|{\bf x},{\bf y},V\right]\right)\\ \nonumber
&&= \left(\frac{p}{q}-\mathbb{E}\left[\mathbb{E}\left[h(\textbf{S})g(\textbf{S})|{\bf x},{\bf y},V\right]\right]\right)^2\mathbb{P}(V=A) + \left(0-\mathbb{E}\left[\mathbb{E}\left[h(\textbf{S})g(\textbf{S})|{\bf x},{\bf y},V\right]\right]\right)^2\mathbb{P}(V=B) \\ \nonumber
&&= \left(\frac{p}{q}-\mathbb{E}\left[h(\textbf{S})g(\textbf{S})|{\bf x},{\bf y}\right]\right)^2q + \left(0-\mathbb{E}\left[h(\textbf{S})g(\textbf{S})|{\bf x},{\bf y}\right]\right)^2(1-q) \\ \nonumber
&&= \left(\frac{p}{q}-p\right)^2q + \left(0-p\right)^2(1-q) \\
&&=p\left(\frac{p}{q}-p\right).
\end{eqnarray}
Putting together Equations~\eqref{eq:var6} and~\eqref{eq:var7} into Equation~\eqref{eq:var5}, we get
\begin{eqnarray}\label{eq:var8} \nonumber
\text{Var}\left(h(\textbf{S})g(\textbf{S})|{\bf x},{\bf y}\right)&>&p\left(\frac{p}{q}-p\right) \\ \nonumber
&>&p(1-p) \\ \nonumber
&=& \text{Var}\left(g(\textbf{S})|{\bf x}+{\bf \Delta x},{\bf y}+{\bf \Delta y}\right).
\end{eqnarray}
The second equality is because we are assuming the case where $p>q$, and the last equality comes from Equation~\eqref{eq:var4}. This concludes the proof.

As a side note, notice that we can give an expression for computing the variance of the random variable $h(\textbf{S})g(\textbf{S})|{\bf x},{\bf y}$ by only reusing the Dirichlet rv at point $({\bf x},{\bf y})$.
\begin{eqnarray}\label{eq:var11} \nonumber
\text{Var}\left(h(\textbf{S})g(\textbf{S})|{\bf x},{\bf y}\right) &=&\mathbb{E}\left[(h(\textbf{S}))^2(g(\textbf{S}))^2|{\bf x},{\bf y}\right]-\mathbb{E}\left[h(\textbf{S})g(\textbf{S})|{\bf x},{\bf y}\right]^2\\
&=&\mathbb{E}\left[(h(\textbf{S}))^2g(\textbf{S})|{\bf x},{\bf y}\right]-p^2
\end{eqnarray}

\section{Demonstration of Lemma~\ref{lem:Dirichlet_Abstencion}}
\label{proof:lem:DIR}
\begin{eqnarray}\label{eq:lemdir1} \nonumber
\text{Cov}\left(\frac{X}{X+Y},Z\right) &=&\text{Cov}\left(\frac{X}{X+Y},1-X-Y\right)\\ \nonumber
&=&-\text{Cov}\left(\frac{X}{X+Y},X+Y\right)\\ \nonumber
&=&-\mathbb{E}[X]+\mathbb{E}[X+Y]\times\mathbb{E}\left[\frac{X}{X+Y}\right]\\
&=&-\frac{a}{a+b+c}+\frac{a+b}{a+b+c}\mathbb{E}\left[\frac{X}{X+Y}\right]
\end{eqnarray}
With respect to the term $\mathbb{E}\left[\frac{X}{X+Y}\right]$, we can compute this by
\begin{eqnarray}  \nonumber
\mathbb{E}\left[\frac{X}{X+Y}\right] &=& \mathbb{E}\left[\frac{X}{1-Z}\right] \\ \nonumber
&=& \int_0^1 \int_0^{1 - z}  \frac{x}{1-z} \cdot f_{X,Y,Z}(x,1-x-z,z) dx dz \\ \nonumber 
&=& \frac{1}{\text{B}(a,b,c)} \int_0^1 \int_0^{1 - z} \frac{x}{1-z} \cdot x^{a-1} (1-x-z)^{b-1} z^{c-1} dx dz \\ \nonumber 
&=& \frac{1}{\text{B}(a,b,c)} \int_0^1  (1-z)^{-1} z^{c-1} \int_0^{1 - z} x^{a} (1-x-z)^{b-1} dx dz \\ \nonumber
&=&  \frac{1}{\text{B}(a,b,c)} \int_0^1  (1-z)^{a+b-1} z^{c-1} \int_0^{1} u^{a} (1-u)^{b-1} du dz \\ \nonumber
&=&  \frac{\text{B}(a,b)}{\text{B}(a,b,c)} \int_0^1  (1-z)^{a+b-1} z^{c-1} dz\int_0^{1} u\frac{u^{a-1} (1-u)^{b-1}}{\text{B}(a,b)} du \\ \nonumber
&=& \frac{\text{B}(a,b)}{\text{B}(a,b,c)} \cdot\text{B}(a+b,c)\cdot \frac{a}{a+b} \\ \nonumber 
&=& \frac{\Gamma(a)\Gamma(b)}{\Gamma(a+b)}\cdot\frac{\Gamma(a+b+c)}{\Gamma(a)\Gamma(b)\Gamma(c)}\cdot\frac{\Gamma(a+b)\Gamma(c)}{\Gamma(a+b+c)} \cdot\frac{a}{a+b} \\ \label{eq:lemdir2}
&=& \frac{a}{a+b}
\end{eqnarray}
In the fifth equality we use the change of variable $u=\frac{x}{1-z}$. Replacing Equation~\eqref{eq:lemdir2} into Equation~\eqref{eq:lemdir1}, we get
\begin{eqnarray}\label{eq:lemdir3} \nonumber
\text{Cov}\left(\frac{X}{X+Y},Z\right) &=&-\frac{a}{a+b+c}+\frac{a+b}{a+b+c}\cdot \frac{a}{a+b} \\ \nonumber
&=& 0
\end{eqnarray}
showing the result for the covariance.

In order to show that $\frac{X}{X+Y} \sim \textbf{Beta}(a,b)$, we compute its pdf. Let $W\coloneqq \frac{X}{X+Y}$, then $(X,Y,Z)=(W(1-Z),(1-W)(1-Z),Z)$. Then consider a $w\in(0,1)$,
\begin{eqnarray*}
f_{W}(w) &=& \int_{0}^{1}\int_{0}^{1-z} \mathbb{1}_{\{x=wx+w(1-x-z)\}}f_{X,Y,Z}(x,1-x-z,z)dxdz \\
&=& \int_{0}^{1}\int_{0}^{1-z} f_{X,Y,Z}(w(1-z),(1-w)(1-z),z)dxdz \\
&=& \int_{0}^{1}(1-z)f_{X,Y,Z}(w(1-z),(1-w)(1-z),z)dxdz \\
&=& w^{a-1} (1-w)^{b-1} \cdot \frac{1}{\text{B}(a,b,c)} \cdot \int_0^1 (1-z)^{a+b-1} z^{c-1} dz \\
&=& w^{a-1} (1-w)^{b-1} \frac{B(a+b,c)}{B(a,b,c)} \\
&=& \frac{w^{a-1} (1-w)^{b-1}}{B(a,b)} 
\end{eqnarray*}
where we used in the sixth equality that $B(a,b,c)=B(a,b)\cdot B(a+b,c)$. The resulting pdf for $W = \frac{X}{X+Y}$ corresponds to a Beta distribution with parameters $(a,b)$, which concludes the proof.

\section{Proof of Theorem~\ref{obs:exp_numb_votes}. }\label{proof:obs:exp_numb_votes}
If the candidates maximize the expected number of votes under MS, candidate $A$'s objective is $\mathbb{E}(\sum_i v_i S^A_i)$, or equivalently $\sum_i v_i \mathbb{E}(S^A_i)$ by linearity of expectation. If now we assume that $\gamma=0$, then $\mathbb{E}(S^A_i) = \frac{x_i + \alpha_i}{x_i + \alpha_i + y_i + \beta_i}$. Therefore, the problem candidate $A$ faces is to maximize $\sum_i v_i \frac{x_i + \alpha_i}{x_i + \alpha_i + y_i + \beta_i}$ with ${\bf x}\in\Delta_n$. Similarly, candidate $B$ maximizes $\sum_i v_i \frac{y_i + \beta_i}{x_i + \alpha_i + y_i + \beta_i}$ with ${\bf y}\in\Delta_n$.

Under EC, the expected number of votes candidate $A$ gets from region $i$ is $\frac{S^A_i}{S^A_i + S^B_i}$. Since for every $i$, $S^A_i$ and $S^B_i$ are components of the same Dirichlet distribution, then its ratio follows a Beta distribution by using Lemma~\ref{lem:Dirichlet_Abstencion}. Namely,  $\frac{S^A_i}{S^A_i + S^B_i}\sim \textbf{Beta}(k(x_{i}+\alpha_{i}),k(y_{i}+\beta_{i}))$. Then, if candidates maximize the expected number of votes, candidate $A$'s objective is $\mathbb{E}\left[\sum_{i}w_{i}\frac{S^A_i}{S^A_i + S^B_i}\right]=\sum_{i}w_{i}\mathbb{E}\left[\frac{S^A_i}{S^A_i + S^B_i}\right]=\sum_i w_i \frac{x_i + \alpha_i}{x_i + \alpha_i + y_i + \beta_i}=\sum_i \theta v_{i} \frac{x_i + \alpha_i}{x_i + \alpha_i + y_i + \beta_i}$ with ${\bf x}\in\Delta_n$, where $\theta\coloneqq w_{i}/v_{i}$ for all $i$ (since we are using the assumption that electoral votes of each state are proportional to the number of popular votes of the respective state). Similarly for candidate $B$, concluding the result.

%, since those are distributed as Dirichlet, this ratio follows a Beta distribution, with parameters $(k(x_i + \alpha_i), k(y_i + \beta_i))$, therefore $\mathbb{E}(\frac{S^A_i}{S^A_i + S^B_i}) = \frac{x_i + \alpha_i}{x_i + \alpha_i + y_i + \beta_i}$, so that the problem candidate $A$ faces now is to maximize $\sum_i w_i \frac{x_i + \alpha_i}{x_i + \alpha_i + y_i + \beta_i}$, $x \in \Delta_n$ (same for $B$). Since we assumed that $v$ and $w$ where proportional, the two problems are the same. 

\section{Proof of Theorem~\ref{obs:prob_winning}}\label{proof:obs:prob_winning}
For ease of notation for the reader, we denote sometimes for this proof $\mathbbm{1}_{\{X\}}$ as $\mathbbm{1}{\{X\}}$. Also, for ease of exposition, we only consider the case where  ${\bf v} \in \mathbb{Z}^n_{+}$ and $\sum_{i}v_i$ is odd and therefore there is no chance of tie under EC. We want to show that $\mathbb{P}(|\mathbbm{1}{\{\sum_{i}v_{i}s_{i}^{A}>\sum_{i}v_{i}s_{i}^{B}\}} - \mathbbm{1}{\{\sum_{i}v_{i}\mathbbm{1}_{\{s_{i}^{A}>s_{i}^{B}\}}>\sum_{i}v_{i}\mathbbm{1}_{\{s_{i}^{A}<s_{i}^{B}\}}\}}>\epsilon)\overset{k\rightarrow0}{\to}0$ for any $\epsilon>0$. Consider $0<\epsilon<1$, then
\begin{eqnarray} \nonumber
&&\mathbb{P}(|\mathbbm{1}{\{\sum_{i}v_{i}s_{i}^{A}>\sum_{i}v_{i}s_{i}^{B}\}} - \mathbbm{1}{\{\sum_{i}v_{i}\mathbbm{1}_{\{s_{i}^{A}>s_{i}^{B}\}}>\sum_{i}v_{i}\mathbbm{1}_{\{s_{i}^{A}<s_{i}^{B}\}}\}}|>\epsilon) \\ \nonumber
&=&  \mathbb{P}(|\mathbbm{1}{\{\sum_{i}v_{i}s_{i}^{A}>\frac{1}{2}\sum_{i}v_{i}\}} - \mathbbm{1}{\{\sum_{i}v_{i}\mathbbm{1}_{\{s_{i}^{A}>1/2\}}>\frac{1}{2}\sum_{i}v_{i}\}}|>\epsilon) \\ \nonumber
&=&  \mathbb{P}(\underbrace{\sum_{i}v_{i}s_{i}^{A}>\frac{1}{2}\sum_{i}v_{i}> \sum_{i}v_{i}\mathbbm{1}_{\{s_{i}^{A}>1/2\}}}_{E_{1}}) + \mathbb{P}(\underbrace{\sum_{i}v_{i}s_{i}^{A}<\frac{1}{2}\sum_{i}v_{i}< \sum_{i}v_{i}\mathbbm{1}_{\{s_{i}^{A}>1/2\}}}_{E_{2}}),
\end{eqnarray}
where the first equality uses the fact of no-abstention. We will define the event $E_{3}$, and show that (i) $E_{3}\cap E_{1},E_{2}=\emptyset$, and (ii) $\mathbb{P}(E_{3})\rightarrow1$ as $k\rightarrow0$. Note that since (i) it holds that $\mathbb{P}(E_{1})+\mathbb{P}(E_{2})+\mathbb{P}(E_{3})\le 1$ and because of (ii) it must be that $\mathbb{P}(E_{1})+\mathbb{P}(E_{2})\rightarrow0$ as $k\rightarrow0$. For what follows, we will denote $s_{i}
^{A}$ as $s_{i}$ to reduce notation. Consider $\delta=1/\left(4\sum_{i}v_{i}\right)$, and let us define the event $E_{3}$ as $E_{3}=\{{\bf s}\in[0,1]^{n}: \exists {\bf z}\in\{0,1\}^{n}\text{ s.t. } \|z-s\|_{\infty}<\delta\}$ or equivalently $E_{3}=\{{\bf s}\in[0,1]^{n}: |s_{i}-\mathbbm{1}_{\{s_{i}>1/2\}}|<\delta \forall i\}$.\\
To show (i), we will first address the case where $E_{3}\cap E_{1}=\emptyset$. By contradiction, consider ${\bf s}\in E_{3}\cap E_{1}$. Then we have that $\sum_{i}v_{i}s_{i}>\frac{1}{2}\sum_{i}v_i>\sum_{i}v_{i}\mathbbm{1}_{\{s_{i}>1/2\}}$, equivalently, $\sum_{i}v_{i}s_{i}>\frac{1}{2}\sum_{i}v_i\ge\sum_{i}v_{i}\mathbbm{1}_{\{s_{i}>1/2\}}+1/2$ since $\sum_{i}v_{i}$ is odd and $v_{i}$ take integer values. Subtracting the right hand side of the previous expression from the left hand side leads to $\sum_{i}v_{i}(s_{i}-\mathbbm{1}_{\{s_{i}>1/2\}})>1/2$. Since ${\bf s}\in E_{3}$, it is easy to see that $\mathbbm{1}_{\{s_{i}>1/2\}}=\mathbbm{1}_{\{z_{i}>1/2\}}=z$, where ${\bf z}$ is the closest ``corner point'' (in $\{0,1\}^{n}$) to ${\bf s}$. Then we have $\sum_{i}v_{i}(s_{i}-\mathbbm{1}_{\{s_{i}>1/2\}})=\sum_{i}v_{i}(s_{i}-z_{i})\le \sum_{i}v_{i}|s_{i}-z_{i}|< \sum_{i}v_{i}\delta=1/4>1/2$ which is a contradiction. The proof of $E_{3}\cap E_{2}=\emptyset$ is analogous, and therefore is omitted. \\
To show (ii), let us define the events $D_{i}=\{{\bf s}\in[0,1]^{n}|\delta\le s_{i}\le 1-\delta\}$. Therefore, it can be seen that $\bigcup_i D_i = E_3^c$. Indeed, if there is some ${\bf s}\in D_{i}$ for some $i$, then clearly there is no ${\bf z}\in\{0,1\}^{n}$ such that $\|{\bf z}-{\bf s}\|<\delta$. Conversely, if there is some ${\bf s}\in E_{3}$, then there is at least one component $k$ such that $\|{\bf z}-{\bf s}\|_{\infty}\ge\delta$ for all ${\bf z}\in\{0,1\}^{n}$; then ${\bf s}\in D_{k}$, thus ${\bf s}\in \bigcup_{i}D_{i}$. Now, the probability of each event $D_{i}$ can be expressed as $\mathbb{P}(D_{i})=F_{S_{i}}(1-\delta)-F_{S_{i}}(\delta)$, where $F_{S_{i}}$ is the CDF of $S_{i}$ (actually $S_{i}^{A}$) which distributes as $\textbf{Beta}(k(x_{i}+\alpha_i),k(y_{i}+\beta_{i}))$. Since $S_{i}$ converges to a Bernoulli when $k\rightarrow0$, it follows that $\mathbb{P}(D_{i})$ as well. Then we have
\begin{eqnarray*}
1-\mathbb{P}(E_{3}) = \mathbb{P}(E_{3}^{c})=\mathbb{P}\left(\bigcup_{i}D_{i}\right)
\le \sum_{i}\mathbb{P}(D_{i}) \overset{k\rightarrow0}{\rightarrow} 0.
\end{eqnarray*}
Then $\mathbb{P}(E_{3})\rightarrow 1$ as $k\rightarrow0$, and therefore $\mathbb{P}(E_{1})+\mathbb{P}(E_{2})\rightarrow 0$ concluding the proof.

\section{Zero-sum game as an LP}\label{app:lpgame}
The optimization game can be expressed as
\begin{equation} \label{eq:app_gamelp}
    \begin{aligned}
        \underset{\boldsymbol{\sigma_B}}{\min\;\;\;\;} & \underset{\boldsymbol{\sigma_A}}{\max}
        & & \boldsymbol{\sigma_B^{T}}\boldsymbol{P} \boldsymbol{\sigma_A} \\
        \text{s.t.}& & & \boldsymbol{e^{T}_A}\boldsymbol{\sigma_{A}} = 1 \\
        & & & \boldsymbol{\sigma_A} \geq 0\\
        & & & \boldsymbol{e^{T}_B}\boldsymbol{\sigma_{B}} = 1 \\
        & & & \boldsymbol{\sigma_B} \geq 0.
    \end{aligned}
\end{equation}
Note that the inner problem of~\eqref{eq:app_gamelp}
\begin{equation} \label{eq:app_gamelp2}
    \begin{aligned}
        \underset{\boldsymbol{\sigma_A}}{\max}
        & & \boldsymbol{\sigma_B^{T}}\boldsymbol{P} \boldsymbol{\sigma_A} \\
        \text{s.t.}& & \boldsymbol{e^{T}_A}\boldsymbol{\sigma_{A}} = 1 \\
        & & \boldsymbol{\sigma_A} \geq 0
    \end{aligned}
\end{equation}
has the following dual
\begin{equation} \label{eq:app_gamelp3}
    \begin{aligned}
        \underset{u}{\min}
        & & u\qquad \\
        \text{s.t.}& & u\boldsymbol{e_A}\ge\boldsymbol{P^{T}}\boldsymbol{\sigma_{B}}.
    \end{aligned}
\end{equation}
Then, we can replace~\eqref{eq:app_gamelp3} into the primal of the inner problem in~\eqref{eq:app_gamelp} and get
\begin{equation} \label{eq:app_gamelp4}
    \begin{aligned}
        \underset{u,\boldsymbol{\sigma_B}}{\min}
        & & u\qquad\quad \\
        \text{s.t.} & & u\boldsymbol{e_A}-\boldsymbol{P^{T}} \boldsymbol{\sigma_B} \ge0  \\
        & & \boldsymbol{e^{T}_{B}}\boldsymbol{\sigma_{B}} = 1 \\
        & & \boldsymbol{\sigma_B} \geq 0
    \end{aligned}
\end{equation}
which is a single LP. The LP for obtaining the first player equilibrium strategies is analogous.%In order to get the optimal decision variables for the first player, we can use CS conditions, or simply solve Optimization Problem~\eqref{eq:app_gamelp2} with the optimal decision variables for $\boldsymbol{\sigma_{B}}$ obtained from Optimization Problem~\eqref{eq:app_gamelp4}.

\section{Demonstration of Claim~\ref{cla:sumq}}\label{proof:cla:sumq}
We have
\begin{eqnarray}\label{eq:sumq}
m=\sum_{i=1}^{n}qx^r_i
=q\sum_{i=1}^{n}x^r_i
=q\left(1-\sum_{i=1}^{n}\frac{\lfloor qx_i\rfloor}{q}\right)=q-\sum_{i=1}^{n}\lfloor qx_i\rfloor=\sum_{i=1}^{n}(qx_i-\lfloor qx_i\rfloor),
\end{eqnarray}
where from the expression before the last equality, it can be seen that the result is a subtraction of integers which results on an integer. From the last expression of Equation~\eqref{eq:sumq} we can see that all the arguments inside the sum are non negative, and are strictly less than one. Therefore, the sum can be at least $0$, and at most $n-1$.

\section{Demonstration of Lemma~\ref{lem:algp}}\label{proof:lem:algp}
Let us denote as ${\bf w^{(k)}}$ the value of the vector ${\bf w}$ at iteration $k$ in Algorithm~\ref{Alg_Discret}, and similar for $t^{(k)}$ with respect to $t$ in the $k$\textsuperscript{th} iteration. Before going into the demonstrations of (i) and (ii), we will show the following lemma:
\begin{lemma}\label{lem:z01}
For each iteration $k$ of Algorithm~\ref{Alg_Discret}, it holds that $z_{j}^{(k)}\neq z_{j}^{(h)}$ for all $h>k$ where $j$ is one of the components that reaches the minimum condition in line 6 of Algorithm~\ref{Alg_Discret}.
\end{lemma}
\proof{Proof.} 
Consider we are in the $k$\textsuperscript{th} iteration of Algorithm~\ref{Alg_Discret}. We have some ${\bf w^{(k)}}$, where we assume that ${\bf w^{(k)}}\in[0,1]^{n}$ and $\sum_{i=1}^{n}w^{(k)}_{i}=m$ (note this is accomplished for the case $k=1$, while we will show this for the case $k+1$). Because ${\bf w^{(k)}}\in[0,1]^{n}$ and $\sum_{i=1}^{n}w^{(k)}_{i}=m$, if ${\bf w^{(k)}}$ has at least one fractional component (which is the interesting case in which we go inside the loop of Algorithm~\ref{Alg_Discret}), it must be the case that ${\bf w^{(k)}}$ has at most $n-m-1$ components with zeros, and at most $m-1$ components with ones. Then, line 4 of the algorithm forces $z^{(k)}_{i}=w^{(k)}_{i}$ for each component $i$ such that $w^{(k)}_{i}\in\{0,1\}$. Then, the Algorithm tries to find the maximum magnitude from which we can move from the vector ${\bf w^{(k)}}$ in the direction ${\bf w^{(k)}}-{\bf z^{(k)}}$ (clearly the latter is non-zero since ${\bf w^{(k)}}$ has at least one fractional component while ${\bf z^{(k)}}\in\{0,1\}^{n}$). We then look for the maximum value of $t$ such that ${\bf w^{(k)}}+t({\bf w^{(k)}}-{\bf z^{(k)}})\in[0,1]^{n}$, this is equivalent to (I) $w^{(k)}_i+t(w^{(k)}_i-z^{(k)}_i)\ge0$ and (II) $w^{(k)}_i+t(w^{(k)}_i-z^{(k)}_i)\le1$ for all components $i$ where ${\bf w^{(k)}}$ is fractional, since for the other components the right term is zero. (I) is equivalent to $\frac{w^{(k)}_{i}}{z^{(k)}_{i}-w^{(k)}_{i}}\ge t$ if $z^{(k)}_{i}>w^{(k)}_{i}$, and $\frac{w^{(k)}_{i}}{z^{(k)}_{i}-w^{(k)}_{i}}\le t$ if $z^{(k)}_{i}<w^{(k)}_{i}$. The latter case always holds, then (I) can be written simply as $\frac{w^{(k)}_{i}}{1-w^{(k)}_{i}}\ge t$ if $z^{(k)}_{i}>w^{(k)}_{i}$. As for (II), this is equivalent to $\frac{1-w^{(k)}_{i}}{w^{(k)}_{i}-z^{(k)}_{i}}\le t$ if $z^{(k)}_{i}>w^{(k)}_{i}$, and $\frac{1-w^{(k)}_{i}}{w^{(k)}_{i}-z^{(k)}_{i}}\ge t$ if $z^{(k)}_{i}<w^{(k)}_{i}$. The former case can be eliminated since always holds, then (II) can be reduced to $\frac{1-w^{(k)}_{i}}{w^{(k)}_{i}}\ge t$ if $z^{(k)}_{i}<w^{(k)}_{i}$. Putting together (I) and (II), we get the expression for $t^{(k)}$ in line 6 of Algorithm~\ref{Alg_Discret}. Note that all the expressions inside the arguments of the $\min\{\}$'s have strictly positive arguments, therefore $t^{(k)}>0$, because $w^{(k)}_{i}\in(0,1)$ for those components. Also, since ${\bf w^{(k)}}$ has some fractional component, $t^{(k)}$ is well defined. We have also shown that ${\bf w^{(k+1)}}\in[0,1]^n$. In addition, as it will be used later, we can show that $\sum_{i=1}^{n}w^{(k+1)}_{i}=m$, since $\sum_{i=1}^{n}w^{(k)}_{i}=\sum_{i=1}^{n}z^{(k)}_{i}=m$. Consider that this holds $j$ to be the index which can follow either one of the two following cases denoted by (a) and (b). In (a), $j$ is such that $t^{(k)}=\frac{w^{(k)}_{j}}{1-w^{(k)}_{j}}$ where $z^{(k)}_{j}=1$ then $w^{(k+1)}_{j}=w^{(k)}_{j}+t(w^{(k)}_{j}-z^{(k)}_{j})=w^{(k)}_{j}+\frac{w^{(k)}_{j}}{1-w^{(k)}_{j}}(w^{(k)}_{j}-1)=0$. Then, in the next iteration, if ${\bf w^{(k+1)}}\in\{0,1\}^{n}$ then $z^{(k+1)}_{j}=0\neq 1=z^{(k)}_{j}$. Contrarily, if ${\bf w^{(k+1)}}\not\in\{0,1\}^{n}$, then there is a fractional component, and by induction of the same arguments given, it will hold that $0=z^{(k+1)}_{j}=w^{(k+1)}_{j}$, and $0=w^{(k+2)}_{j}=w^{(k+1)}_{j}$. Therefore $z^{(h)}_{j}\neq z^{(k)}_{j}$ for all $h>k$. In case (b), $j$ is the component such that $t^{(k)}=\frac{1-w^{(k)}_{j}}{w^{(k)}_{j}}$ where $z^{(k)}_{j}=0$ then $w^{(k+1)}_{j}=w^{(k)}_{j}+t(w^{(k)}_{j}-z^{(k)}_{j})=w^{(k)}_{j}+\frac{1-w^{(k)}_{j}}{w^{(k)}_{j}}(w^{(k)}_{j}-0)=1$. Then, in the next iteration, if ${\bf w^{(k+1)}}\in\{0,1\}^{n}$ then $z^{(k+1)}_{j}=1\neq 0=z^{(k)}_{j}$. On the contrary, if ${\bf w^{(k+1)}}\not\in\{0,1\}^{n}$, then there is a fractional component, and by induction of the same arguments given, it will hold that $1=z^{(k+1)}_{j}=w^{(k+1)}_{j}$, and $1=w^{(k+2)}_{j}=w^{(k+1)}_{j}$. Therefore $z^{(h)}_{j}\neq z^{(k)}_{j}$ for all $h>k$.
\Halmos
\endproof 

Proof of (i). As shown in the proof of Lemma~\ref{lem:z01}, at each iteration $k$ inside the loop of Algorithm~\ref{Alg_Discret}, one fractional component of ${\bf w^{(k)}}$ is fixed to either $0$ or $1$ when moving to the next ${\bf w^{(k+1)}}$. Since ${\bf w^{(k)}}\in\mathbb{R}^{n}$, we can have at most $n$ iterations, and therefore $|\mathcal{Z}|\le n$.

Proof of (ii). From line 7 of Algorithm~\ref{Alg_Discret} we can see that ${\bf w^{(k+1)}}={\bf w^{(k)}}+t^{(k)}({\bf w^{(k)}}-{\bf z^{(k)}})$, which is equivalent to ${\bf w^{(k)}}=\frac{1}{1+t^{(k)}}{\bf w^{(k+1)}}+\frac{t^{(k)}}{1+t^{(k)}}{\bf z^{(k)}}$. Then, we have that for $k=1$, it holds that ${\bf w^{(1)}}=\frac{1}{1+t^{(1)}}{\bf w^{(2)}}+\frac{t^{(1)}}{1+t^{(1)}}{\bf z^{(1)}}$, where ${\bf w^{(1)}}={\bf y}$, the input point. Then,  ${\bf w^{(1)}}=\frac{1}{1+t^{(1)}}\left(\frac{1}{1+t^{(2)}}{\bf w^{(3)}}+\frac{t^{(2)}}{1+t^{(2)}}{\bf z^{(2)}}\right)+\frac{t^{(1)}}{1+t^{(1)}}{\bf z^{(1)}}=\frac{1}{1+t^{(1)}}\left(\frac{1}{1+t^{(2)}}\left(\frac{1}{1+t^{(3)}}{\bf w^{(4)}}+\frac{t^{(3)}}{1+t^{(3)}}{\bf z^{(3)}}\right)+\frac{t^{(2)}}{1+t^{(2)}}{\bf z^{(2)}}\right)+\frac{t^{(1)}}{1+t^{(1)}}{\bf z^{(1)}}$ and so on until the last point which if we denote this by ${\bf w^{(|\mathcal{Z}|)}}$, in which case the equation is ${\bf w^{(1)}}={\bf w^{(|\mathcal{Z}|)}}\prod_{j=1}^{|\mathcal{Z}|}\frac{1}{1+t^{(j)}}+\sum_{k=1}^{|\mathcal{Z}|-1}{\bf z^{(k)}}\frac{t^{(k)}}{1+t^{(k)}}\prod_{j=1}^{k-1}\frac{1}{1+t^{(j)}}$. Finally, to show that the weights are strictly positive, it suffices to note that $t^{(k)}>0$ for all $k\in\{1,\dots,|\mathcal{Z}|\}$. The proof of the latter was done in particular in the proof of Lemma~\ref{lem:z01}, concluding the proof. %We will use the fact that $\sum_{i=1}^{n}w^{(k)}_{i}=m$ for all $k\in\{1,\dots,|\mathcal{Z}|\}$, this can be easily seen since $w^{(k)}$ are build as convex combinations of elements in $\mathcal{Y}$, and the starting point $y$. Then, at each iteration, inside the loop of Algorithm~\ref{Alg_Discret}, it must be the case that $w$ has at most $n-m-1$ zeros, $m-1$ ones, and at least one fractional component. Then, when solving line 4 of Algorithm~\ref{Alg_Discret}, all components of $w$ with zeros will end up in the same components of the resulting $z$ with zeros. Similarly, components with ones in $w$ will result in components with ones in $z$. Because of the latter, in the computation of $t$ (line 6 of the algorithm), there will never be a component $i$ where $w_{i}=0$ and $z_{i}=1$, or $w_{i}=1$ and $z_{i}=0$ which implies that all the elements analyzed inside the $\min$ arguments in line 6 of Algorithm~\ref{Alg_Discret} are strictly positive. Note also that since $w$ has at least one fractional component, $t$ is well defined. As a consequence, $t>0$. Then $t^{(k)}>0$ for every $k\in\{1,\dots,|\mathcal{Z}|\}$, which implies that all weights of the convex combination are strictly positive, concluding the proof.

\section{Showing that output points in $\mathcal{\overline{Z}}$ are \textit{minimal}}\label{app:minimal}

We want to formally show that given ${\bf x}\in\Delta_n$ and $q\in\mathbb{Z}_{+}$, and apply Algorithm~\ref{Alg_Discret} using as input ${\bf y}=q{\bf x^{r}}=q{\bf x}-\lfloor q{\bf x}\rfloor$ (where the floor function is applied to each component) from which we obtain $\mathcal{Z}$ and construct $\mathcal{\overline{Z}}$, there can not exist a set $\mathcal{W}\neq\mathcal{\overline{Z}},\subseteq D^{q}(\Delta_n)$ such that (a) ${\bf x}\in \text{Conv}(\mathcal{W})$ and (b) $\text{Conv}(\mathcal{W})\subsetneq \text{Conv}(\mathcal{\overline{Z}})$. Let us assume by contradiction that such set $\mathcal{W}$ exists. The latter implies the following: (i) $\mathcal{W}\subseteq\mathcal{\overline{Y}}$ (where $\mathcal{\overline{Y}}\coloneqq\{\frac{\lfloor q{\bf x}\rfloor}{q}+\frac{{\bf y}}{q}|{\bf y}\in\mathcal{Y}\}$), (ii) $D^q(\Delta_n)\cap(\text{Conv}(\mathcal{\overline{Z}})\setminus\mathcal{\overline{Z}})=\varnothing$, (iii) $\mathcal{W}\subsetneq\mathcal{\overline{Z}}$, (iv) elements in $\mathcal{\overline{Z}}$ are linearly independent, and (v) ${\bf x}\not\in\text{Conv}(\mathcal{W})$ which contradicts (a). For ease of notation, let us define ${\bf x^b}\coloneqq\frac{\lfloor q{\bf x}\rfloor}{q}$. To show (i), it must be the case that all elements ${\bf w}\in\mathcal{W}$ are such that $w_{i}\in\left\{x^b_i,x^b_i+\frac{1}{q}\right\}$ (which is accomplished by each $i$\textsuperscript{th} components of the points in $\mathcal{\overline{Z}}$). If not, then (b) can not be true since there would be a point in $\mathcal{W}$ which can not be generated as a convex combination of points in $\mathcal{\overline{Z}}$. Using a similar argument, it must be the case that all elements in $\mathcal{W}$ have $m$ components equal to the respective component of $1/q+{\bf x^b}$ and $n-m$ components with the value equal to the respective component of ${\bf x^b}$ (recall that all vectors in $\mathcal{Z}$ have $m$ ones and $n-m$ zeros, where $m=\sum_{i=1}^{n}qx^r_i$, and so all elements in $\mathcal{\overline{Z}}$ have $m$ components with value equal to the respective component of $1/q+{\bf x^b}$ and $n-m$ components equal to the respective component ${\bf x^b}$). If otherwise, then for any {\boldmath$\mu$}$ \in\Delta_{|\mathcal{\overline{Z}}|}$ it holds that $\sum_{i=1}^{n}\sum_{k=1}^{|\mathcal{\overline{Z}}|}\mu_{k}z^{(k)}_{i}=\sum_{k=1}^{|\mathcal{\overline{Z}}|}\mu_{k}\sum_{i=1}^{n}z^{(k)}_{i}=\sum_{k=1}^{|\mathcal{\overline{Z}}|}\mu_{k} m=m\sum_{k=1}^{|\mathcal{\overline{Z}}|}\mu_{k}=m$ (where ${\bf z^{(k)}}$ denotes the $k$\textsuperscript{th} element added to the set $\mathcal{\overline{Z}}$ when running Algorithm~\ref{Alg_Discret}), i.e., any convex combination of the points in $\mathcal{\overline{Z}}$ will have components that add up to $m$. Therefore, if (b) holds, it must be the case that all elements in $\mathcal{W}$ have $m$ components equal to the respective components of $1/q+{\bf x^{b}}$, and $n-m$ components equal to the respective component of ${\bf x^b}$, thus $\mathcal{W}\subseteq{\mathcal{Y}}$. For (ii), note that if by contradiction there is ${\bf v}\in D^q(\Delta_n)\cap(\text{Conv}(\mathcal{\overline{Z}})\setminus\mathcal{\overline{Z}})$, then it must exist {\boldmath  $\mu$}$\in\Delta_{|\mathcal{\overline{Z}}|}$ such that ${\bf v}=\sum_{k=1}^{|\mathcal{\overline{Z}}|}\mu_{k}{\bf z^{(k)}}$ where at least there are two indexes, $l,h\in\{1,\dots,|\mathcal{\overline{Z}}|\}$ for which $\mu_{l},\mu_{h}\in(0,1)$ (otherwise there would be a single index $j$ where $\mu_{j}=1$ while $\mu_{i}=0$ $\forall i\neq j$, and ${\bf v}$ would be equal to ${\bf z^{(j)}}$ which can not be possible). Consider a component $i$ such that $z^{(l)}_{i}\neq z^{(h)}_{i}$ (notice that either $(z^{(l)}_{i},z^{(h)}_{i})=(x^{b}_{i},1/q+x^{b}_{i})$ or $(z^{(l)}_{i},z^{(h)}_{i})=(1/q+x^{b}_{i},x^{b}_{i})$), then it must be the case that $\sum_{k=1}^{|\mathcal{\overline{Z}}|}\mu_{k}z^{(k)}_{i}\in(x^{b}_{i},1/q+x^{b}_{i})$, which is a contradiction since ${\bf v}\in D^{q}(\Delta_n)$, which shows (ii). Note that (ii) states that every convex combination of the points in $\mathcal{\overline{Z}}$ which is not an extreme point of $\text{Conv}(\mathcal{\overline{Z}})$ (i.e. the convex combination weights can not be one), the resulting vector has at least one component $i$ with value in the interval $(x^b_i,1/q+x^b_i)$, and therefore the convex combination does not belong to the simplex lattice $D^q(\Delta_n)$. Then it must holds that $\mathcal{W}\subseteq\mathcal{\overline{Z}}$, since otherwise there would exist ${\bf w}\in{\mathcal{W}}\setminus\mathcal{\overline{Z}}$. Note that because of (b) we have $\mathcal{W}\subseteq\text{Conv}(\mathcal{W})\subseteq\text{Conv}(\mathcal{\overline{Z}})$, and then it holds that ${\bf w}\in\text{Conv}(\mathcal{\overline{Z}})\setminus\mathcal{\overline{Z}}$. Adding the fact that $\mathcal{W}\subseteq D^q(\Delta_n)$, leads to ${\bf w}\in D^q(\Delta_n)\cap(\text{Conv}(\mathcal{\overline{Z}})\setminus\mathcal{\overline{Z}})$, which contradicts (ii). Then (iii) follows from $\mathcal{W}\subseteq\mathcal{\overline{Z}}$ and $\mathcal{W}\neq\mathcal{\overline{Z}}$. For (iv), if it is assumed by contradiction that the vectors in $\mathcal{\overline{Z}}$ are linearly dependent, then there would exist {\boldmath $\mu$}$\in\mathbb{R}^{|\mathcal{\overline{Z}}|}\setminus\{{\bf 0}\}$ such that $\sum_{k=1}^{|\mathcal{\overline{Z}}|}\mu_{k}{\bf z^{(k)}}=0$. Consider $h\coloneqq\min\{k\in\{1,\dots,|\mathcal{\overline{Z}}|\}|\mu_{k}>0\}$. It holds that ${\bf z^{(h)}}$ can be written as a linear combination of vectors in $\mathcal{\overline{Z}}$ of higher indexes. This last cannot be possible due to Lemma~\ref{lem:z01}, since we know that there is a component, say $j$, such that $z^{(h)}_{j}\neq z^{(l)}_{j}$ for all $l>h$, concluding (iv). Finally, to show (v), if we look at the convex combination of ${\bf x}$ with respect to the vertices in $\mathcal{\overline{Z}}$, from Lemma~\ref{lem:algp} we have that all weights are positive, and from (iv) we have that this is the only way to generate ${\bf x}$. As a result, if we remove any element in $\mathcal{\overline{Z}}$, then we cannot generate ${\bf x}$ as a convex combination. In other words, because of (iii), we cannot generate {\bf $x$} as a convex combination of the elements in $\mathcal{Y}$, i.e. ${\bf x}\not\in\text{Conv}(\mathcal{W})$, which concludes the proof.

\section{Majority System Unbounded Equilibrium}\label{app:num_res_ub}

As an hypothetical exercise, we computed the unbounded equilibrium of the game under the MS. Recall that this equilibrium can be computed in closed form solution by using Proposition~\eqref{pro:CUB} with $\mathcal{I}^{*}=\mathcal{I}$ (see Equations~\eqref{eq:x_CUB} and \eqref{eq:y_CUB}). The results of the unbounded equilibrium are computed for the same instance presented in Table~\ref{tab:eqPSdet}, and its results are illustrated in Figure~\ref{fig:Equilibrio_UB}. In this, region' biases ({\boldmath $\alpha$},{\boldmath $\beta$}) are depicted with empty circles whereas the regions' biases plus candidates' strategies ({\boldmath $\alpha$}$+{\bf x}$,{\boldmath $\beta$}$+{\bf y}$) are depicted with filled circles.
%In particular, the plot in Figure~\ref{fig:Equilibrio_UB} shows the candidates efforts as the differences between $(x+\alpha,y+\beta)$ (represented by filled circles) and the bias parameters $(\alpha,\beta)$ (represented by circles). 
The arrows in Figure~\ref{fig:Equilibrio_UB} represent the effort put by both candidates, where the x-component (y-component) corresponds to candidate $A$ ($B$). It can be seen that only the three largest regions receive positive effort. On the contrary, smaller regions have negative efforts, therefore acting as \textit{lenders} to the former regions. 
%Therefore, the arrows in Figure~\ref{fig:Equilibrio_UB} represent for each region the (unbounded) equilibrium efforts of both candidates. It can be seen that regions with more effort, the ones with lower index (or equivalently the ones with more votes), have more investments by the candidates. On the contrary, the regions with less votes have negative efforts, acting as \textit{lenders} to the regions with more votes. 
It is interesting to observe the perfect linear relation between the efforts plus bias parameters of one candidate with respect to the other (see the straight line in Figure~\ref{fig:Equilibrio_UB}). More specifically, the linear relation $y_i+\beta_i=(x_i+\alpha_i)\left(\frac{1+\sum_{j}\beta_j}{1+\sum_{j}\alpha_j}\right)$ holds for every region $i$. %Unfortunately, candidates' efforts in the unbounded equilibrium are negative in some regions, therefore this does not match the equilibrium of the game with non negativity constraints.

%\vspace{-1cm}

\begin{figure}[h]
\begin{center}
\addtolength{\leftskip}{-1cm}
\includegraphics[scale = 0.4]{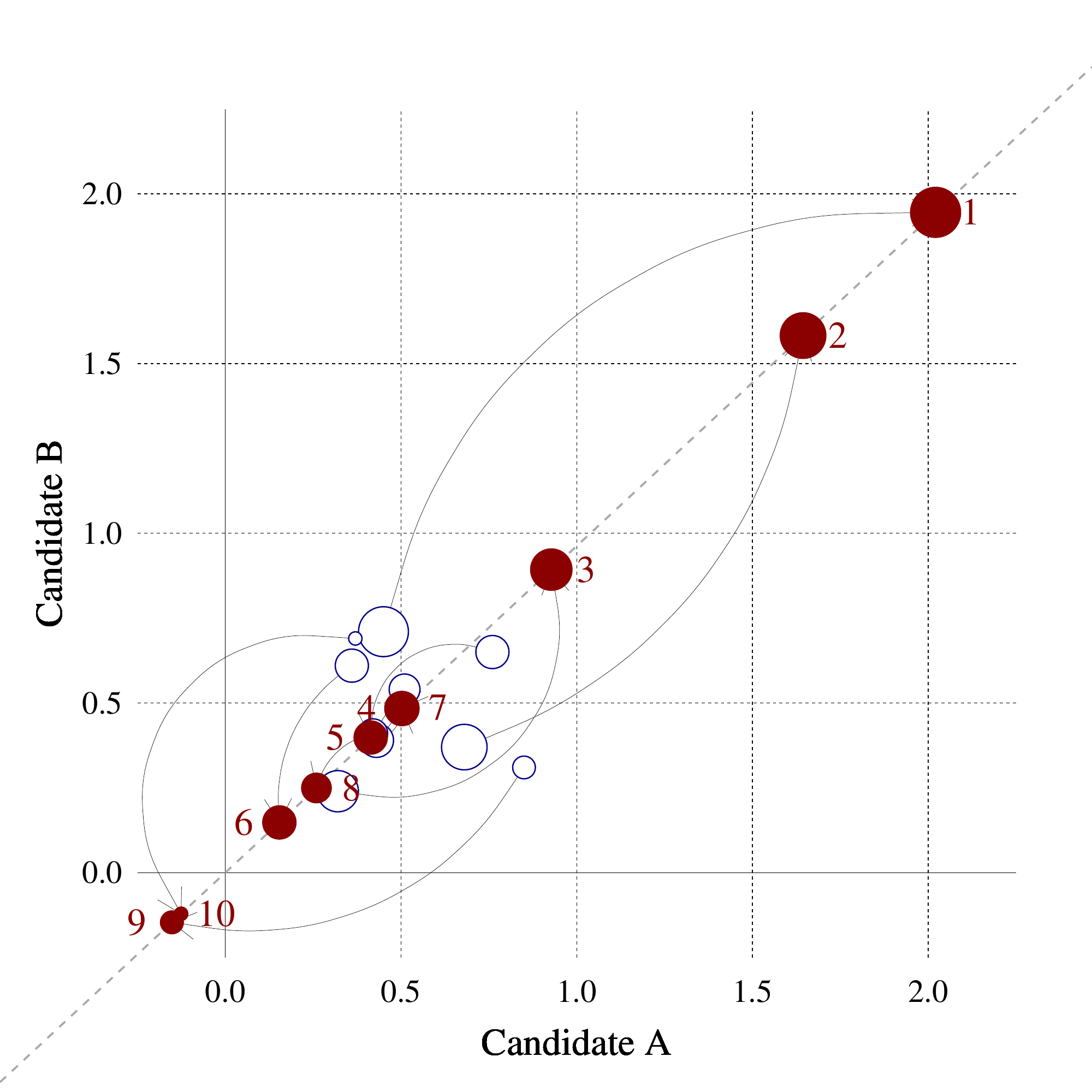}
\end{center}
\caption{Unbounded equilibrium for instance in Table~ \ref{tab:eqPSdet}. \textbf{Circles}, ``$\circ$'', represent the bias parameters $(\alpha,\beta)$; \textbf{filled circles}, ``$\bullet$'', represent $(x+\alpha,y+\beta)$ at the unbounded equilibrium. Sizes are proportional to the number of votes of the region. The dotted diagonal represents the diagonal with slope $\frac{1 + \sum \beta}{1 + \sum \alpha}$.}
\label{fig:Equilibrio_UB}
\end{figure}

\end{APPENDICES}
%%%%%%%%%%%%%%%%%
\end{document}